\documentclass{aa}
\usepackage{natbib}
\usepackage[varg]{txfonts}
\usepackage{graphicx}
\usepackage{multirow}

\usepackage[usenames,dvipsnames]{color}
\definecolor{mygreen}{rgb}{0,0.5,0}
\definecolor{myorange}{rgb}{0.5,0.5,0}
\definecolor{myred}{rgb}{0.5,0,0}

\def\ms{\hbox{\,m\,s$^{-1}$}}         
\def\m2s2{\hbox{\,m$^{2}$\,s$^{-2}$}} 

\begin{document}

\title{Accounting for stellar activity signals in radial-velocity data by using change point detection techniques  \thanks{Based on observations collected at the La Silla Paranal Observatory,
ESO (Chile), with the HARPS spectrograph at the 3.6-m telescope.}}

\titlerunning{CPD exoplanets}
\authorrunning{U. Simola, A. Bonfanti, X. Dumusque, J. Cisewski-Kehe, S. Kaski and J. Corander}

\author{U. Simola \inst{1}
            \thanks{\email{umberto.simola@gmail.com}}
            \and A. Bonfanti \inst{2}
            \and X. Dumusque\inst{3}
            \thanks{Branco Weiss Fellow--Society in Science (url: \protect\url{http://www.society-in-science.org})}    
            \and J. Cisewski-Kehe\inst{4}
             \and S. Kaski\inst{5}
             \and J. Corander \inst{1,6}
            }
\institute{Department of Mathematics and Statistics, University of Helsinki, Helsinki, Finland
              \and Space Research Institute, Austrian Academy of Sciences, Schmiedlstrasse 6, A-8042 Graz, Austria
              \and Observatoire de Gen\`eve, Universit\'e de Gen\`eve, 51 ch. des Maillettes, CH-1290 Versoix, Switzerland 
              \and Department of Statistics, University of Wisconsin - Madison, USA
              \and Department of Computer Science, Aalto University, Espoo, Finland
              \and Department of Biostatistics, University of Oslo, Oslo, Norway
              }
\date{Received XXX; accepted XXX}

\abstract
{Active regions on the photosphere of a star have been the major obstacle for detecting Earth-like exoplanets using the radial velocity (RV) method. A commonly employed solution for addressing stellar activity is to assume a linear relationship between the RV observations and the activity indicators along the entire time series, and then remove the estimated contribution of activity from the variation in RV data (overall correction method). However, since active regions evolve on the photosphere over time, correlations between the RV observations and the activity indicators will correspondingly be anisotropic.}
{We present an approach that recognizes the RV locations where the correlations between the RV and the activity indicators significantly change in order to better account for variations in RV caused by stellar activity.}
{The proposed approach uses a general family of statistical breakpoint methods, often referred to as change point detection (CPD) algorithms; several implementations of which are available in \texttt{R} and \texttt{python}. A thorough comparison is made between the breakpoint-based approach and the overall correction method. To ensure wide representativity, we use measurements from real stars that have different levels of stellar activity and whose spectra have different signal-to-noise ratios.}
{When the corrections for stellar activity are applied separately to each temporal segment identified by the breakpoint method, the corresponding residuals in the RV time series are typically much smaller than those obtained by the overall correction method. Consequently, the generalized Lomb-Scargle periodogram contains a smaller number of peaks caused by active regions. The CPD algorithm is particularly effective when focusing on active stars with long time series, such as $\alpha$ Cen B. In that case, we demonstrate that the breakpoint method improves the detection limit of exoplanets  by 74\% on average with respect to the overall correction method.}
{CPD algorithms provide a useful statistical framework for estimating the presence of change points in a time series. Since the process underlying the RV measurements generates anisotropic data by its intrinsic properties, it is natural to use CPD to obtain cleaner signals from RV data. We anticipate that the improved exoplanet detection limit may lead to a widespread adoption of such an approach. Our test on the HD\,192310 planetary system is encouraging, as we confirm the presence of the two hosted exoplanets and we determine orbital parameters consistent with the literature, also providing much more precise estimates for HD\,192310 c.}


\maketitle

\section{Introduction} \label{intro}

The radial velocity (RV) method \citep[e.g.,][]{mayor1995jupiter, lovis2010radial, hatzes2016radial} is one of the most successful techniques for detecting extrasolar planets \citep[e.g.,][]{fischer2016state}. Nevertheless, perturbations in the RV data caused by different types of stellar signals such as stellar oscillations, granulations, and the presence of active photospheric regions continuously plague the exoplanet community \citep[e.g.,][]{Saar-1997b, Queloz-2001, Lindegren-2003, Desort-2007, Meunier-2010a, Dumusque-2011a, Dumusque-2016a, meunier2017variability, dumusque2018measuring}. In particular, stellar activity in the form of spots and faculae represents the main limitation in the full characterization of Earth-like exoplanets in RV surveys \citep[e.g.,][]{Saar-1997b, Meunier-2010a, Dumusque-2014b, dumusque2018measuring}. New generation spectrographs such as the EXtreme PREcision Spectrometer \citep[EXPRES,][]{jurgenson16} and the Echelle SPectrograph for Rocky Exoplanet and Stable Spectroscopic Observations \citep[ESPRESSO,][]{Pepe-2014} have recently been constructed to solve the issue of instrumental signal, and therefore to provide data only affected by stellar and planetary signals. However, disentangling the perturbations due to stellar activity from the RV variations caused by small size exoplanets remains the most crucial challenge \citep[e.g.,][]{davis2017insights, dumusque2018measuring, reiners2018carmenes, simola2019measuring}, as the RV variations induced by active regions are an order of magnitude larger than those expected from Earth-like exoplanets \citep[e.g.,][]{dumusque2018measuring}.

Efforts have been made to model the signals caused by different sources of stellar variability within the RV time series \citep[e.g.,][]{tuomi2013habitable,rajpaul2015ghost,davis2017insights, simola2019measuring}. Several solutions have been successfully proposed in order to deal with stellar oscillations and granulation phenomena, such as: calculating stellar evolution sequences \citep[e.g.,][]{christensen1995modeling}; fitting a  two-level structure tracking (TST) algorithm based on a two-level representation of granulation \citep{del2004solar}; using daytime spectra of the Sun in order to measure the solar oscillations \citep[e.g.,][]{kjeldsen2008amplitude, lefebvre2008variations}; and characterizing the statistical properties of magnetic activity cycles focusing on HARPS observations \citep[e.g.,][]{Pepe-2011, Dumusque-2011c}. However, properly modeling the other sources of stellar activity remains extremely challenging \citep[e.g.,][]{nava2019exoplanet}. In the present work, we deal with the cross correlation function (CCF) that is derived from the stellar spectrum \citep[e.g.,][]{Hatzes1996, hatzes2000radial, fiorenzano2005line}. As it is well known, the CCF barycenter estimates the RV of the star. The asymmetry and the full width at half maximum (FWHM) of the CCF give a strong indication for stellar activity, meaning that variations in RV are caused by active regions rather than by an exoplanet \citep[e.g.,][]{Hatzes1996, Queloz-2001, Boisse-2011, Figueira-2013,  simola2019measuring}. Several solutions have been successfully proposed for mitigating stellar activity perturbations when working with RV measurements, including: decorrelating RV data against activity indicators such as $\log R'_{HK}$ \citep[e.g.,][]{wilson1968flux, noyes1984rotation} or $H\alpha$ \citep[e.g.,][]{Robertson-2014}; modeling stellar activity by fitting Gaussian processes \citep[GPs,][]{rasmussen05, haywood2014planets, rajpaul2015ghost}; or moving averages \citep[e.g.,][]{tuomi2013habitable} to the RV data.
A common statistic employed for identifying changes in the shape of the CCF is the bisector span \citep[e.g.,][]{Hatzes1996, Queloz-2001}.

As already pointed out, disentangling the signal caused by stellar active regions and the signal due to an Earth-like planet is extremely challenging, but some differences in the characteristics of the two signals may be of help. Active regions produced by spots and faculae evolve on the star's photosphere and can generate RV variations that spread from a few days to several weeks, depending on the rotation period of the star and the intensity of the magnetic cycle \citep[e.g.,][]{noyes1984rotation, Dumusque-2014b, davis2017insights, dumusque2018measuring, nava2019exoplanet}. While exoplanets produce a Doppler-shift on the CCF {without} modifying its shape or its width, active regions produce variations in both the asymmetry and the FWHM of the CCF, whose effect is a variation in the barycenter of the CCF, and therefore in the estimated median of the RV CCF \citep[see e.g.,][]{Hatzes1996, Queloz-2001, Boisse-2011, Figueira-2013, Dumusque:2017aa, simola2019measuring, thompson2020spectral}, hereafter indicated as $\overline{RV}$. We refer the reader to Table~\ref{tab:RVdef} for the main RV-related symbols we adopt throughout paper. Moreover, while an exoplanet produces a persistent Keplerian signal, the signal produced by active regions is not persistent as it waxes, wanes, and changes as a function of time \citep[e.g.,][]{fischer2016state, davis2017insights, dumusque2018measuring, thompson2020spectral}. 

\begin{table}
 \caption{Legend of the RV-related symbols adopted in this paper. See the text for further details.}
 \label{tab:RVdef}
\centering
\begin{tabular}{lc}
\hline\hline
Symbol & Definition  \\
\hline                        
RV & Radial Velocity   \\
$\overline{RV}$  &  Median of the RV CCF based on an SN fit  \\
$RV^*_{\text{activity}}$  &  $\beta_{0} + \beta_{1} A + \beta_{2} \gamma + \beta_{3} \mathrm{FWHM_{SN}} +\epsilon$   \\
$\Delta RV^*_{bp}$  &  $\overline{RV} - \sum_{k=1}^{D+1} RV^*_{\text{activity, } k}$ \\
$\Delta RV^*_{oc} $  &  $\overline{RV} - RV^*_{\text{activity}}$   \\
$\Delta RV_{bp}^p$ & $(\overline{RV} + RV_K) - \sum_{k=1}^{D+1} RV^*_{\text{activity, } k}$ \\
$\Delta RV_{oc}^p$ & $(\overline{RV} + RV_K) - RV^*_{\text{activity}}$ \\
\hline
\end{tabular}
\end{table}

By using the stellar activity parameters estimated from the study of the CCF, a linear model is often proposed in order to decorrelate the RV data against the RV variations due to stellar activity \citep[e.g.,][]{Dumusque:2017aa, Feng:2017aa, simola2019measuring}. Rather than using a normal fit to the CCF and then calculating the bisector span, \cite{simola2019measuring} used a skew normal (SN) fit to the CCF. The SN distribution allows a parameter to be specified, hereafter $\gamma$, which describes the asymmetry of the CCF \citep[e.g.,][]{simola2019measuring, adcock2020selective}. \cite{simola2019measuring} suggest using the median $\overline{RV}$ as the barycenter of the SN fit to the CCF.

Given the benefits of fitting an SN to the CCF as emphasized by \citet{simola2019measuring}, the model we employed in the analyses of this work is defined as follows:
\begin{equation}
RV^{*}_{\text{activity}}= \beta_{0} + \beta_{1} A + \beta_{2} \gamma + \beta_{3} \mathrm{FWHM_{SN}} +\epsilon,
\label{eq:RV:correction}
\end{equation}
where $\beta_{0}$ is the coefficient corresponding to the intercept, $A$ is the contrast parameter of the CCF \citep[Fig. 2 in][]{Dumusque-2014b}, $\gamma$ is the aforementioned CCF asymmetry parameter, FWHM$_\text{SN}$ is the FWHM of the CCF fitted with the SN, and $\epsilon$ is the random error having a multivariate normal distribution with vector of means equal to \textbf{0} and variance-covariance matrix equal to $\sigma^{2}I$ (with $I$ being the identity matrix). 

The temporal evolution of active regions is reflected by the change of the activity indicators obtained from the CCF (i.e., the already defined $A$, $\gamma$, and FWHM$_\text{SN}$) and depends on the rotation period of the star and on its magnetic cycle \citep[e.g.,][]{dewarf10, Dumusque-2011c, Borgniet-2015}. It seems therefore reasonable to assume that stellar activity is not stationary, but rather ``piecewise stationary''. By the term ``piecewise stationary'', we mean that the stellar activity does not change significantly within certain properly selected temporal segments. If we were able to detect the bounds of those segments, we could properly split the RV time series and correct for stellar activity by applying Eq. \eqref{eq:RV:correction} to each segment. This represents the core of the change point detection (CPD) methods to be compared with the overall correction ({oc}) method, where the correction model based on Eq. \eqref{eq:RV:correction} is applied over the entire RV time series.

In this paper, we propose the \emph{breakpoint (bp) method} using a CPD technique to correct for variations within an RV time series that are induced by active stellar regions.
The bp method is compared to the oc method using four different stars, some of which have known exoplanets.
We demonstrate that the bp method is better able to correct the RV time series variations, suggesting that the class of CPD methods may be helpful for detecting low-RV planetary signals such as
the ones induced by Earth-like exoplanets.

The paper is organized as follows. In Sec. \ref{sec:met} we introduce the CPD methods, which constitute the statistical framework we used to perform our analyses. In Sec. \ref{sec:analysis} we compare the performances of the CPD method in use and the {oc} method, both using real observational data and developing a proper simulation study to quantify the threshold of detection of exoplanets. The discussion of the results and our conclusions are outlined in Secs.~\ref{sec:discu} and ~\ref{sec:conclu}, respectively.

\section{Change point detection methods} \label{sec:met}

CPD methods are widely used when the goal is to find changes and variations in a time series \citep[e.g.,][]{truong2020selective}. The presence of change points in a time series is a strong indication that the data generating process has changed \citep[e.g.,][]{van2020evaluation}. The first CPD method (the so-called ``intercept-only'' method) was originally introduced by \cite{page1954continuous} in order to identify changes in the mean of an industrial quality control variable.  CPD methods have since become a reliable and widely used solution in bioinformatics \citep[e.g.,][]{guedon2013exploring, hocking2013learning, truong2018ruptures}, climatology \citep[e.g.,][]{reeves2007review, verbesselt2010detecting, maidstone2016efficient}, financial analyses \citep[e.g.,][]{lavielle2006detection, frick2014multiscale}, medicine \citep[e.g.,][]{liu2018change}, network data traffic analysis \citep[e.g.,][]{levy2009detection, lung2011homogeneity}, signal processing analysis \citep[e.g.,][]{lavielle2007adaptive, jandhyala2013inference, haynes2017computationally}, and speech processing \citep[e.g.,][]{angelosante2012group}.

Following \cite{truong2020selective}, a CPD method may be classified as either {online} or {offline}. The {online} methods are used when seeking real-time changes in the data generating process \citep[e.g.,][]{adams2007bayesian, sahki2018change}. Instead, the so-called {offline} (referred to also as ``{a posteriori}'') methods are based on the CPD algorithms designed to estimate change points in the generative process of a time series when the collecting data process is over \citep[e.g.,][]{truong2020selective, van2020evaluation}. If only a single-parameter change is monitored, then we talk about {univariate} processes and {univariate} CPD methods being employed; otherwise we deal with {multivariate} methods \citep[e.g.,][]{aminikhanghahi2017survey}.
 
\subsection{The CPD statistical framework}\label{sec:framework} 
 
In order to introduce the CPD methods, we start by defining the object of interest as an univariate time series $y = y_{1, \dots, T} = \{y_i\}_{i=1}^{T}$ made of $T$ data points. The $y$ time series is assumed to be ``piecewise stationary'' \citep[e.g.,][]{chen2011parametric, truong2020selective, van2020evaluation}, which means that the temporal behavior of $y$ changes at $D$ different and (generally) unknown locations. The goal of the CPD methods is to estimate the set of indices $l$ corresponding to the $D$ locations, $l = \{l_1, \dots, l_D\}$, which delimit different stationary regions; by convention $l_0 \equiv 1$ and $l_{D+1} \equiv T$.

From a statistical standpoint, CPD methods fall into the model selection problem, because the overall goal of estimating the vector $l$ of change point locations is equivalent to retrieving the best segmentation of the time series among all the possibilities \citep[e.g.,][]{chen2011parametric, truong2020selective, van2020evaluation}.

In order to retrieve the best possible segmentation, CPD methods rely on the maximization of the penalized log-likelihood function $\mathcal{L}_P(D;y)$, which is defined as:
\begin{equation}
\begin{split}
\mathcal{L}_P(D; y) &= \mathcal{L}(D; y)  - \lambda P(T) \\
                    &= {\sum_{k=1}^{D+1} \log L(y_{[l_{k-1}:l_k -1]})} - \lambda P(T),
\end{split}
\label{eq:loss.function}
\end{equation} 
where $k$ is the index used to identify the $D+1$ segments of $y$ bounded by $l_0, \dots, l_{D+1}$, $\log L (y_{[l_{k-1}:l_k -1]})$ is the $\log$-likelihood function evaluated at the $k$-th segment, $\lambda$ is a positive pre-selected constant, and $P(T)$ is a penalty function that may be added to balance out the goodness-of-fit term given by the log-likelihood function $\mathcal{L}(D;y)\equiv\sum_{k=1}^{D+1} \log L(y_{[l_{k-1}:l_k -1]})$ \citep[e.g.,][]{zeileis2001strucchange, bai2003computation, truong2020selective}.
After evaluating the $\hat{D}$ value for which $\mathcal{L}(D; y)$ is maximum (i.e., $\mathcal{L}(\hat{D}; y)=\hat{\mathcal{L}}(y)=\max_D{\mathcal{L}(D; y)}$), $\lambda P(T)$ (which is, in general, a function of $\hat{\mathcal{L}}(y)$, see e.g., Eq. (\ref{eq:penalty})) is subtracted from $\mathcal{L}(\hat{D};y)$ to obtain $\mathcal{L_P}(\hat{D}; y)$. The best partition $\hat{l}=\{\hat{l_k}\}$ for that given $\hat{D}$ is the one corresponding to the maximum value of the penalized log-likelihood function, that is $\hat{\mathcal{L}_P}=\max_{l}\mathcal{L}_P(\hat{D},y_{\{l\}})$. In other words, the different $\mathcal{L}_P(\hat{D}; y_{\{l\}})$ values are used to address the model selection problem (i.e., the choice of the best segmentation), and $\hat{\mathcal{L}_P}$ identifies the optimal partition of the $y$ time series made of $\hat{D}+1$ segments. If the total number of change points is assigned a priori and only their locations along $y$ is unknown, no penalty is added in Eq. \eqref{eq:loss.function}. 

\subsection{The breakpoint method}\label{sec:bp}

Given the multiple sources of stellar activity, the most adequate methods to treat the changes in RV time series are the {offline multivariate} CPD methods, which allow a linear relationship to be expressed between the response variable $y$ and a set of covariates (which may be arranged in a matrix, $X$), as the one defined in Eq. \eqref{eq:RV:correction}. Using those methods, the set of locations $l = l_1, \dots, l_D$ is estimated not only by evaluating the changes in the mean and in the variance of $y$, but also by considering the relation between $y$ and $X$ \citep[e.g.,][]{hackl1989statistical, bai1997estimating, bai1997estimation, bai1998estimating, zeileis2001strucchange, zeileis2003testing, bai2003computation}. In the analyses presented in the following Secs., we use the CPD method proposed by \cite{bai2003computation}, known as the {breakpoint} ({bp}) method.

In order to introduce the {bp} method, we start by providing the definition of the multivariate linear regression model made of $p$ covariates for the vector of the observations $y$, which generalizes the model proposed in Eq. \eqref{eq:RV:correction}. In matrix form, this is:
\begin{equation}
y = X \boldsymbol{\beta} + \epsilon = \beta_0 \mathbf{1} + \beta_1 X_1 + \dots + \beta_p X_p + \epsilon , 
\label{eq:lm}
\end{equation}
where $X$ is the matrix of covariates, $\boldsymbol{\beta}$ is the vector of parameters to be determined according to the CPD method in use\footnote{Since $y$ is a vector of dimension $T \times 1$, the matrix of covariates has dimensions $T \times (p+1)$, where $p$ is the number of covariates. The columns of $X$ are $X_1, \dots, X_p$ plus an initial extra column $X_0$ padded with a vector of ones that is indicated as $\mathbf{1}$. Consequently, the length of the parameters' vector $\boldsymbol{\beta}$ is $p+1$.}, and $\epsilon$ is the already defined random error distributed according to a multivariate normal distribution having vector of means equal to \textbf{0} and variance-covariance matrix equal to $\sigma^{2}I$. Assuming the whole time series $y$ as stationary implies that all the coefficients $\boldsymbol{\beta} = (\beta_0, \dots, \beta_p)$ are constant over the entire temporal range of observations. Conversely, assuming piecewise stationarity, $y$ can be divided in $D+1$ segments over each of which the coefficients are constant, so that there is one set of coefficients per segment, for a total of $D+1$ vectors $\boldsymbol{\beta}_k$ ($k=1, \dots, D+1$). With the piecewise stationarity assumption, following \cite{bai2003computation, zeileis2003testing}, we may update Eq. \eqref{eq:lm} as:
\begin{equation}
y_{[l_{k-1}:l_k -1]} = X_{[l_{k-1}:l_k -1],\cdot} \boldsymbol{\beta}_k + \epsilon_k
\label{eq:lm.piecewise}
\end{equation}

The index $k = 1, \dots, D+1$ indicates the temporal segment over which the vector of the $\boldsymbol{\beta}_k$ coefficients is constant.

\cite{bai2003computation} retrieve the simultaneous estimation of multiple breakpoints $l_1, \dots, l_D$ by calculating the residual sum of  squares (RSS) of Eq. \eqref{eq:lm.piecewise}. As $\epsilon$ is distributed according to a multivariate normal distribution, we recall that the least squares solution proposed by \cite{bai2003computation} is equivalent to the maximum likelihood solution retrieved by maximizing Eq. \eqref{eq:loss.function}, as long as the same penalty is used in both procedures. In particular, following the implementation of \citet{bai2003computation}, $\lambda=1$ and the penalty function $P(T)$ is given by the Bayesian information criterion \citep[BIC,][]{schwarz1978estimating}, leading to
 \begin{equation}
  P(T) = (p+1) (D+1) \ln(T) - 2 \ln{\hat{\mathcal{L}}(y)},
  \label{eq:penalty}
 \end{equation}
where the multiplicative term $(p+1)(D+1)$ is equal to the overall number of parameters used by the {bp} method. According to Eq.(1), $p+1=4$ parameters (i.e., $\beta_0, \dots, \beta_3$) are used inside each segment. 

Since $D > 1$ is unknown, the optimization problem might be computationally challenging. In order to tackle the computational burden of estimating $\hat{D}$, \cite{bai2003computation} used the dynamic programming algorithm originally proposed by \cite{fisher1958grouping} and extended by \cite{bellman1969curve} and \cite{guthery1974partition}. The core idea of the dynamic programming approach is to first compute a set of triangular RSS matrices for each $D=1,\dots, D_{\mathrm{max}}$ value (where $D_{\mathrm{max}}$ is an integer number tuned by the researcher, indicating the maximum number of breaks allowed in that analysis\footnote{We set $D_{\mathrm{max}}=8$ throughout the analyses of this paper, which turned out to be high enough to avoid putting a strong a priori constrain on the number of breakpoints. In fact, the {bp} method usually converges only up to $\bar{D}<D_{\mathrm{max}}$, and therefore the set of RSS matrices is computed only for those models.}). In other words, within a given set of triangular RSS matrices, the number of segments $D$ is fixed, while the starting and ending locations of each segment change. Then, the {bp} method returns that partition, which minimizes the RSS of Eq. (\ref{eq:lm.piecewise}) (and hence maximizes $\mathcal{L}_P(D; y)$ of Eq. (\ref{eq:loss.function})) among all the previously created matrices. The main challenge of the dynamic programming approach is given by the computational effort needed to compute the sets of triangular RSS matrices. Further details about the dynamic programming algorithm and the {bp} method can be found in \cite{bai2003computation, zeileis2001strucchange, zeileis2003testing}. In particular, this work uses the version proposed by \citet{zeileis2003testing}.

\section{Data analysis}\label{sec:analysis}
\subsection{Preliminary considerations}
To test the performances in modeling and removing stellar activity from RV time series, we carried out several analyses of RV time series to compare the {bp} method presented in Sec. \ref{sec:bp} with the {oc} method that corrects for stellar activity over the entire temporal range considered as a whole segment.
We considered real RV observations of four different stars, namely $\alpha$ Cen B, HD\,215152, HD\,10700, and HD\,192310, which were selected because of their high signal-to-noise ratio (S/N) and long-baseline coverage with an intensive observational cadence. 

In particular, \citet{rajpaul2015ghost} analyzed the 459 HARPS RV data points of $\alpha$ Cen B obtained by \citet{dumusque2012earth} (a subsample of the RV observations used in this work). After properly modeling the RV time series, they concluded that all the significant variations in the time series are compatible with stellar activity. Therefore, as $\alpha$ Cen B does not likely host any planet, by applying both the {bp} and {oc} methods to its RV time series, we may test the performances of the two methods in removing stellar sources of noise and retrieving a "cleaned" time series. Only if a further signal is present will it then be worth wondering whether it is exoplanetary in origin.

We considered RV data points derived only from spectra with $S/N\geq 10$ at 550 nm. The average S/N values at 550 nm, the number of CCF data points, and the rotation period $P_{\mathrm{rot}}$ of the four stars analyzed in this work are listed in Table.~\ref{tab:starComparison}.
\begin{table}
\caption{Average signal to noise ratio (S/N) of the available spectra, number of RV data points (\# CCFs), and stellar rotational periods ($P_{\mathrm{rot}}$) for each of the four targets of interest.}
\label{tab:starComparison}
\centering
\resizebox{\columnwidth}{!}{%
\begin{tabular}{llcccc}
\hline\hline
 & & $\alpha$ Cen B & HD\,215152 & HD\,10700 & HD\,192310 \\
\hline                        
 S/N @\,550\,nm &    &  339  &  141  & 273  & 207  \\
 CCFs   & [\#]    &  16451  &  273  & 9243 & 1348 \\
 $P_{\mathrm{rot}}$ & [d] & 39\tablefootmark{(a)} &  42\tablefootmark{(b)} & 34\tablefootmark{(c)}   & 48\tablefootmark{(d)}   \\
\hline
\end{tabular}
}
\tablefoot{$P_{\mathrm{rot}}$ sources: \tablefoottext{a}{\citet{dewarf10}}; \tablefoottext{b}{\citet{Delisle:2018aa}}; \tablefoottext{c}{\citet{baliunas1996magnetic}}; \tablefoottext{d}{\citet{Pepe-2011}}}
\end{table}

All the results were obtained using the statistical software \texttt{R} \citep{Rcite}. In particular, the vector listing the locations of change points within the {bp} method was estimated by using the \texttt{R} package \texttt{strucchange}\footnote{\url{https://cran.r-project.org/web/packages/strucchange/}. The same method is also available in \textsc{Python} through the \texttt{ruptures} package \citep{truong2018ruptures}.}. Since the presence of outliers in the data could have led to an exaggerated number of change point locations \citep[e.g.,][]{fearnhead2019changepoint}, before applying the {bp} method and the {oc} method to the data,  we preliminarily rejected all those measurements falling off the range between the $5^{\mathrm{th}}$ and $95^{\mathrm{th}}$ percentile for each considered variable \citep[e.g.,][]{ghosh2012outliers, pollet2017remove}. 

\subsection{Analyses of a real star: $\alpha$ Cen B} \label{sec:alpha}

The RV data of $\alpha$ Cen B consist of $T=17627$ CCFs taken from the end of February 2008 to the end of May 2013 by the HARPS spectrograph. Firstly we used the SN fit to the CCF in order to estimate the parameters $\overline{RV}$, $A$, $\gamma$, and FWHM$_\text{SN}$. Secondly, following \citet{dumusque2012earth}, all the measurements were preliminarily corrected from the contamination that was induced by the close-in $\alpha$ Cen A. Finally, the outliers were clipped as specified above, which resulted in a final set of $T=16451$ measurements.

After cleaning the data, the response variable becomes $y= \overline{RV}$, while the matrix of the covariates is $X=(\mathbf{1}, A, \gamma, \mathrm{FWHM_{SN}})$. Since $X$ also contains the intercept term $X_0\equiv\mathbf{1}$, we emphasize again that the parameter vector $\boldsymbol{\beta}$ is made of $p+1=4$ elements, where $p$ is the number of covariates. 

\begin{figure*}[htbp]
\centering
\includegraphics[width=\textwidth]{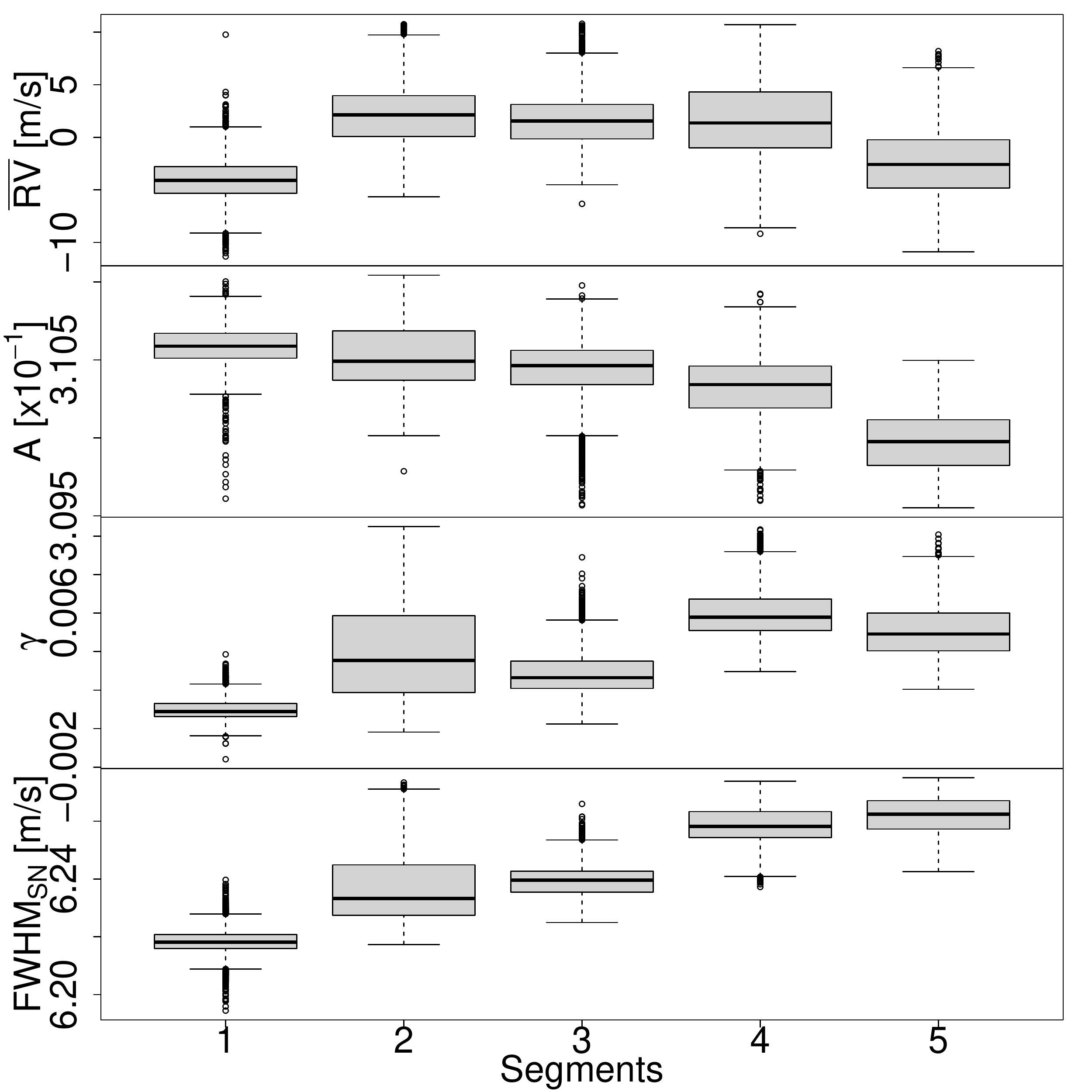}
\caption{Boxplots of the response variable $\overline{RV}$ and of the activity indicators $A$, $\gamma$, and FWHM$_\text{SN}$ for each segment of the optimal partition found by the {bp} method when applied to the $\alpha$ Cen B RV time series. We recall that a boxplot is a graphical tool that visualizes a distribution; the median is shown by the thick horizontal  line, while the box extension defines the interquartile range (IQR), being bounded by the first ($Q_1$) and third ($Q_3$) quartiles. The upper and lower limits of the vertical dashed lines (the whiskers) are defined as $\min{\{\max{(X)},Q_3+1.5\cdot\mathrm{IQR}\}}$ and $\max{\{\min{(X)},Q_1-1.5\cdot\mathrm{IQR}\}}$, respectively, where $X$ indicates the sample to be represented. Outliers beyond the whiskers (if any) are represented as empty dots.}
\label{fig.boxplots}
\end{figure*}

We applied both the {oc} and {bp} methods to the $y=\overline{RV}$ time series, further assuming as input the linear regression model of Eq. \eqref{eq:RV:correction} that specifies the relationship between the response variable $y$ and the matrix of covariates X. The {bp} method found $\hat{D}=4$ change points (i.e., $\hat{D}+1 = 5$ piecewise stationary segments), whose locations are summarized in Table \ref{tab:change.points.alphacentb}, where the error bars at the$1\sigma$ level highlight the precision of the {bp} method. The variability of the activity indicators inside each segment is extremely small, while the variability between each segment is larger, suggesting indeed that the activity of the star changes where a new segment starts. To highlight the locality, the spread, and the skewness of the data synthesized in Table \ref{tab:change.points.alphacentb}, we produced the boxplots \citep{dutoit2012graphical} shown in Fig.~\ref{fig.boxplots}. For each segment, Fig.~\ref{fig.boxplots} displays the boxplots of the response variable $\overline{RV}$ and of the $3$ covariates $A$, $\gamma$, and FWHM$_\text{SN}$. The difference between contiguous boxplots describing the covariate distributions in contiguous segments visually marks the change in the correlation between $\overline{RV}$ and the activity indicators. From a quantitative point of view, the Mann-Whitney test \citep{mcknight2010mann} -- which assumes that the two samples to be compared are not statistically different as null hypothesis -- confirms the differences of the covariates distributions between contiguous segments. In fact, for all the covariates, it gives \text{p-values}\,$\sim$\,$10^{-8}$ for each pair of contiguous segments. Moreover, we note that $A$ globally decreases, while $\gamma$ and FWHM$_\text{SN}$ increase by moving from one segment to the next, which suggests a significant temporal change in the stellar activity level. Overall, Table \ref{tab:change.points.alphacentb} and Figure \ref{fig.boxplots} show the characteristics and the quality of the optimal segmentation determined by the {bp} method for $\alpha$ Cen B.

\begin{table*}
\caption{Locations of the change points (CPL)  estimated by the {bp} method for $\alpha$ Cen B. For each location, the JD and its corresponding date in the Gregorian Calendar format is displayed. For each segment, the number of data points (\# CCFs), the response variable $\overline{RV}$, and the estimates of the detrending parameters are shown by reporting their median value and error bars at the $1\sigma$ level.}
\label{tab:change.points.alphacentb}
\centering
\begin{tabular}{lcccccccc}
\hline\hline
CPL & JD & Date & Time span & CCFs & $\overline{RV}$ & $A$ & $\gamma$ & FWHM$_\text{SN}$ \\
 & & & [d] & [\#] & [\ms] & [$\cdot10^{-1}$] & & [\ms] \\
\hline                        
$l_0$ & $2454524.90655$         &       28 Feb 2008 & \multirow{2}*{$500.69$} & \multirow{2}*{$3146$} & \multirow{2}*{$-4.1_{-1.9}^{+2.1}$} & \multirow{2}*{$3.1059_{-0.0011}^{+0.0012}$} & 
\multirow{2}*{$0.0009_{-0.0005}^{+0.0007}$} & \multirow{2}*{$6.2182_{-0.0037}^{+0.0046}$} \\
$l_1$ & $2455025.59597$         &       13 Jul 2009 & \multirow{2}*{$333.86$} & \multirow{2}*{$2779$} & \multirow{2}*{$2.1_{-1.3}^{+2.7}$} & \multirow{2}*{$3.1049_{-0.0016}^{+0.0028}$} & \multirow{2}*{$0.0035_{-0.0020}^{+0.0037}$} & \multirow{2}*{$6.2332_{-0.0077}^{+0.0174}$}     \\
$l_2$ & $2455359.45222 $        &       11 Jun 2010  & \multirow{2}*{$396.21$} & \multirow{2}*{$3923$} & \multirow{2}*{$1.5_{-0.6}^{+2.5}$} & \multirow{2}*{$3.1046_{-0.0019}^{+0.0015}$} & \multirow{2}*{$0.0026_{-0.0007}^{+0.0015}$} & \multirow{2}*{$6.2396_{-0.0064}^{+0.0045}$}  \\
$l_3$ & $2455755.65986$ &       13 Jul 2011 & \multirow{2}*{$363.80$}  & \multirow{2}*{$2673$}  & 
\multirow{2}*{$1.4_{-0.5}^{+4.0}$} & \multirow{2}*{$3.1034_{-0.0022}^{+0.0018}$} & \multirow{2}*{$0.0058_{-0.0010}^{+0.0016}$} & \multirow{2}*{$6.2582_{-0.0053}^{+0.0076}$}     \\
$l_4$ & $2456119.45503$ &       10 Jul 2012 &\multirow{2}*{$332.34$} & \multirow{2}*{$3930$} &  
\multirow{2}*{$-2.6_{-3.2}^{+1.7}$} & \multirow{2}*{$3.0998_{-0.0022}^{+0.0019}$} & \multirow{2}*{$0.0049_{-0.0011}^{+0.0015}$} & \multirow{2}*{$6.2624_{-0.0071}^{+0.0063}$}     \\
$l_5$ & $2456441.79243$ &       29 May 2013 &  & & & & &        \\
\hline
\end{tabular}
\end{table*}

\begin{figure*}[htbp]
\centering
\includegraphics[height=.31\textheight]{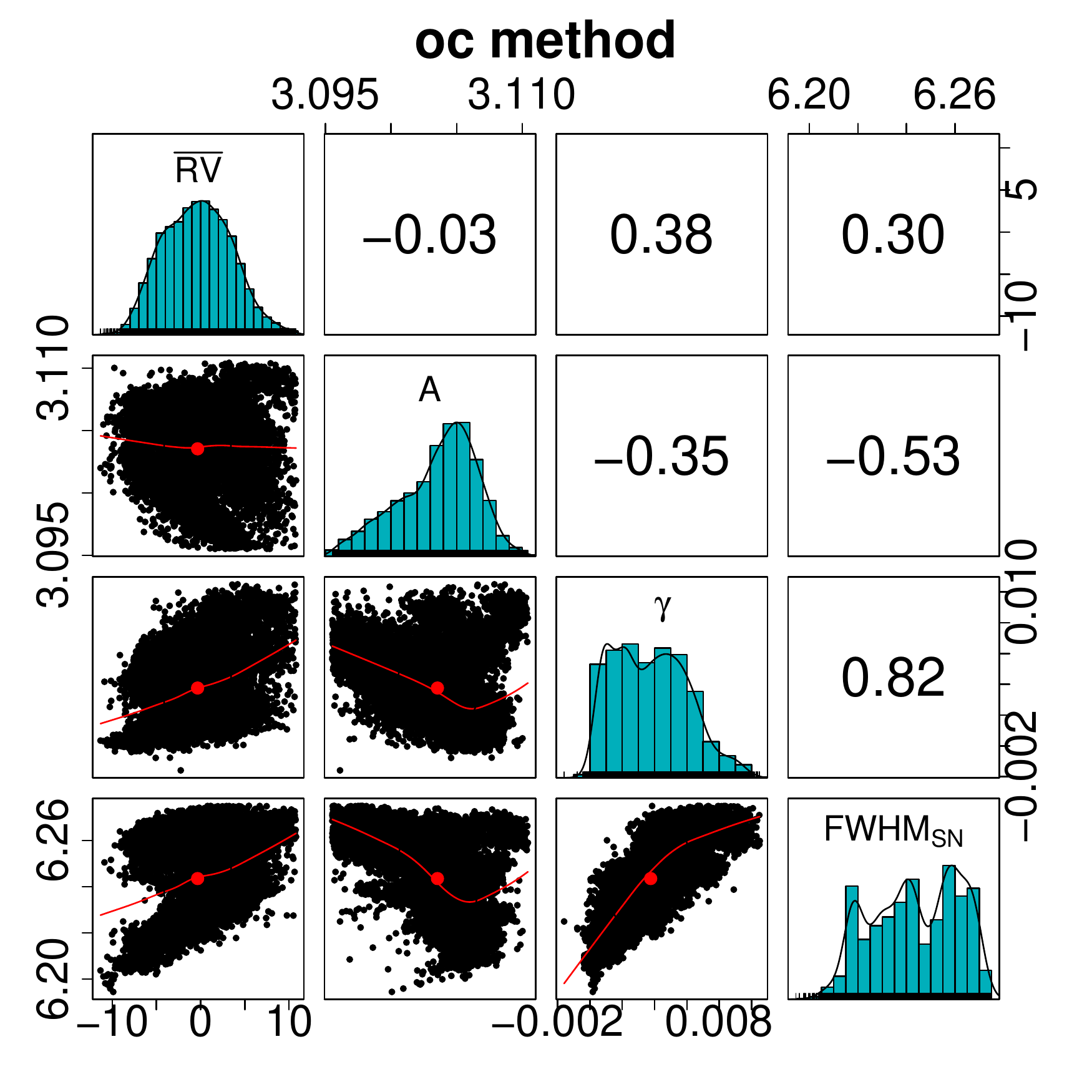} 
\includegraphics[height=.31\textheight]{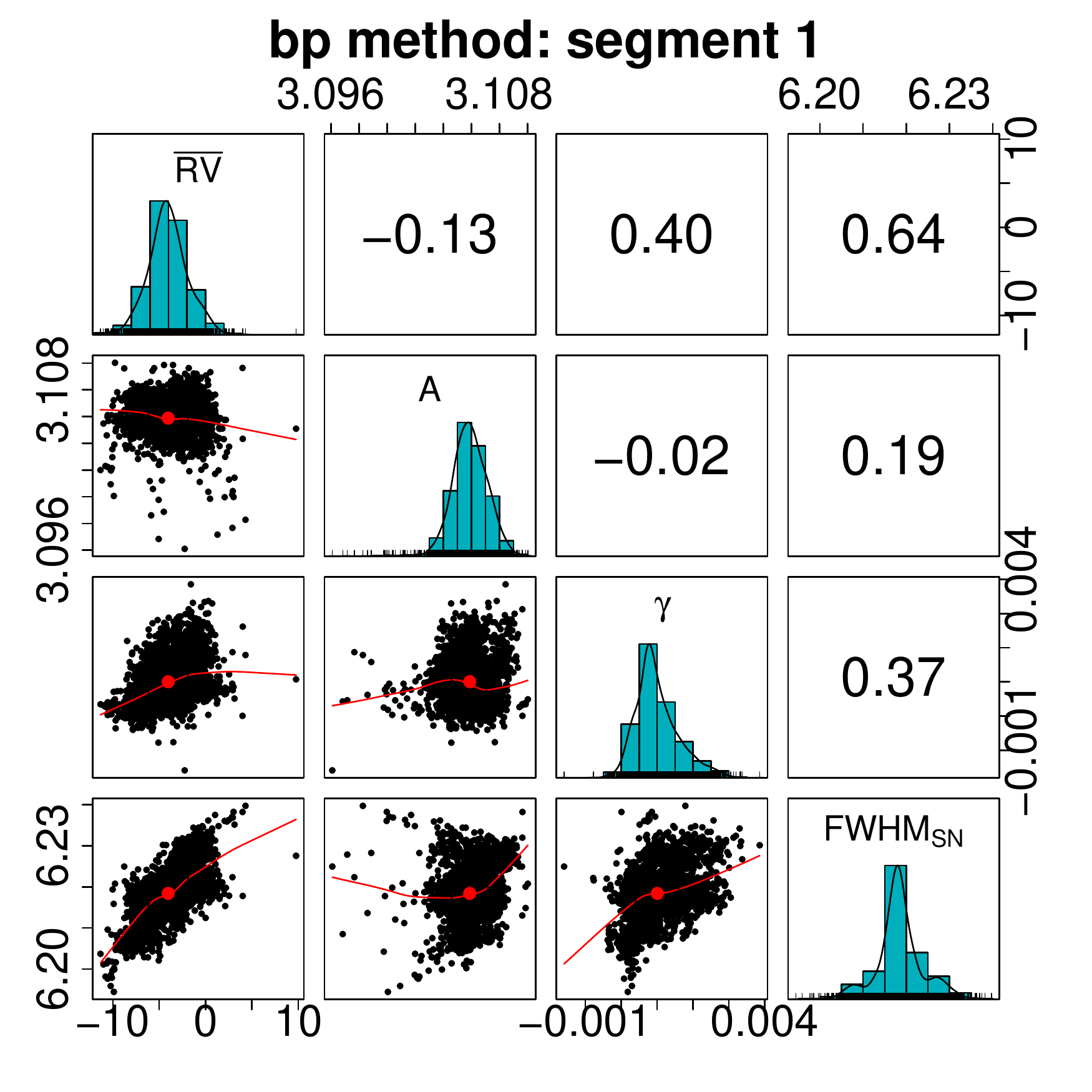} 
\includegraphics[height=.31\textheight]{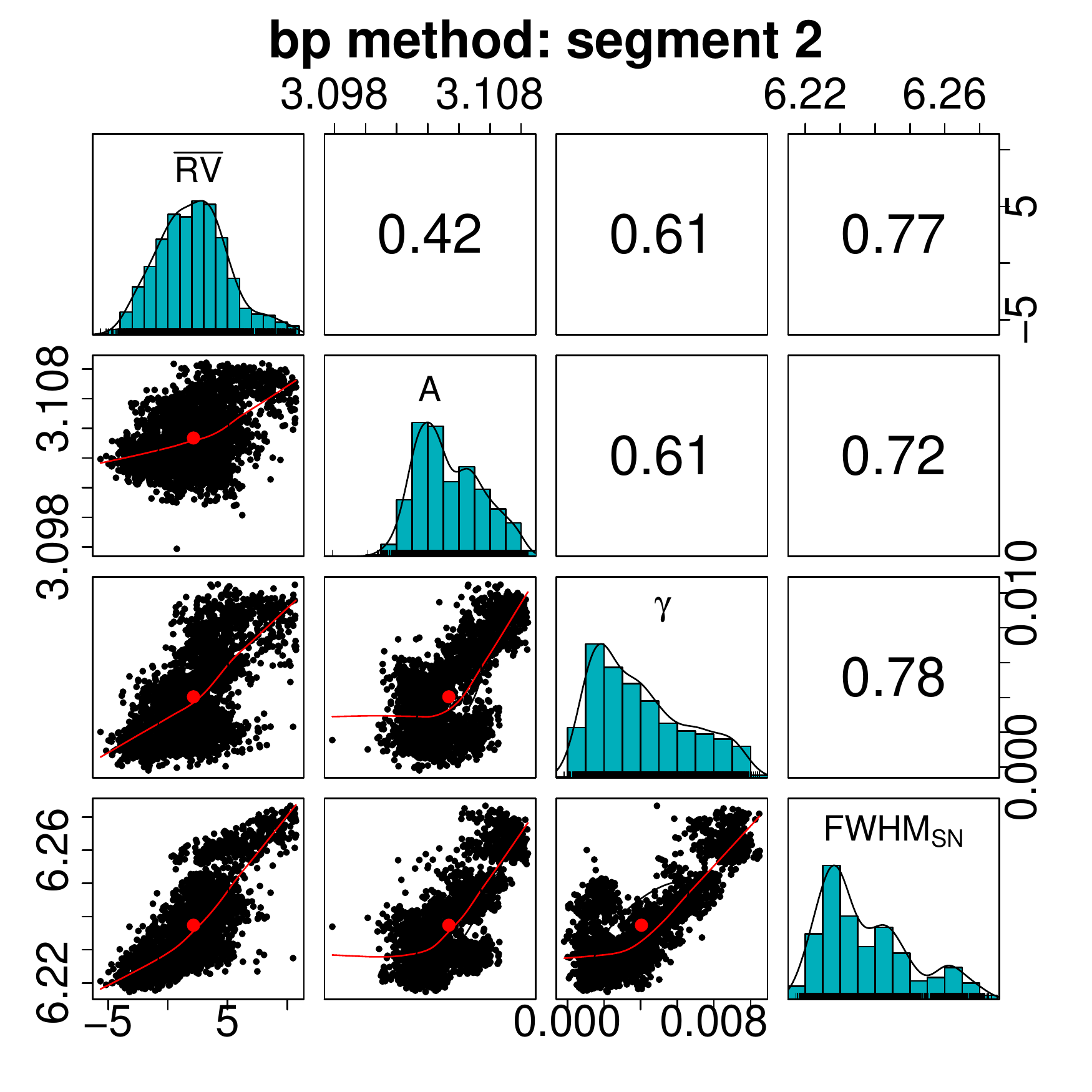} 
\includegraphics[height=.31\textheight]{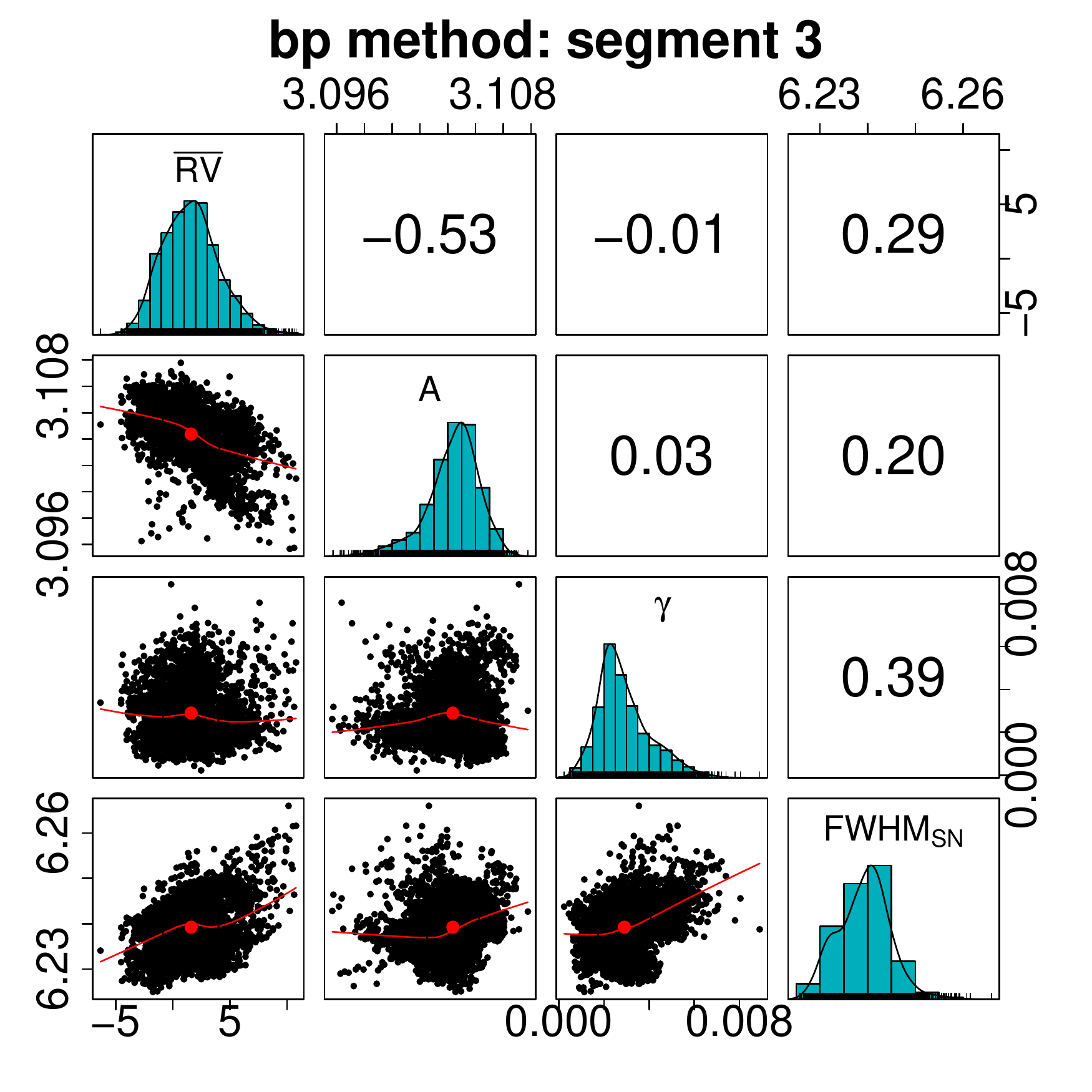}
\includegraphics[height=.31\textheight]{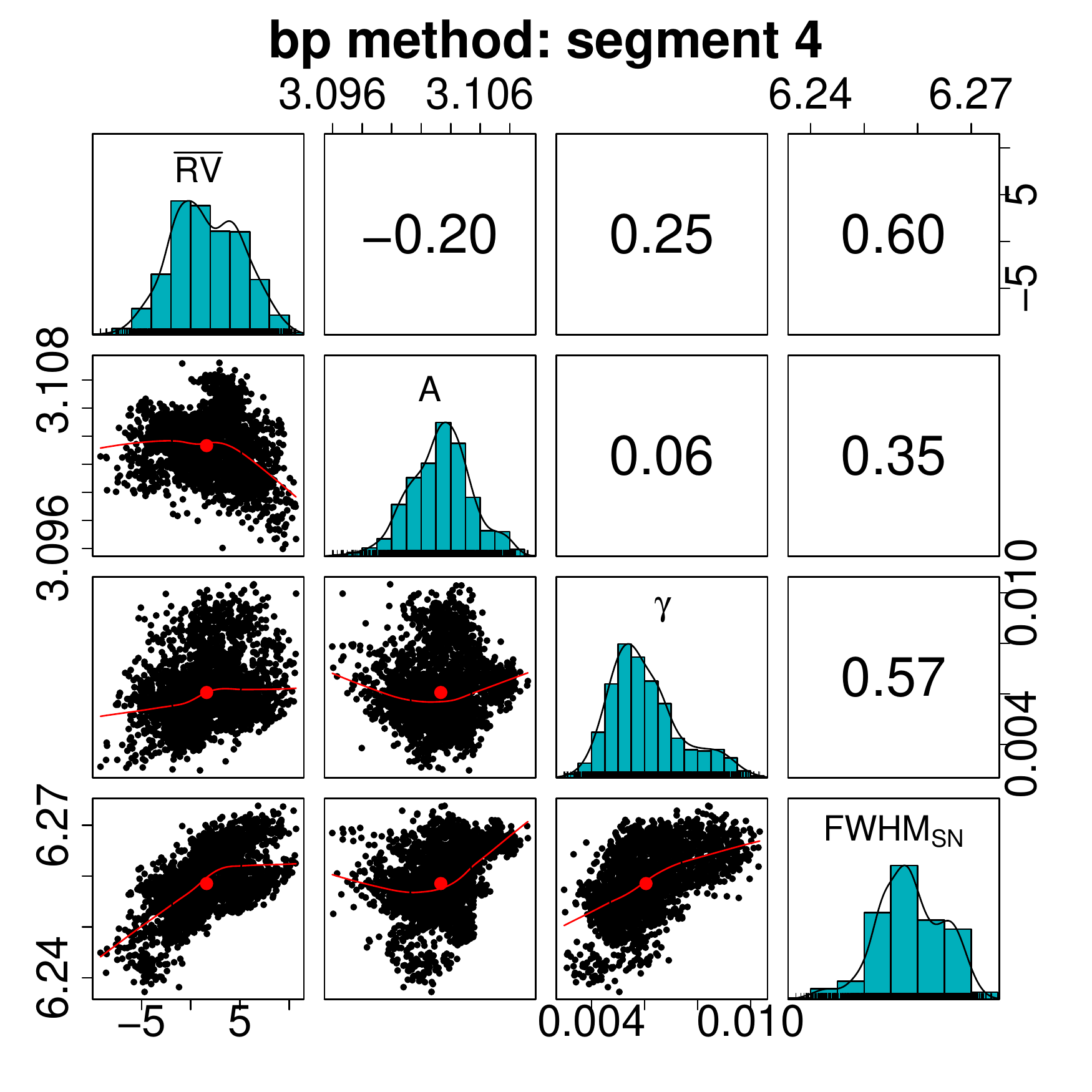}
\includegraphics[height=.31\textheight]{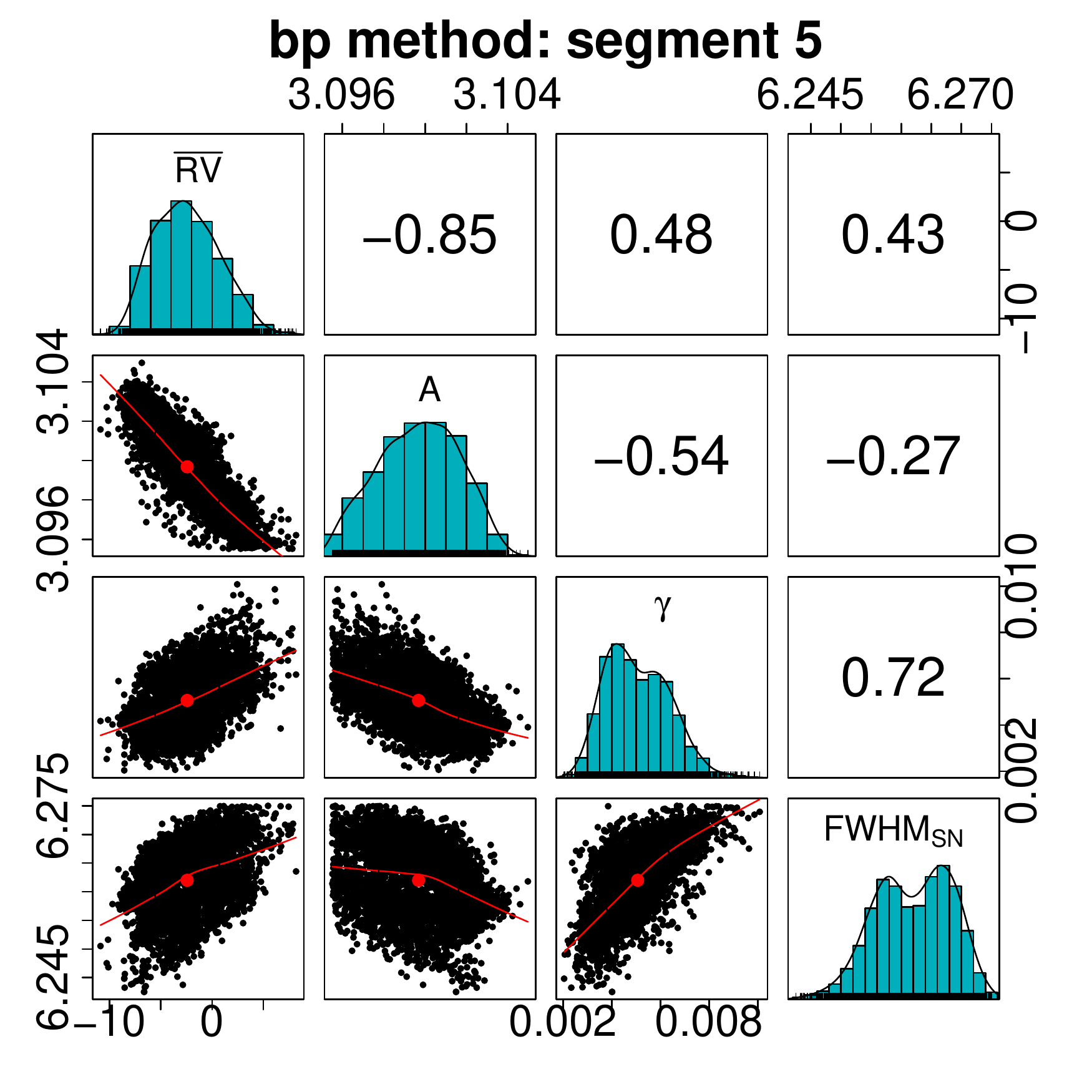}
\caption{$\alpha$ Cen B: correlation plots between $\overline{RV}$ and the activity indicators ($A$, $\gamma,$ and FWHM$_\text{SN}$); the contrast parameter $A$ is shown after having been multiplied by a factor ten. {Top left panel}: the entire data are taken into account as a result of the {oc} method. {Other panels}: correlations are shown for each segment found by the {bp} method. A nonparametric local regression for each pair of considered variables (continuous red line) and its barycenter (i.e., the median, red dot) are also shown.}
\label{fig:alphacentb_pairsx}
\end{figure*}

Another instrument used to evaluate the quality of the solution recommended by the {bp} method is the correlation--pairs plot. Fig. \ref{fig:alphacentb_pairsx} displays the correlation plots between $\overline{RV}$ and the activity indicators when the {oc} method is applied to the data (top left panel) that is to be compared to the results obtained on each of the five segments found by the {bp} method (other panels). The {oc} method is not able to catch all the variations over the entire time series. In fact, the correlation parameters sensibly differ from one segment to the other, suggesting that the piecewise stationary assumption of $\overline{RV}$ as a function of  $A$, $\gamma$, and FWHM$_\text{SN}$ is reasonable, as it is also highlighted in Table \ref{tab:correlations.bp.alphacentb}. %

\begin{table*}
   \caption{Correlation $\rho$ between $\overline{RV} $ and the activity indicators of $\alpha$ Cen B for each of the piecewise-stationary segments detected by the {bp} method. A 95\% confidence interval for each correlation coefficient $\rho$ is specified within parentheses.}
   \label{tab:correlations.bp.alphacentb}
\centering
\begin{tabular}{lccccc}
\hline\hline
 & $1$ & $2$ & $3$ & $4$ & $5$  \\
\hline                        
$\rho(\overline{RV}  ; A)$ & $-0.13$  & $0.42$ & $-0.53 $ & $-0.20$ & $-0.85$  \\
& $(-0.17, -0.10)$ & $(0.39, 0.45)$ & $(-0.55, -0.51)$ & $(-0.24 -0.17)$ & $(-0.86 -0.84)$ \\
\hline 
$\rho(\overline{RV}  ; \gamma)$  & $0.40$ & $0.61$ & $-0.01$ & $0.25$ & $0.48$   \\
& $(0.37, 0.43)$ & $(0.58, 0.63)$ & $(-0.037, 0.026)$ & $(0.21, 0.28)$ & $(0.46 0.50)$ \\
\hline 
$\rho(\overline{RV}  ; \mathrm{FWHM_{SN}})$ & $0.64$ & $0.77$ & $0.29$ & $0.60$ & $0.43$  \\
& $(0.62, 0.66)$ & $(0.76, 0.79)$ & $(0.26, 0.32)$ & $(0.57, 0.62)$ & $(0.41, 0.46)$ \\
\hline
\end{tabular}
\end{table*}

\begin{table*}
\caption{Variability of the $\overline{RV}$, $RV^{*}_{\text{activity}}$, and $\Delta RV^*$ time series of $\alpha$ Cen B quantified through the rms. We recall that the $\overline{RV}$ statistic depends on the SN fit, and hence the rms values are the same for both the {bp} and {oc} methods, which are applied afterward. The last row shows the BIC coefficients corresponding to the {bp} and {oc} models.
}
\label{tab:bp.vs.oc.alphacentb}
\centering
\begin{tabular}{llccccccc}
\hline\hline
\multicolumn{2}{c}{rms} & {bp} segment 1 & {bp} segment 2 & {bp} segment 3 & {bp} segment 4 & {bp} segment 5 & {bp} overall & {oc} overall  \\
\hline                        
$\overline{RV}$ & [\ms] & $2.15$ & $2.94$ & $2.48$ & $3.53$ & $3.13$ & $3.79$  & $3.79$ \\
$RV^{*}_{\text{activity}}$ & [\ms] & $1.52$ & $2.35$ & $1.71$  & $2.71$ & $2.78$ & $3.37$ & $1.51$ \\
$\Delta RV^*$ & [\ms] & $1.52$  & $1.77$ & $1.80$ & $2.27$ & $1.43$ & $1.75$  &  $3.48$ \\
\hline
BIC & & & & & & & $65332$ & $87771$  \\
\hline
\end{tabular}
\end{table*}

We can further stress the benefits of using the {bp} method by comparing $\Delta RV^*_{bp}$ and $\Delta RV^*_{oc}$, which are defined as the residuals obtained using the {bp} and {oc} methods, respectively, when the estimated activity level of the star $RV^*_{\text{activity}}$ is subtracted from the $\overline{RV}$ time series:
\begin{equation}
\Delta RV^*_{bp}= \overline{RV} - \sum_{k=1}^{D+1} RV^*_{\text{activity, } k,}
\label{eq:RV_bp}
\end{equation}
\begin{equation}
\Delta RV^*_{oc}= \overline{RV}  - RV^*_{\text{activity.}}
\label{eq:RV_oc}
\end{equation}

\begin{figure*}[htbp]
\centering
\includegraphics[width=\textwidth]{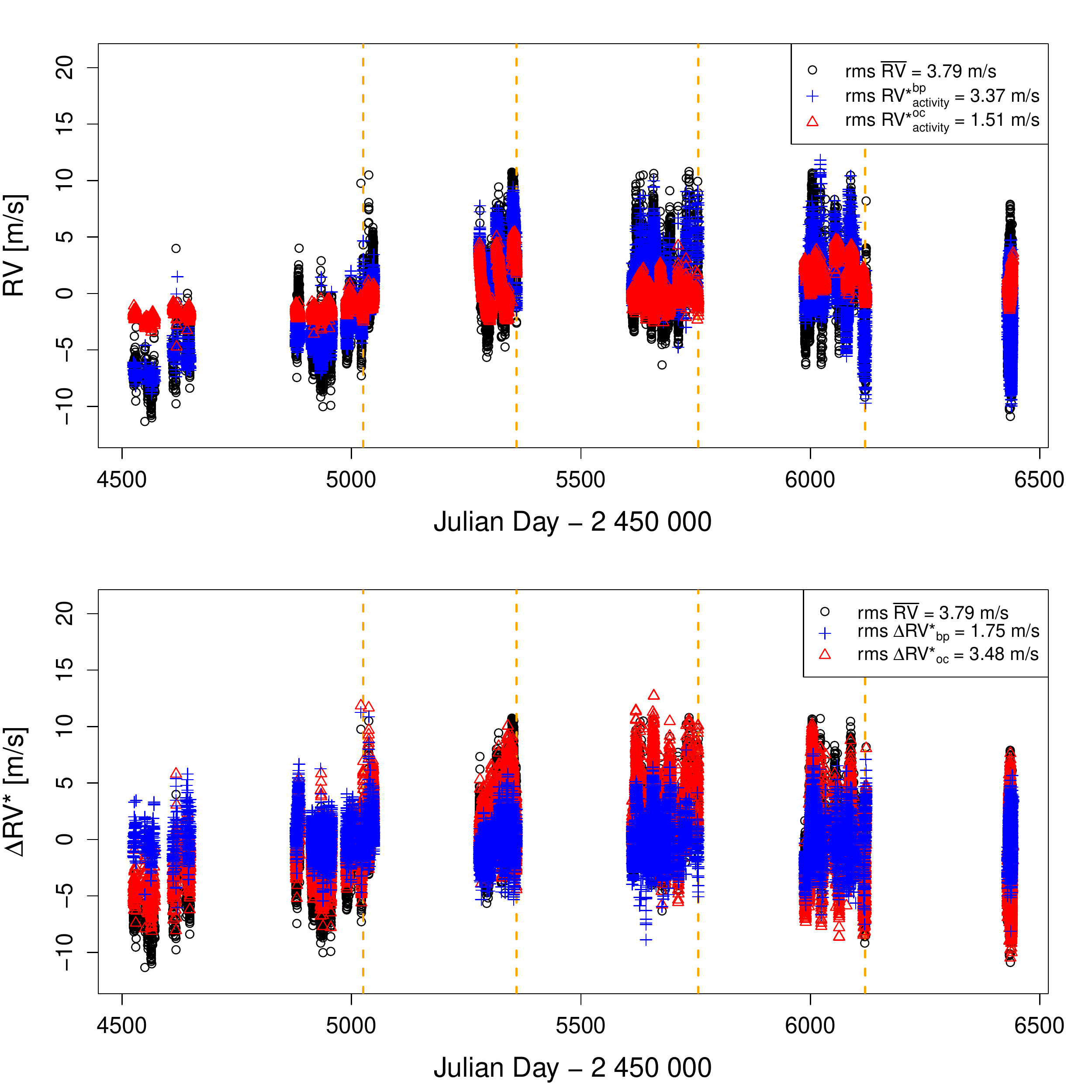}
\caption{Modeling of the stellar activity of $\alpha$ Cen B, as performed by the {bp} and {oc} methods. {Top panel}: signal due to stellar activity as estimated by the {bp} method (blue crosses) and by the {oc} method (red triangles) superimposed to the $\overline{RV}$ time series (black circles) of $\alpha$ Cen B. {Bottom panel}: $\Delta RV^{*}$ as computed by the {bp} method (blue crosses) and by the {oc} method (red triangles). The black circles on the background show the original $\overline{RV}$ time series, enabling a visual comparison of the improvement achieved by the {bp} method in the data correction. The change point locations (Table \ref{tab:change.points.alphacentb}) are displayed in both plots as vertical orange dashed lines.}
\label{fig:alphacentb}
\end{figure*}

The rms of $\Delta RV^*_{oc}$ vs. $\Delta RV^*_{bp}$ are $3.48 \ms$ and $1.75 \ms$ respectively, so we reach similar conclusions in preferring the piecewise linear regression strategy, which lowers the rms by 50\%. In addition, by comparing the rms of the modeled stellar activity with the rms of the $\overline{RV}$ signal, it turns out that the {bp} method explains 89\% of the variability of the $\overline{RV}$ signal in terms of stellar activity, while the {oc} method is able to model only 40\% of the $\overline{RV}$ signal (see Fig. \ref{fig:alphacentb}). Since we expect the $\overline{RV}$ time series to be entirely produced by stellar activity, adopting Eq. \eqref{eq:RV:correction} in each of the $D+1$ segments for the {bp} method significantly improves the detrending performances. 

The detailed rms values of the relevant quantities for both methods are listed in Table~\ref{tab:bp.vs.oc.alphacentb}. In particular, the $\Delta$ BIC between the {oc} and the {bp} methods is $+22439$, thus favoring the {bp} method according to the BIC minimization criterion. We recall that HARPS is built to obtain RV precision of the order of 1 m/s.

\begin{figure*}[htbp]
\centering
\includegraphics[width=\textwidth]{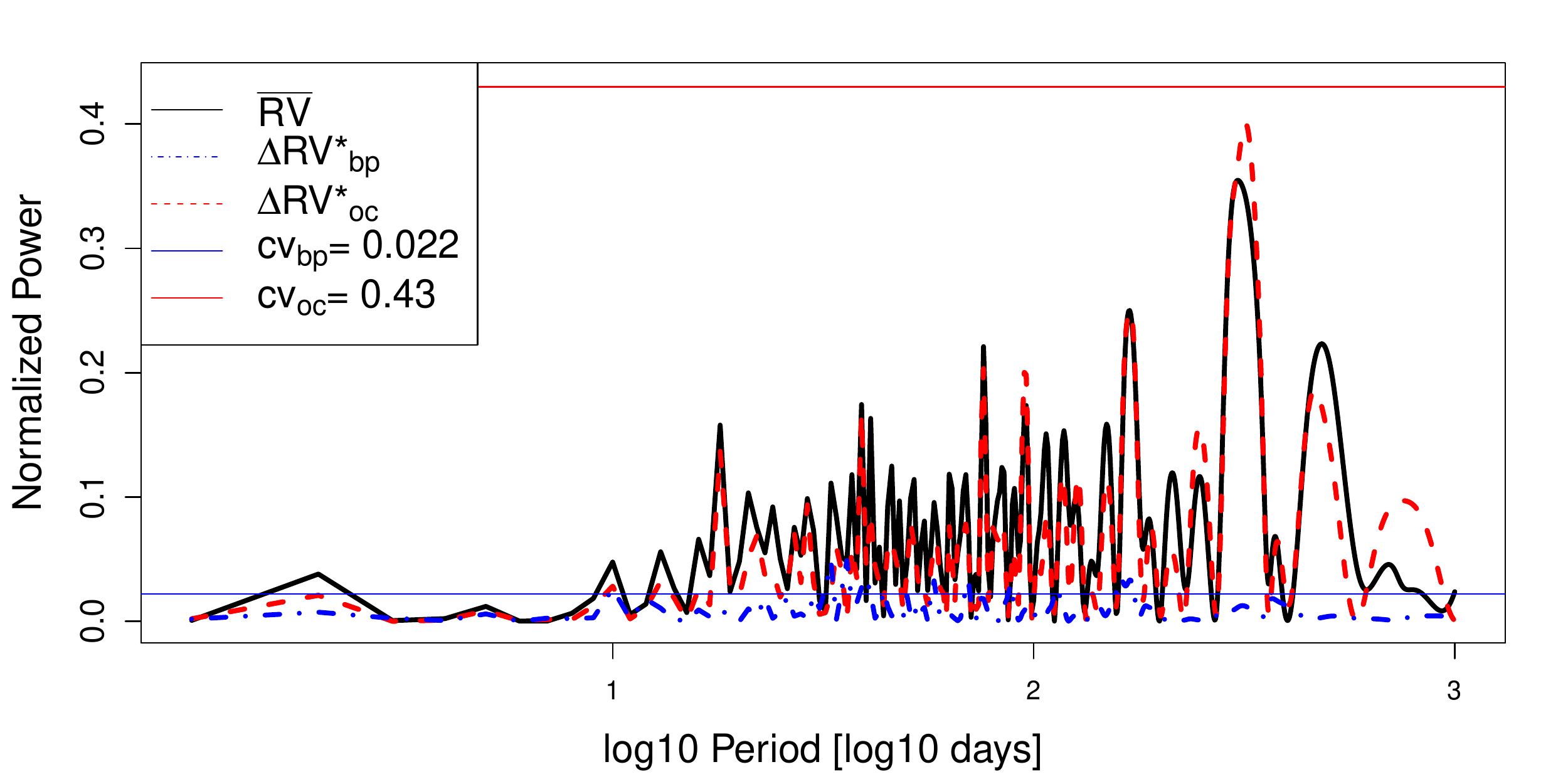}
\caption{GLS periodograms of $\alpha$ Cen B derived from the original $\overline{RV}$ time series (solid black line), $\Delta RV^*_{oc}$ (dashed red line), and $\Delta RV^*_{bp}$ (dash-dotted blue line). The x-axis uses the base-$10$ logarithmic transformation of the period in order to improve the readability of the plot. The GLS periodogram obtained from $\Delta RV^*_{bp}$ does not show any of the periodical peaks caused by stellar signals, which are instead present in the GLS periodogram obtained from $\Delta RV^*_{oc}$. The cv values for $\Delta RV^*_{oc}$ and $\Delta RV^*_{bp}$ are displayed as solid red  and blue lines, respectively.}
\label{fig:alphacentb_GLS}
\end{figure*}

A fourth approach used to compare the performances of the two methods is the evaluation of the generalized Lomb-Scargle (GLS) periodograms \citep[e.g.,][]{Lomb-1976a, Scargle-1982, Zechmeister-2009} on both the $\Delta RV^*_{oc}$ and $\Delta RV^*_{bp}$ residuals. The GLS periodograms for both methods are presented in Fig. \ref{fig:alphacentb_GLS}.
Thanks to the {bp} method, the majority of the peaks caused by stellar activity are successfully removed from the GLS periodogram (dash-dotted blue line in Fig. \ref{fig:alphacentb_GLS}).

Conversely, the {oc} method is not able to properly tackle the peaks caused by active regions (dashed red line in Fig. \ref{fig:alphacentb_GLS}), especially for periods longer than $100$ days.

We used the Cramér-von-Mises (CvM) distance minimization criterion \citep{cramer1928composition}, which defines the measure  we refer to as the critical value (cv), to further compare the GLS periodograms obtained from the two different methods. Going into further detail, after the GLS is obtained, the main goal is to check whether any periods are significant. A period is considered significant if its GLS periodogram peak statistically differs from the distribution that would result from the absence of periodic fluctuations (null hypothesis). To determine the significance level of the peak, this distribution needs to be known or estimated. A Beta distribution $\mathcal{B}(\alpha, \beta)$ whose parameters are $\alpha=\frac{m-1}{2}$ and $\beta=\frac{n-m}{2}$, where $n$ is the number of data points and $m$ is the number of GLS parameters, is usually assumed \citep[see e.g.,][]{schwarzenberg1998distribution,seber2003polynomial,gupta2004handbook}. However, adopting those $\alpha$ and $\beta$ values is not flexible enough, as some period values identified as significant may turn out to be false positives \citep[see e.g.,][]{thieler2014robuste, thieler2016robper}. In order to avoid the selection of those false positive periods, \citet{thieler2014robuste} and \citet{thieler2016robper} propose relaxing the assumption on the Beta distribution, by not defining its parameters ``a priori'', which makes the distribution more flexible. Once the assumptions on the Beta distribution are relaxed, \citet{thieler2014robuste} and \citet{thieler2016robper} suggest estimating the parameters of the $\mathcal{B}(\alpha, \beta)$ by minimizing the following CvM distance:
\begin{equation}
\begin{split}
\text{CvM}(\theta) &= \int_{0}^{+ \infty} \left(F_n(u) - F_{\theta}(u) \right)^2 d F_{\theta}(u) \\
&= \frac{1}{n} \sum_{i=1}^{n} \left( F_{\theta}(u_{(i)}) - \frac{i - 0.5}{n} \right)^2 + \frac{1}{12 n^2},
\end{split}
\end{equation}
where $\theta$ is the parameter space ($\theta=(\alpha, \beta)$ in our case), $u = u_{(1)}, \dots, u_{(n)}$ is the vector of the ordered set of the observations (data points), $F_n(u)$ is the empirical distribution function (i.e., the empirical cumulative distribution function built from the data), and $F_{\theta}(u)$ is the theoretical distribution function (the Beta distribution in our case). Once an estimate for $\theta$ is found (i.e., $\hat{\theta}=(\hat{\alpha}, \hat{\beta})$), the cv is calculated as the $\sqrt[q]{0.95}$--quantile of the CvM-fitted Beta distribution $\mathcal{B}(\hat{\alpha}, \hat{\beta})$, where $q$ indicates the number of period grid-points used to build the GLS periodogram. According to this criterion, the GLS peaks above the computed cv value are significant and deserve further investigation. In this work, we used the cv rather than the well known false alarm probability (FAP) because the cv is more robust \citep[e.g.,][]{thieler2014robuste,thieler2016robper}.

\begin{figure}[htbp]
\centering
\includegraphics[width=\columnwidth]{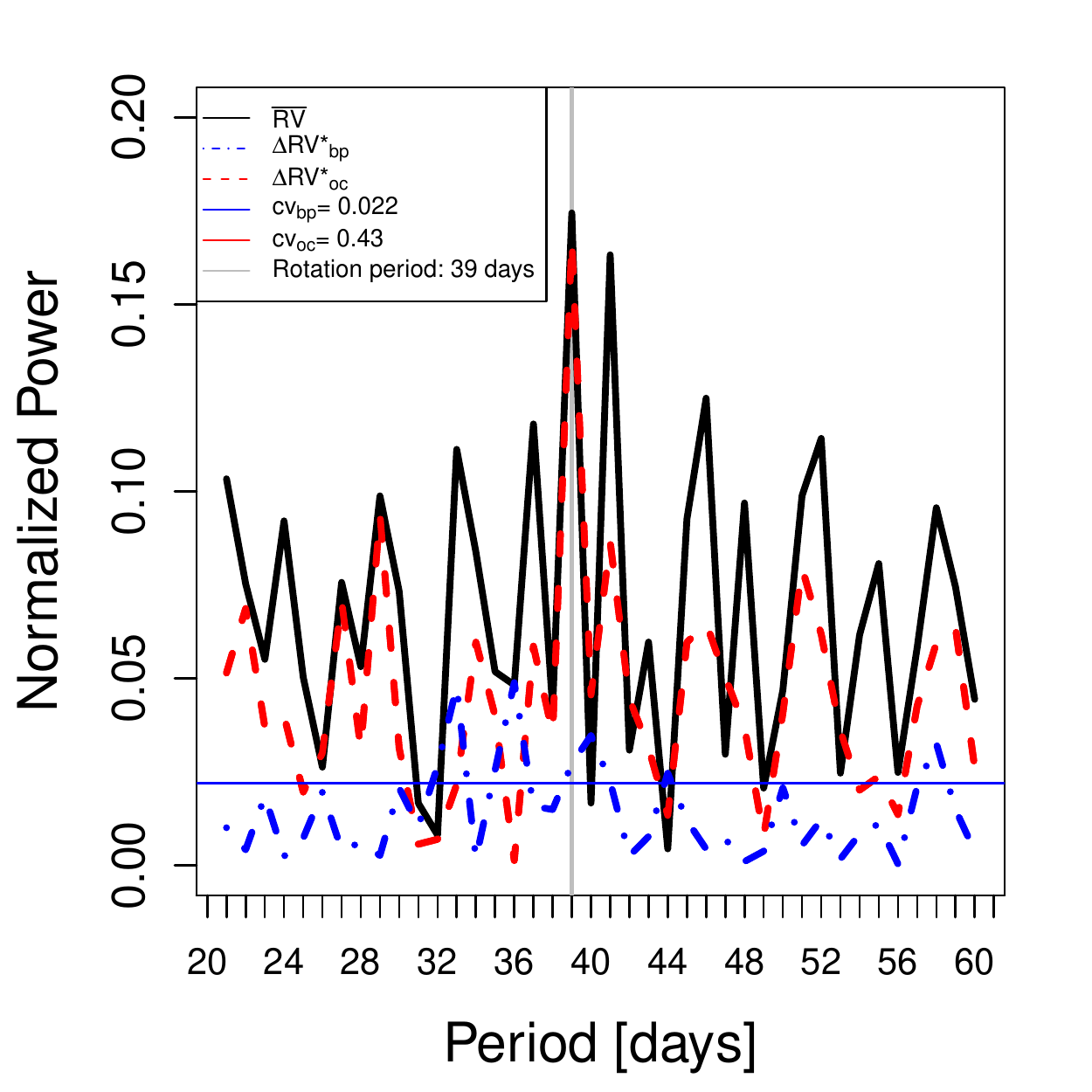}
\caption{GLS periodograms of $\alpha$ Cen B derived from the original $\overline{RV}$ time series (solid black line), $\Delta RV^*_{oc}$ (dashed red line), and $\Delta RV^*_{bp}$ (dash-dotted blue line), focusing on a subset of the epochs including the rotation period of the star equal to 39 days (highlighted by the vertical gray line). The 39-day peaks have normalized powers $\mathcal{P}_{\overline{RV}}=0.173$ and $\mathcal{P}_{oc}=0.167$ in the original and in the {oc}-corrected GLS, respectively. The normalized power of this peak is reduced at $\mathcal{P}_{bp}=0.027$ for the {bp}-corrected GLS. The cv value for $\Delta RV^*_{bp}$ is displayed as the horizontal blue line and is equal to 0.022; we recall that the cv for  $\Delta RV^*_{oc}$ is 0.43.}
\label{fig:alphacentb_GLS_focus39days}
\end{figure}

The cv computation for   $\Delta RV^*_{oc}$ and  $\Delta RV^*_{bp}$ yields $\text{cv}_{oc}=0.43$ and $\text{cv}_{bp}=0.022$. Peaks in the GLS below the cv are to be considered as caused by stellar activity, while more detailed analyses might be needed for peaks above the cv to discover their nature.
All in all, the high $\text{cv}_{oc}$ value suggests quite a noisy GLS, which prevents an effective detection of exoplanetary signals. Instead, the much lower $\text{cv}_{bp}$ indicates a cleaner GLS, meaning that  it is worth investigating even low-powered peaks, which in principle might reveal the presence of small exoplanets.
We specifically checked the behavior of the GLS periodogram, by inspecting the period interval that includes the rotation period of $\alpha$ Cen B. Looking at Fig. \ref{fig:alphacentb_GLS_focus39days}, we note that the {oc} method is still unable to remove the peak at $P$\,$\sim$\,39 days, which is close to $P_{\mathrm{rot}}$ of $\alpha$ Cen B \citep[e.g.,][]{dewarf10}. In fact, the 39-day peaks in the original and in the {oc}-corrected GLS periodograms  both have normalized powers $\mathcal{P}_{\overline{RV}}$\,$\approx$ \,$\mathcal{P}_{oc}$\,$\approx$\,0.17. Instead, in the {bp} method, the normalized power of this peak is sensibly reduced to $\mathcal{P}_{bp}$\,$\approx$\,0.027.
Both $\mathcal{P}_{oc}$ and $\mathcal{P}_{bp}$ cannot bring us to immediately postulate the existence of a candidate exoplanet with an orbital period of 39 days. In fact $\mathcal{P}_{oc}<\text{cv}_{oc}=0.43$, while $\mathcal{P}_{bp}$ is quite close to $\text{cv}_{bp}=0.022$, although greater. Regardless, as our main goal is cleaning the original $\overline{RV}$ time series from stellar activity, it is better to deal with the {bp} method because it recognizes and reduces the GLS-$\mathcal{P}(P_{\mathrm{rot}})$ (although the corrected peak is higher than $\text{cv}_{bp}$, which will encourage further investigatations to better understand its nature), rather than dealing with the {oc} method where the unchanged GLS-$\mathcal{P}(P_{\mathrm{rot}})$ (which is below $\text{cv}_{oc}$) may lead to a false negative case. As such, since $\mathcal{P}_{bp}>\text{cv}_{bp}$, we refer the reader to Table.~\ref{tab:peaks.alphacentb} and to our discussion in Sec.~\ref{sec:discu}, where we investigate the significance of all the $\Delta RV^*_{bp}$-peaks whose normalized powers are greater than the cv threshold. We anticipate that no Keplerian-like signals have been detected.

\begin{figure}[htbp]
\centering
\includegraphics[width=\columnwidth]{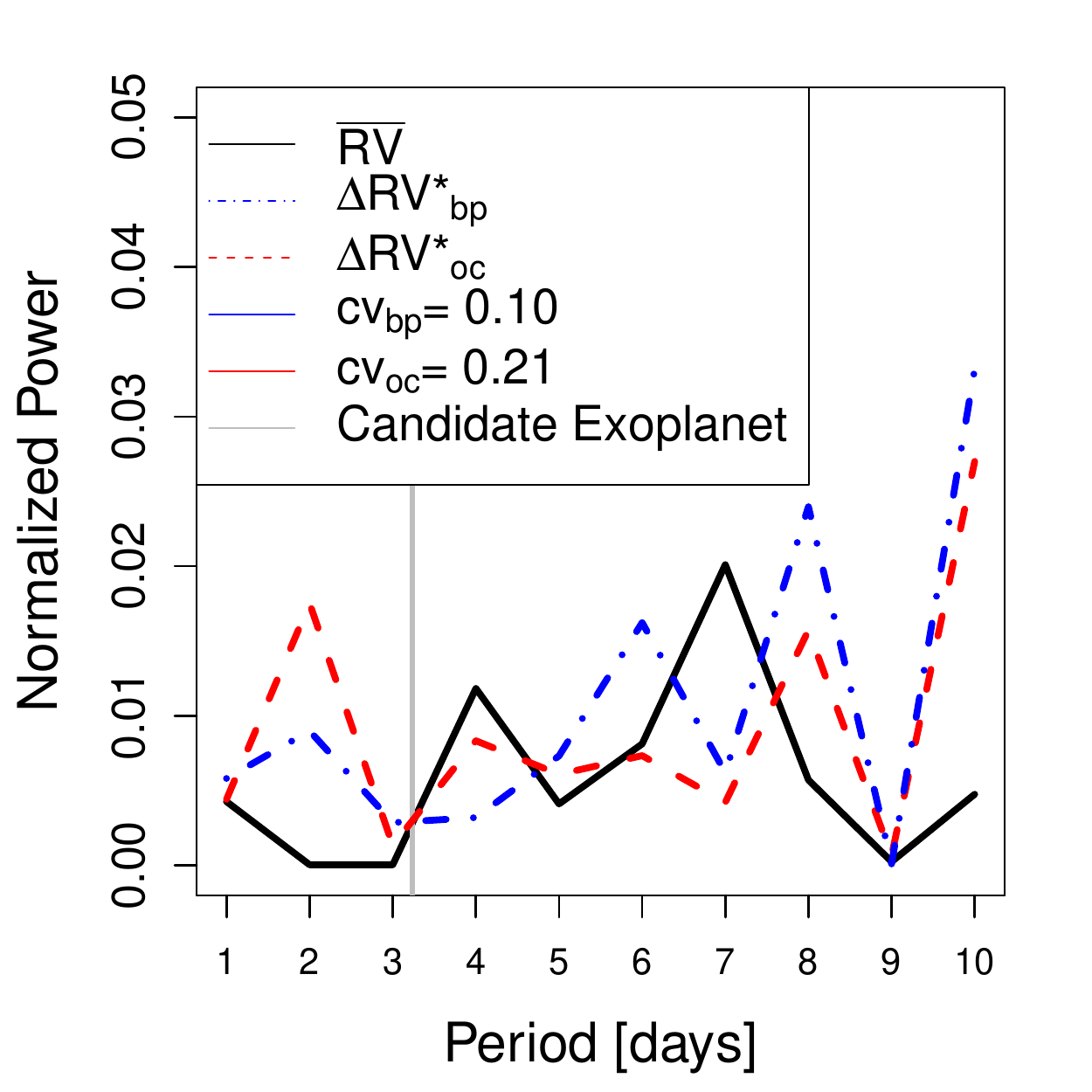}
\caption{GLS periodograms of $\alpha$ Cen B derived from the subset of 459 RVs, also analyzed by \citet{dumusque2012earth, rajpaul2015ghost}. In detail, the periodograms have been computed from the original $\overline{RV}$ time series (solid black line), $\Delta RV^*_{oc}$ (dashed red line), and $\Delta RV^*_{bp}$ (dash-dotted blue line). The plot focuses on the neighborhood of 3.2357 days (highlighted as a vertical gray line), which corresponds to the orbital period of the exoplanet announced by \citet{dumusque2012earth} and then disproved by \citet{rajpaul2015ghost}. We note that all the GLS peaks are below the cv thresholds for both of the methods.}
\label{fig:alphacentb_GLS_459}
\end{figure}

Finally, to test the strength of the {bp} method, we also applied it to the RV data set analyzed by \citet{dumusque2012earth} and then by \citet{rajpaul2015ghost}. As already noted, that data set made of 459 CCFs is a subset of the RV time series considered in this work. The GP framework presented in \citet{rajpaul2015ghost} contains significantly fewer free parameters than the model by \citet{dumusque2012earth} (14 vs. 23). Using our {bp} method, we found 3 piecewise stationary segments. This makes our model comparable to the GP model of \citet{rajpaul2015ghost} in terms of free parameters (12 in our case), with the advantage that the linear correction we propose in Eq. \eqref{eq:RV:correction} is simpler than any GP framework. According to our analyses, whose consequent GLS periodograms are displayed in Fig.~\ref{fig:alphacentb_GLS_459}, there is not any statistical evidence showing the presence of an exoplanet having an orbital period of $\sim$\,3.2 days, disproving the discovery by \citet{dumusque2012earth} and confirming the conclusions presented in \citet{rajpaul2015ghost}. In fact, our GLS normalized powers are well below the respective $\text{cv}_{bp}$ and $\text{cv}_{oc}$ thresholds for a wide neighborhood of 3 days (see Fig.~\ref{fig:alphacentb_GLS_459}). Therefore, we conclude that the GLS signal at $\sim$\,3.2 days visible in Fig.~4 of \citet{dumusque2012earth} and announced as an exoplanet is instead likely caused by an overfitting issue.

\subsection{Comparison between the optimal and the suboptimal solutions of the {bp} method} 
\label{sec:bp.comparison}

The {bp} method does not only return the best $\hat{D}$ and the corresponding change point locations in the time series; the {bp} method also returns the best partition for each $D=1,\dots, D_{\mathrm{max}}$. We can interpret those solutions as the best segmentations for each fixed $D$. When working with $\alpha$ Cen B, we found $\hat{D}=4$ as the best number of breakpoints, by applying the RSS criterion to Eq. (4), combined with the BIC-based penalty function of Eq. (5). Indeed the $\hat{D}=4$-model has the lowest BIC among all the other models, as emphasized by the $\Delta\text{BIC}$ values reported in Table~\ref{tab:bp.vs.bp.alphacentb}.

\begin{figure*}[htbp]
\centering
\includegraphics[width=\textwidth]{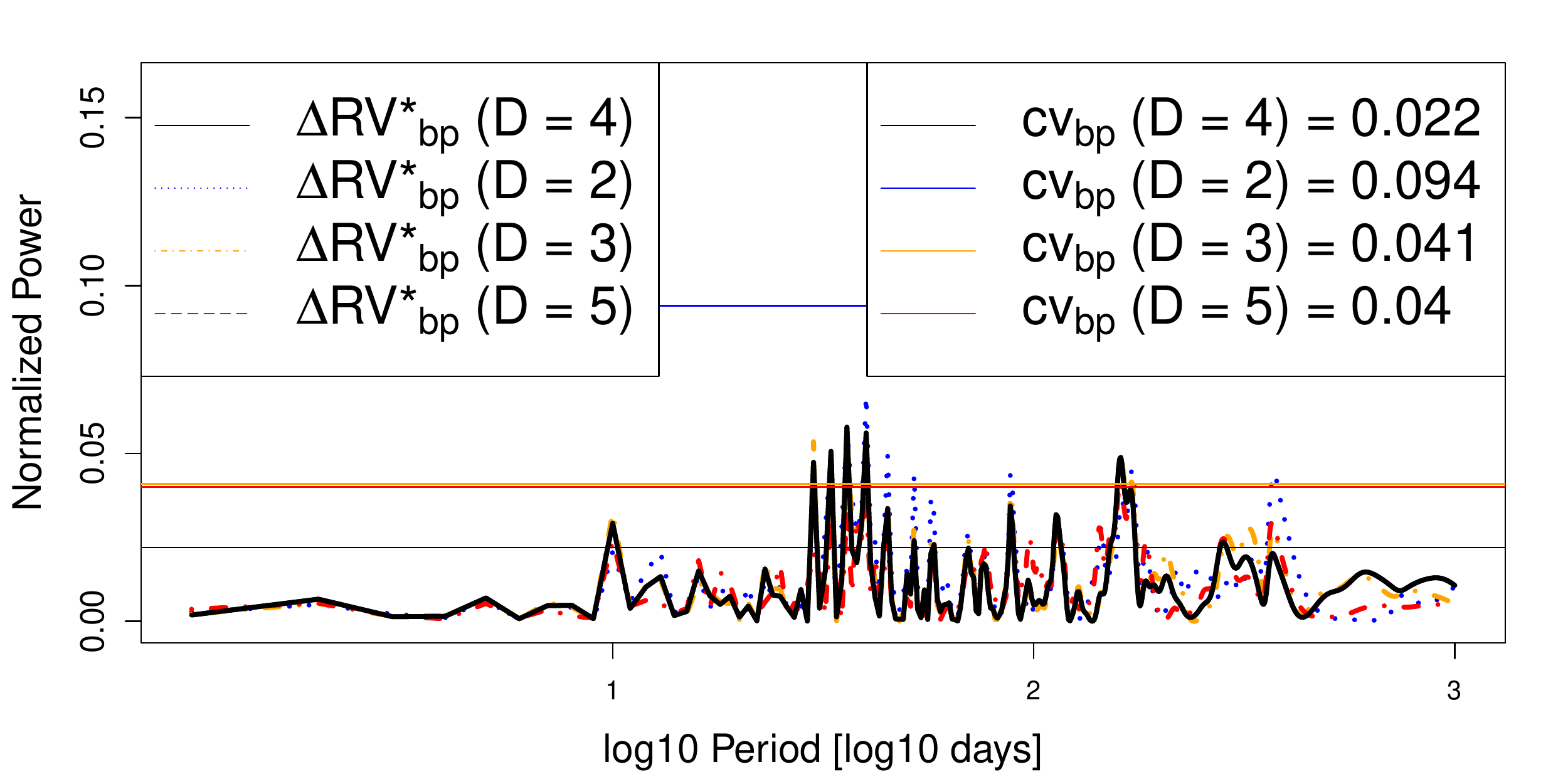}
\caption{
Generalized Lomb-Scargle (GLS) periodograms of $\alpha$ Cen B derived from the $\Delta RV^*_{bp}$ when using $\hat{D}=4$ (solid black line; this solution is the same reported in Figure \ref{fig:alphacentb_GLS} as the dash-dotted blue line). The GLS periodograms derived from the $\Delta RV^*_{bp}$ when using $D=2,3,$ and $5$ are displayed as a dotted blue line, a dash-dotted orange line and a dashed red line, respectively. The $x$-axis uses the base-$10$ logarithmic transformation of the period in order to improve the readability of the plot. The cv values for $\Delta RV^*_{bp}$ are displayed as black, blue, orange and red lines for the cases $D=4,2,3$, and $5$, respectively. Although the rms of $\Delta RV^*_{bp}$ for the considered cases are not statistically different, the {bp}-optimal solution ($\hat{D}=4$) leads to the smallest cv.
}
\label{GLSbp}
\end{figure*}

We decided to further compare the results obtained from the best partition (i.e., the optimal solution $\hat{D}=4$) with the other suboptimal solutions derived from the other $D$ values. From a statistical standpoint, we found that the solutions for $D=2$, $D=3$ or $D=5$ did not differ noticeably from the optimal one in terms of rms computed on $\Delta RV^*_{bp}$; the results are presented in Table \ref{tab:bp.vs.bp.alphacentb}. In particular, we used the Wilcoxon test \citep[e.g.,][]{zimmerman1993relative} to compare the rms of $\Delta RV^*_{bp}$ for $\hat{D}=4$ with the the rms of $\Delta RV^*_{bp}$ for the other available cases (null hypothesis: the compared samples are not statistically different). The p-values inferred from the Wilcoxon tests were $0.27$, $0.83,$ and $0.68$, obtained by comparing the $\hat{D}=4$ optimal solution with the suboptimal solutions derived from $D=2,3,$ and $5$, respectively. The results of the Wilcoxon tests suggest that the solutions for $D=2,3,4,$ and $5$ are not statistically different for the commonly employed significance levels: $\alpha_{\mathrm{fixed}}=(0.01, 0.05, 0.1)$. However, the solution for $\hat{D}=4$ leads not only to the smallest rms of the $\Delta RV^*_{bp}$ time series, but also to the smallest cv among all the considered solutions, as shown in Figure \ref{GLSbp}. We further investigated the robustness of the estimated cv by performing a bootstrap simulation study \citep[e.g.,][]{efron1993introduction}. The results confirmed that the cv found when using $\hat{D}=4$ is significantly smaller than the cv retrieved by using suboptimal solutions; in fact the standard deviation of each cv is of the order of $10^{-3}$. Since all the GLS peaks below the cv threshold are ignored, dealing with low cv values means not cancelling out low GLS peaks, which might represent putative low-mass planets. As a consequence, from an astronomical standpoint, the optimal solution retrieved by the {bp} method is always preferable, especially when seeking super--Earths or Earth--like exoplanets.

\begin{table*}
\caption{
rms statistic of $\alpha$ Cen B RV time series for different partitions ($D$ is the number of breakpoints); we recall that $\text{rms}(\overline{RV})=3.79$\,\ms. The second to last row lists the p-values of the Wilcoxon tests, which were used to compare the rms of $\Delta RV^*_{bp}$ for $\hat{D}=4$ (the optimal solution) with the rms of $\Delta RV^*_{bp}$ for the other choices of $D$. Finally, the last row reports the BIC differences of the various $D$-models with respect to the optimal $\hat{D}$-model.
}
\label{tab:bp.vs.bp.alphacentb}
\centering
\begin{tabular}{llcccccc}
\hline\hline
\multicolumn{2}{c}{$\alpha$ Cen B, {bp} method} & $\hat{D}=4$ & $D=1$ & $D=2$ & $D=3$ & $D=5$ \\
\hline
rms $ RV^{*}_{\text{activity}}$ & [\ms] & $3.37$ & $2.43$  & $3.36$ & $3.36$  & $3.37$ \\
rms $\Delta RV^*_{bp}$ & [\ms] & $1.75$  & $2.97$ & $1.90$ & $1.78$ & $1.80$  \\
\multicolumn{2}{l}{Wilcoxon test p-value} & & $0.015$ & $0.27$ & $0.83$ & $0.68$  \\
\hline
\multicolumn{2}{l}{$\Delta\text{BIC}=\text{BIC}_D-\text{BIC}_{\hat{D}=4}$} &  & $+10360$ & $+1829$ & $+447$ & $+853$  \\
\hline
\end{tabular}
\end{table*}

\subsection{Exoplanets detection limit} \label{sec:det.lim}

\subsubsection{Simulation study based on $\alpha$ Cen B data} \label{sec:sim.study.alphacenb}

After the results were obtained by analysing real RV data of $\alpha$ Cen B, we decided to perform a simulation study to test how well the {bp} method works in detecting exoplanets. As already stressed, the RV time series of $\alpha$ Cen B is essentially made of stellar activity signals since no exoplanets have been detected \citep[e.g.,][]{rajpaul2015ghost}. Therefore, we added artificial Keplerian signals ($RV_K$) that would be generated by exoplanets in circular orbits to the original RV time series, we applied both the {bp} and {oc} methods to clean the data set for stellar activity (obtaining $\Delta RV_{bp}^p$ and $\Delta RV_{oc}^p$, respectively), and we checked to which extent  the two methods are able to find the simulated planets, distinguishing the planetary signals from stellar activity. Any Keplerian signal added to the RV time series translates in a shifting of the CCFs that changes according to the period, the amplitude, and the phase that is imposed. In particular, we used the following grid of values:  periods from 1 to 500 days, by steps of $1$ day for periods shorter than $50$ days, and then by steps of $10$ days for longer orbital periods, semiamplitudes from $0.1$ to $15$ \ms\, by steps of $0.1$ \ms and $51$ evenly sampled phases between $0$ and $2\pi$, for a total of $726750$ simulated planets.

Similar simulation studies have already been proposed by \cite{howard2010occurrence, Mayor-2011} and  \cite{simola2019measuring}.  In this work, the detection of an exoplanet was tested by inspecting the GLS produced from the $\Delta RV_{bp}^p$ and $\Delta RV_{oc}^p$ data.
This was implemented by first flagging as a positive finding those GLS features that are above the cv previously calculated for $\Delta RV^*_{oc}$ ($\text{cv}_{oc}=0.43$) and for $\Delta RV^*_{bp}$ ($\text{cv}_{bp}=0.022$)\footnote{Despite the addition of the Keplerian signal, we noted that the $\Delta RV^p$ cvs are basically the same as the cvs computed from $\Delta RV^*_{bp}$ and $\Delta RV^*_{oc}$, and therefore decided to use the already computed $\text{cv}_{oc}=0.43$ and $\text{cv}_{bp}=0.022$ throughout the following analysis.}. Then,  we checked that the normalized power at $P_p$ (i.e., $\mathcal{P}_p=\mathcal{P}(P_p)$) was statistically comparable to the theoretical normalized power $\hat{\mathcal{P}}_p$. This would appear from the time series only made of the Keplerian signal of the synthetic planet that had been perturbed with a normal distribution having a mean equal to $0$ and a standard deviation computed by subtracting in quadrature the rms induced by $RV_K$ from the rms of the $\Delta RV^p$ time series.
Because the theoretical normalized powers produced by any given exoplanet $\hat{\mathcal{P}}_p$ follow a normal distribution having constant standard deviation equal to $\sigma_{\hat{\mathcal{P}}}=0.028$ (which can be interpreted as the inner variability of the GLS periodogram), $\mathcal{P}_p$ is compared to $\hat{\mathcal{P}}_p$ with the commonly used Wald test \citep[e.g.,][]{gourieroux1982likelihood}. In particular, for a given exoplanet, the test checks whether $\mathcal{P}_p$ belongs to the $99\%$ confidence interval inferred from $\mathcal{N}(\hat{\mathcal{P}}_p,\sigma_{\hat{\mathcal{P}}})$. This test enables us to establish whether the expected and observed RV semiamplitudes are consistent, so to declare the synthetic exoplanet as detected.

\begin{figure*}
\begin{center}
\includegraphics[width=\textwidth]{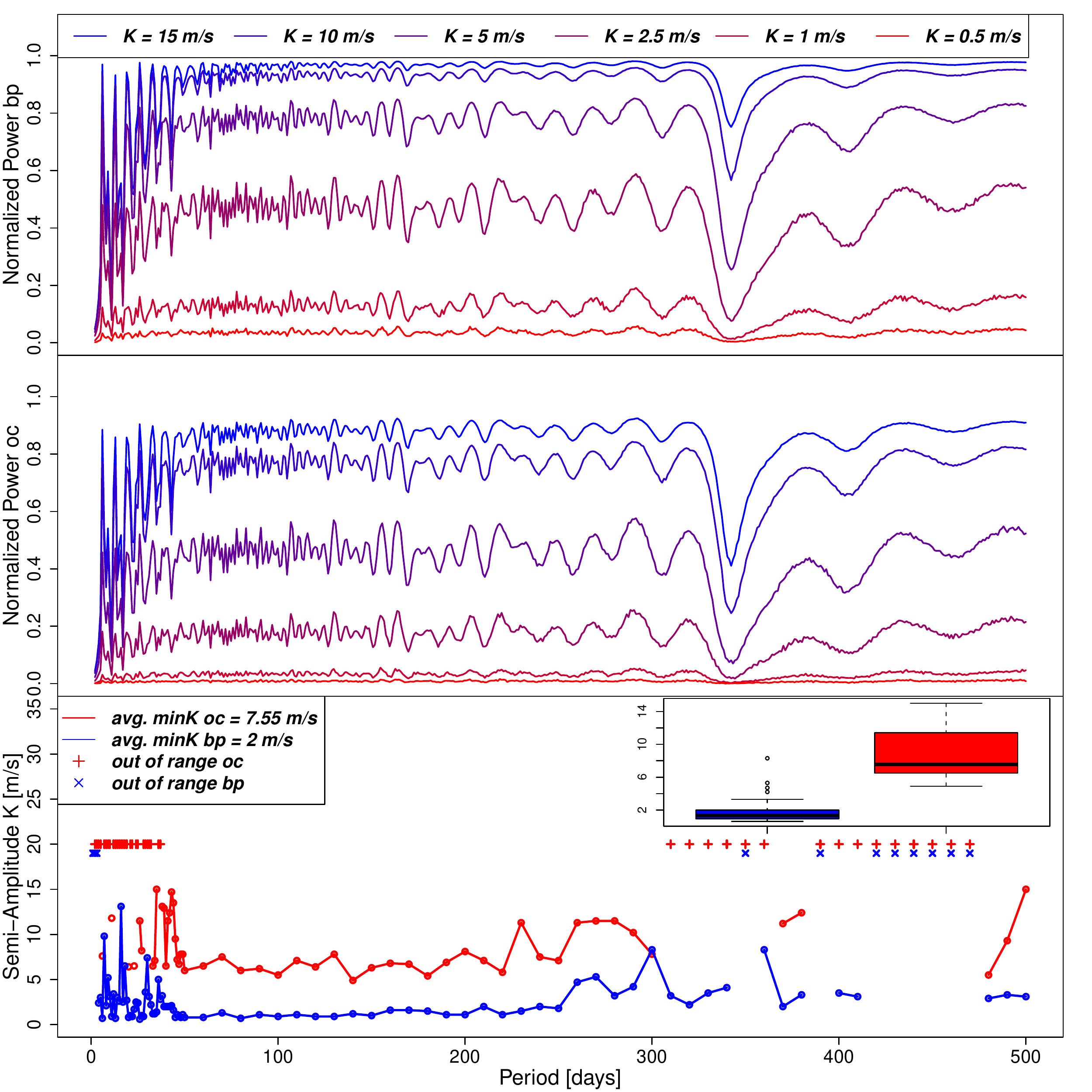}
\caption{{Upper panels}: GLS periodograms derived from synthetic Keplerian RV signals of $K= 15, 10, 5, 2.5, 1,$ and $0.5 \ms$, which were perturbed by injecting white noise whose amplitude follows a normal distribution with standard deviation of $1.75 \ms$ ({top}) and $3.48 \ms$ ({middle}). The unevenly sampled temporal baseline (i.e., the temporal baseline of $\alpha$ Cen B) is responsible for the power decrease at $365$ days, as pointed out in Sec. \ref{sec:K.gls}. {Bottom panel}: Detection thresholds of synthetic exoplanets injected into the $\alpha$ Cen B RV time series. It is the minimum semiamplitude $K$ at which a synthetic exoplanet of a given period is recovered from the two different GLS periodograms. The minimum $K$ values are inferred from the $\Delta RV_{bp}^p$ time series ({bp} method, dashed blue line) and from the $\Delta RV_{oc}^p$ time series ({oc} method, solid red line). Blue crosses ({bp} method) or red crosses ({oc} method) highlight the null detection of any exoplanet having $K \le 15$\ms. For both methods, a null detection occurs at $\sim$\,365 days, where the GLS normalized power drops. Overall, by correcting for stellar activity on each of the five temporal segments found by the {bp} method, the detection threshold of an exoplanet lowers  by $74\%$ on average. The two boxplots represented in the top right inset show the distributions of the minimum $K$ for the two methods. The two distributions are statistically different as quantified by the Wilcoxon test through its p-value\,=\,$2.22\cdot10^{-16}$. The detection threshold of an exoplanet lowers by $78\%$ on average, when focusing on planets up to an orbital period of $250$ days.}
\label{fig:alphacentb_detection.limit}
\end{center}
\end{figure*}

This procedure is repeated for each of the $726750$ simulated planets. In order to quantify the $K$ detection threshold, we search for the minimum RV amplitude at which, for a given orbital period, at least $90\%$ of the planets having different phases are detected. The results of the simulation study, presented in Fig. \ref{fig:alphacentb_detection.limit}, confirm the intuition that properly dividing the $\overline{RV}$ data into segments (where each segment is piecewise stationary) significantly improves the chances of detecting exoplanets that have smaller amplitudes with respect to the {oc} method. In fact, by applying the correction for stellar activity based on the {bp} method, the detection threshold is on average lower by $74 \%$ with respect to the detection threshold estimated using the {oc} method. The median threshold for the {bp} method is $2 \ms$, while the median threshold for the {oc} method is $7.55 \ms$.

 A statistical test to compare the two vectors of minimum RV amplitudes was also carried out. Assuming as null hypothesis that the two groups are not statistically different, the Wilcoxon test estimated a p-value equal to $2.22\cdot10^{-16}$, which means that the null hypothesis is strongly rejected and that the two groups are statistically different. The detection threshold of an exoplanet lowers by $78\%$ on average when focusing on planets up to an orbital period of $250$ days. Fig. \ref{fig:alphacentb_detection.limit} displays the orbital periods for which the {oc} method and the {bp} method were not able to detect any planet having $K$ lower than our grid upper limit. In order to explain this situation, we further explored the behavior of the GLS, highlighting our findings in Sec. \ref{sec:K.gls}.

\subsubsection{Behavior of the GLS periodogram} \label{sec:K.gls}

We investigated the behavior of the GLS periodogram in order to understand the inability to detect exoplanets. In particular, we were interested in understanding if the unavailability of detections at a given $P_{\mathrm{orb}}$ were caused by the criteria we employed for our simulation study or rather by an issue related to the GLS periodogram. We generated several Keplerian signals sampled upon the epochs of observations of $\alpha$ Cen B using the following grid of values\footnote{This grid produces fewer synthetic exoplanets than before. This made the present simulation quicker, without penalizing the strength of our conclusions.}: periods $P_{\mathrm{orb}}$ from 1 to 500 days, by steps of $1$ day, semiamplitudes $K$ from $0.1$ to $15$ \ms\, by steps of $0.1$ \ms and a phase value, randomly sampled from the [0, 2$\pi$] interval.
Each vector of RV was perturbed by using a normal distribution having a mean equal to 0 and a standard deviation of $1 \ms$, which is the error expected by HARPS. At this stage, as vector of RV, just the Keplerian signal and the error term are used.

\begin{figure*}
\centering
\includegraphics[width=\textwidth]{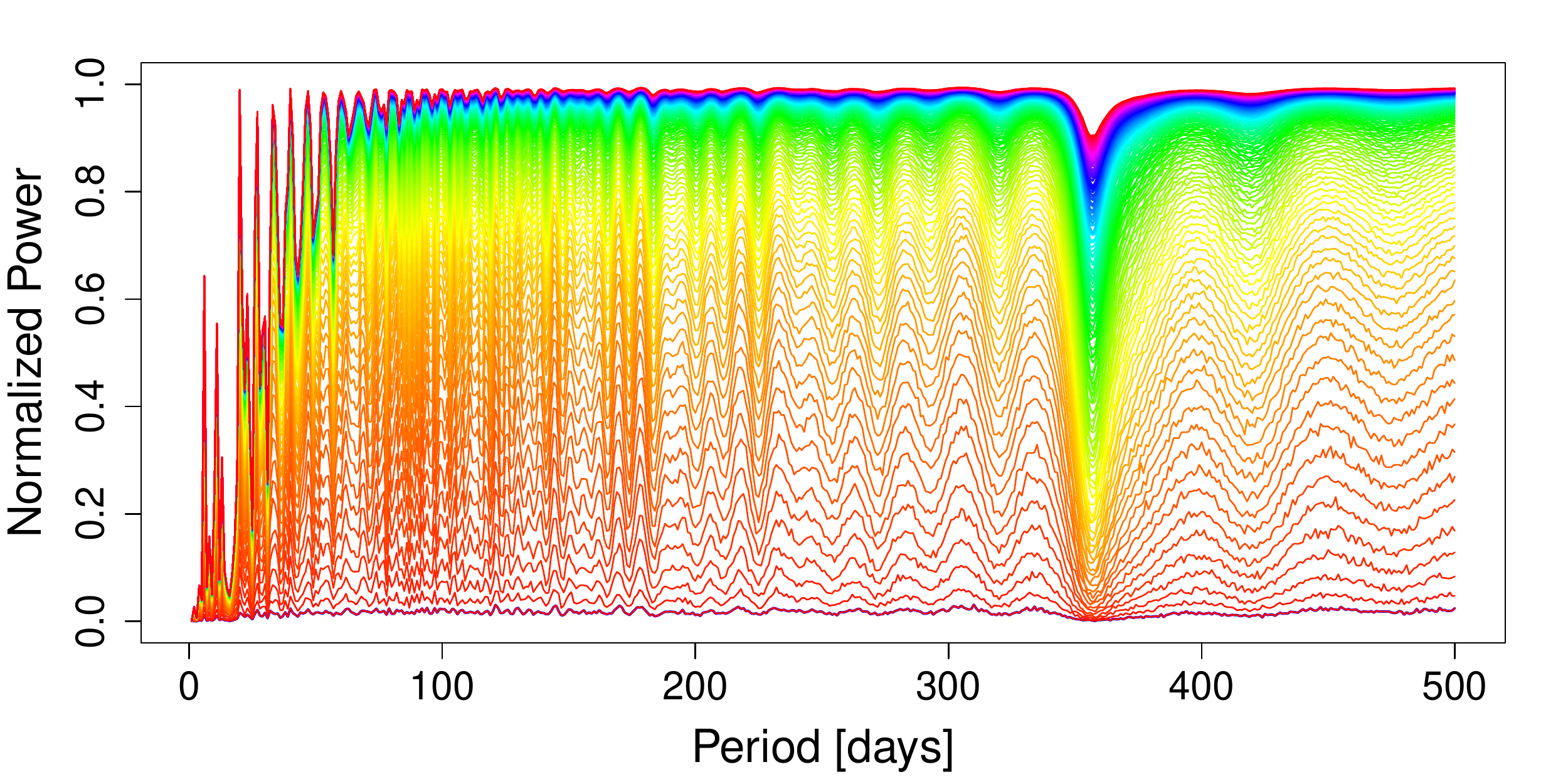}
\caption{Relation between the peak normalized power $\mathcal{P}$ of the GLS periodogram obtained from a synthetic Keplerian RV signal and the period of that signal (i.e., the orbital period of the synthetic exoplanet). The different colors represent the different $K$ values imposed on the Keplerian signals, which increase from bottom to top. The ruggedness of the lines is due to the white noise that was also injected into the RV data series. Regardless of the considered $K$ (i.e., the color), $\mathcal{P}$ decreases at $\sim$\,365 days (close to the revolution period of the Earth), as expected in ground-based observations.}
\label{fig:h_P.rel}
\end{figure*}

When plotting $\mathcal{P}$ as a function of $P_{\mathrm{orb}}$ for any given value of $K$, rather than finding constant behavior, there is a problematic behavior when considering short orbital periods. Moreover, we see a systematic decrease of $\mathcal{P}$ at some specific $P_{\mathrm{orb}}$ values, which is caused by the data sampling due to $\alpha$ Cen B visibility. In particular, a clear decrease occurs at $P_{\mathrm{orb}}$\,$\sim$\,365 days, as expected in ground-based observations.
These conclusions are confirmed in our simulation study that aims to establish the detection threshold of the {bp} and {oc} methods, as both the methods have some issues in detecting planets that have a short orbital period, and both methods are unable to detect planets with $P_{\mathrm{orb}}$\,$\sim$365 days (Fig. \ref{fig:h_P.rel}).

\subsection{Comparison of the oc and the bp methods applied to HD\,215152, HD\,10700, and HD\,192310}\label{sec:otherStars}

As done for the RV time series of $\alpha$ Cen B, we repeated the comparison between the {oc} and the {bp} methods for three other main sequence stars: HD\,215152 \citep[K3V,][]{Delisle:2018aa}, HD\,10700 \citep[G8V,][]{Feng:2017ac}, and HD\,192310 \citep[K2V,][]{Pepe-2011}. 
Unlike $\alpha$ Cen B, which does not host any known planet, HD\,215152 and HD\,10700 host four planets \citep[][each]{Delisle:2018aa,Feng:2017ac}, while HD\,192310 hosts two planets \citep{Pepe-2011}.
Therefore, to obtain $\Delta RV^*$-like indicators that are consistent with those derived for $\alpha$ Cen B, we should also remove the exoplanetary signals from these RV time series, after applying the correction for stellar activity.
Since the discovered exoplanets orbiting HD\,215152 and HD\,10700 are at the level of instrumental precision, removing those planetary signals would have led to biased results. For this reason, in the following analyses we decided to remove the planetary signals only for the HD\,192310 case. 

\begin{figure*}[htbp]
\centering
\includegraphics[width=\textwidth]{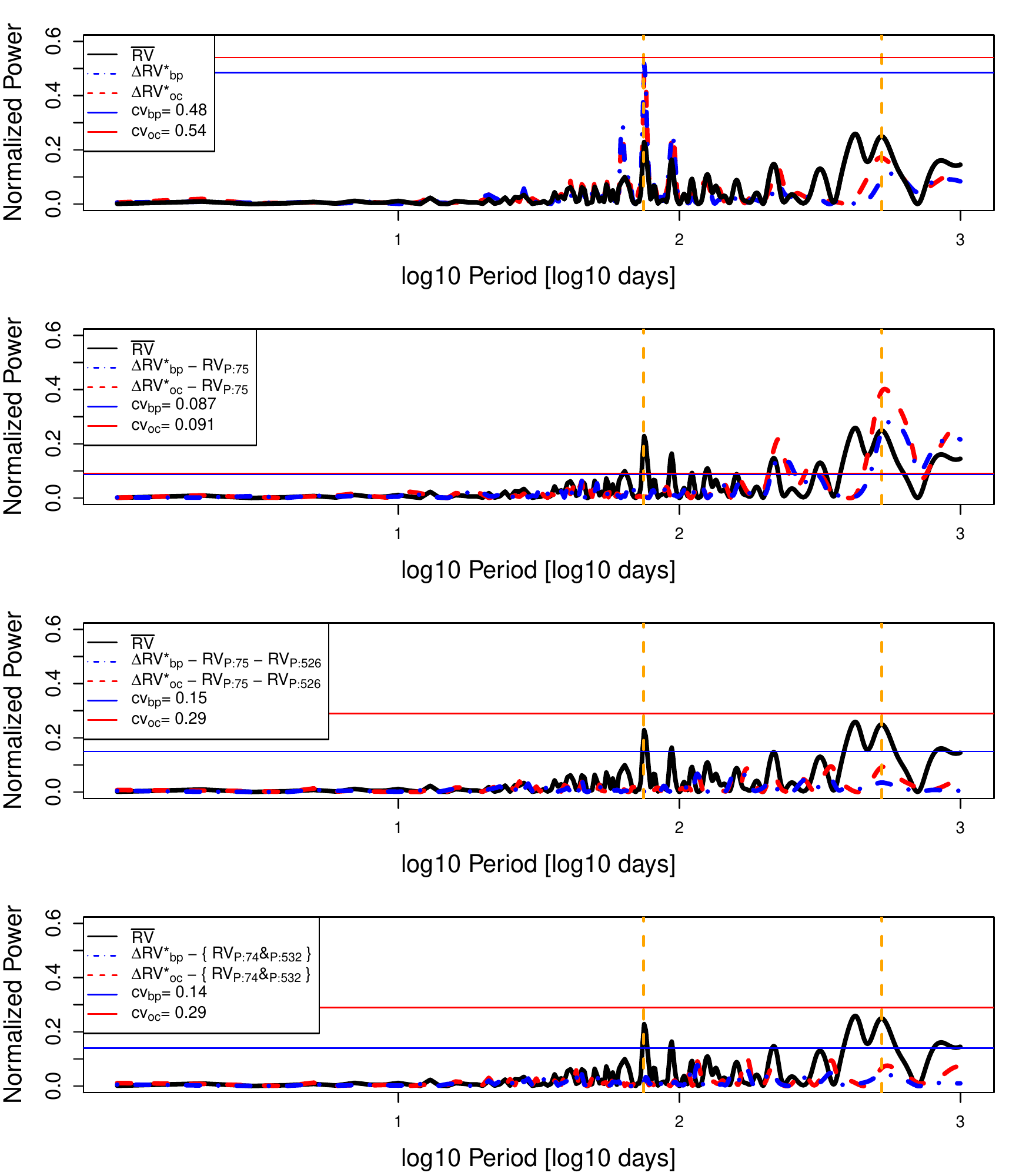}
\label{fig:hd192310.GLS}
\caption{Similar to Fig. \ref{fig:alphacentb_GLS}, but for HD\,192310. The vertical dashed orange lines at $\sim$\,75 days and $\sim$\,526 days highlight the orbital periods of the two Neptune-mass planets found by \cite{Pepe-2011}. 
{Top panel}: As a result of the stellar activity correction, the {bp}-peak at $\sim$\,75 days is above the respective cv value, revealing the planetary signal. {Second panel}: Once the $P$\,$\sim$\,75 days planetary signal ($RV_{P:75}$) is removed from the original time series, another peak corresponding to the $P$\,$\sim$\,526 days planetary signal pops up (this time both the {bp}- and {oc}-peaks are above the respective cv values). {Third panel}: After removing also the second planetary signal ($RV_{P:526}$), the GLS periodogram does not show any other significant peak. {Bottom panel}: From the activity-corrected RV time series, we subtracted the Keplerian signals of the two planets ($RV_{P:74 \& P:532}$) inferred from the global fit of the joint A-MCMC analysis. It turned out that $P_b$\,$\sim$\,75 days and $P_c$\,$\sim$\,532 days. The main orbital parameters are summarized in Table~\ref{tab:HD192310planetParameters}.
}
\label{fig:hdhd192310.GLS_detection}
\end{figure*}

We used the adaptive Markov Chain Monte-Carlo (A-MCMC) algorithm\ \citep[][]{haario2005componentwise} to estimate the orbital parameters of the two known exoplanets orbiting HD\,192310. In detail, following a step-by-step procedure, first we corrected the original $\overline{RV}$ time series by using both the {oc} and the {bp} methods. As a result, a clear peak due to the exoplanet with $P_{\mathrm{orb}}\approx75$ days appeared in the two GLS periodograms (although only in case of the {bp} method $\mathcal{P}_{bp}(75\,d)>\text{cv}_{bp}$; top panel of Fig.~\ref{fig:hdhd192310.GLS_detection}). Therefore, we estimated the orbital parameters of this exoplanet by using the A-MCMC algorithm, and removed its signal ($RV_{P:75}$) from the $\Delta RV^*$ time series. After producing new GLS periodograms from the $\Delta RV^*-RV_{P:75}$ time series, for both methods we saw another peak above the respective cv values, which was compatible with $P_{\mathrm{orb}}\approx526$ days of the second exoplanet (second panel of Fig.~\ref{fig:hdhd192310.GLS_detection}). Then we used the A-MCMC algorithm again to estimate the orbital parameters of the exoplanet with $P_{\mathrm{orb}}\approx526$ days, we removed its signal ($RV_{P:526}$) from the previously used RV time series, and we produced new GLS periodograms based on the $\Delta RV^*-RV_{P:75}-RV_{P:526}$ time series. The inspection of these GLS periodograms did not show any other significant peak (third panel of Fig.~\ref{fig:hdhd192310.GLS_detection}) and we concluded that only the {bp} method finds both the two known exoplanets following the cv-threshold criterion. Finally, we launched a global A-MCMC run accounting for both planets at the same time to obtain unbiased exoplanetary parameters, which were used to build our final GLS periodograms, where both the planetary signals were subtracted from $\Delta RV^*$ (bottom panel of Fig.~\ref{fig:hdhd192310.GLS_detection}).

Our outcomes for the HD\,192310 planetary system are listed in Table.~\ref{tab:HD192310planetParameters} and compared with the results obtained by \citet{Pepe-2011}. The precision we got for the planet $b$ parameters is comparable with that obtained by \citet{Pepe-2011} and all those parameters are consistent within $\sim$\,1.5$\sigma$, except for the orbital period $P_b$. Our $P_b=74.06_{-0.09}^{+0.10}$ days is 4.8$\sigma$ away from the \citet{Pepe-2011} estimate. However, it is in excellent agreement with the recent revision by \citet{rosenthal21}, who found a period of $74.062\pm0.085$ days.
Regarding planet $c$, instead, we sensibly improve the precision on the determined parameters (from a factor of two up to a factor of four) with respect to the results provided by \citet{Pepe-2011}. All the respective estimates are within 2$\sigma$, except for $K_c$, for which we infer a higher value after applying our {bp} method for cleaning the RV time series. Even if a detailed analysis of the orbital parameters is beyond the scope of this paper, we note that our estimates for HD\,192310 $c$ are the most precise available in the literature so far\footnote{According to the Nasa Exoplanet Archive (\url{https://exoplanetarchive.ipac.caltech.edu/overview/HD192310}), where only the set of parameters by \citet{Pepe-2011} is provided.}.
\begin{table*}
\caption{
Planetary parameters as derived in this work and by \citet{Pepe-2011} of both HD\,192310 $b$ and HD\,192310 $c$. The relative uncertainties are specified within brackets, while the $\Delta_{\sigma}$ column refers to the statistical tension between the generic pair $(x_1\pm\sigma_1;x_2\pm\sigma_2)$, that is $\Delta_{\sigma}=\frac{|x_1-x_2|}{\sqrt{\sigma_1^2+\sigma_2^2}}$.
}
\label{tab:HD192310planetParameters}
\centering
\resizebox{\textwidth}{!}{%
\begin{tabular}{llcccccccccc}
\hline\hline
\multicolumn{2}{c}{Parameters} &\multicolumn{5}{c}{HD\,192310 $b$} & \multicolumn{5}{c}{HD\,192310 c} \\
&& \multicolumn{2}{c}{This work} & \multicolumn{2}{c}{\citet{Pepe-2011}} & $\Delta_{\sigma}$ & \multicolumn{2}{c}{This work} & \multicolumn{2}{c}{\citet{Pepe-2011}} & $\Delta_{\sigma}$ \\
\hline      
$K$ & [\ms] & $2.74_{-0.16}^{+0.14}$ & (5.5\%) & $3.00\pm0.12$ & (4.0\%) & 1.4 & $3.08\pm0.18$ & (5.8\%) & $2.27\pm0.28$ & (12\%) & 2.4 \\
$P$ & [d] & $74.06_{-0.09}^{+0.10}$ & (0.13\%) & $74.72\pm0.10$ & (0.13\%) & 4.8 & $531.6_{-3.2}^{+3.9}$ & (0.7\%) & $525.8\pm 9.2$ & (1.7\%) & 0.6 \\
$M_p\sin{i}$ & [$M_{\oplus}$] & $15.0\pm0.8$ & (5.3\%) & $16.9\pm0.9$ & (5.3\%) & 1.6 & $31.5_{-1.6}^{+1.6}$ & (5.1\%) & $24\pm5$ & (21\%) & 1.4 \\
$a$ & [AU] & $0.320\pm0.003$ & (0.9\%) & $0.320\pm0.005$ & (1.6\%) & 0.0 & $1.19_{-0.0047}^{+0.0057}$ & (0.4\%) & $1.180\pm0.025$ & (2.1\%)  & 0.4 \\
$e$ & & $0.23\pm0.05$ & (22\%) & $0.13\pm0.04$ & (31\%) & 1.6 & $0.35_{-0.05}^{+0.04}$ & (13\%) & $0.32\pm0.11$ & (34\%) & 0.3 \\
$\omega$ & [deg] & $145_{-17}^{+13}$ & (10\%) & $173\pm20$ & (12\%) & 1.1 & $64_{-10}^{+9}$ & (15\%) & $110\pm21$ & (19\%) & 2.0 \\
\hline
\end{tabular}}
\end{table*}
 Coming to the effectiveness of modeling the stellar activity, when compared to the rms of the $\Delta RV^*_{oc}$ time series, the rms of $\Delta RV^*_{bp}$ is lower by $0.21 \ms$ (HD\,215152), $0.06 \ms$ (HD\,10700), and $0.1 \ms$ (HD\,192310, after the planets removal). This means that using Eq.~(\ref{eq:RV_bp}) rather than Eq.~(\ref{eq:RV_oc}) for correcting $\overline{RV}$, we better explain the $\overline{RV}$-variability by gaining 9.5\%, 4.3\%, and 6.6\% for HD\,215152, HD\,10700, and HD\,192310, respectively.
 The {bp} method has to be preferred also following the BIC minimization criterion for model selection; in fact, $\Delta\text{BIC}=\text{BIC}_{oc}-\text{BIC}_{bp}=+79$, $+503$, and $+174$  for HD\,215152, HD\,10700, and HD\,192310, respectively.

The number of available data points per time series is significantly smaller if compared to the $\alpha$ Cen B time series, especially for HD\,215152 and HD\,192310.
It seems therefore reasonable that the {bp} method finds a lower number of piecewise-stationary segments as optimal solutions for those stars ($\hat{D}=1$ and 2 for HD\,215152 and HD\,192310, respectively). Instead, concerning HD\,10700, a larger number of observations spread on a longer baseline is available, which yields to $\hat{D}=5$ optimal breakpoints.

As a final note, since HD\,10700 is a quiet star, the {bp} vs. {oc} improvement in modeling the stellar activity is lower (i.e., we registered a lower gain when comparing the $\Delta RV^*_{bp}$ vs. $\Delta RV^*_{bp}$ rms, as expected). However, the {bp} method still models the stellar activity better than the {oc} method and the $\Delta$BIC confirms a strong preference for the {bp} method. Overall, the obtained results confirm our assumption that the CPD algorithm is particularly effective when focusing on active stars.
Relevant tables and plots synthesizing our results can be found in Appendix \ref{appendixStars}.

\section{Discussion} \label{sec:discu}

We tested the {bp} method by using real measurements taken from four stars, carrying out comparisons with the {oc} method that considers a single correction for stellar activity on the entire time series. Before performing the comparisons, the data were properly cleaned by removing the outliers from the set of $\overline{RV}$ values. The results suggest that properly dividing the $\overline{RV}$ time series into segments (where each segment is piecewise stationary) is a helpful operation when trying to account for higher variations in $\overline{RV}$ caused by stellar activity. For all the considered stars, the rms on $\Delta RV^*_{bp}$ are smaller than the rms on the corresponding $\Delta RV^*_{oc}$. 

The stronger the activity signals within the time series, the more effective the {bp} method is at modeling the $\overline{RV}$ variations. In addition, the longer the time series, the more likely sensible variations in stellar activity have occurred, and so the {bp} method is particularly suitable to detrend RV data. This is especially the case of $\alpha$ Cen B, which is characterized by strong solar-like activity signals \citep[e.g.,][]{Thompson-2017, Dumusque:2018aa}. By dividing the RV measurements of $\alpha$ Cen B into five piecewise stationary segments, we reach an rms of $\Delta RV^*_{bp}$ of $1.75 \ms$ (to be compared with an rms of $\Delta RV^*_{oc}$ of 3.48 \ms resulting from the application of the {oc} method). Also, the activity indicators' variability within each segment is smaller than the variability between each segment, as displayed in Figure \ref{fig.boxplots} and summarized by Table \ref{tab:change.points.alphacentb}, which provides strong evidence that the segmentation proposed by the {bp} method captures those time series locations where the stellar activity changes significantly. As a consequence, the {bp} method is able to interpret a larger fraction of $\overline{RV}$ variability in terms of stellar activity: $89\%$ vs. $40\%$ for the {oc} method. 

The $\alpha$ Cen B GLS periodogram obtained with the {bp} method shows a much smaller number of peaks caused by active regions with respect to the GLS periodogram obtained by performing the {oc} method. When producing the GLS periodogram from the $\Delta RV^*_{oc}$ time series, we found that there were no peaks above $\text{cv}_{oc}$. Conversely, when considering the $\Delta RV^*_{bp}$ time series, $11$ peaks in the GLS were above the critical value of $\text{cv}_{bp}$ (see Table~\ref{tab:peaks.alphacentb}). We  further investigated the nature of those peaks to check whether they could be produced by exoplanets. For a given exoplanet candidate, starting from $\Delta RV^*_{bp}$ and assuming a Keplerian model, we used the A-MCMC algorithm to retrieve the posterior distribution of the parameters of interest. Then, we used the marginal posterior means for each parameter in order to estimate the RV-variations caused by the candidate planet. Finally, we compared the rms of $\Delta RV^*_{bp}$ with the new rms obtained by subtracting to $\Delta RV^*_{bp}$ the RV Keplerian signal that would be caused by the candidate planet. We did not find any strong signal suggesting the possible presence of an exoplanet, confirming the conclusions provided in \cite{rajpaul2015ghost}.
\begin{table}[htbp]
   \caption{Periods $P$ and corresponding normalized powers $\mathcal{P}$ found in the $\Delta RV_{bp}^*$ GLS periodogram of $\alpha$ Cen B, whose peaks are above the critical value of $0.022$.}
 \label{tab:peaks.alphacentb}
\centering
\begin{tabular}{lc}
\hline\hline
$P$ [d] & $\mathcal{P}$ \\
\hline                        
$10$ & $0.026$  \\
$33$ & $0.048$          \\
$36$ & $0.049 $      \\
$40$ & $0.035 $      \\
$44$ & $0.025 $      \\
$58$ & $0.033 $      \\
$70$ & $0.033 $      \\
$88$ & $0.030 $      \\
$113$ & $0.027 $             \\
$161$ & $0.035 $             \\
$169$ & $0.033 $             \\
\hline
\end{tabular}
\end{table}

When repeating the comparative analyses for HD\,215152, HD\,10700, and HD\,192310, the {bp} method produces $\Delta RV^*_{bp}$ time series having rms lower by $0.21 \ms$ for HD\,215152, $0.06 \ms$ for HD\,10700, and $0.1 \ms$ for HD\,192310 when compared to the rms of $\Delta RV^*_{oc}$. In other words, the {bp} method increases the fraction of $\overline{RV}$ variability that can be explained in terms of stellar activity by $9.5\%$, $4.3\%$, and $6.6\%$ for HD\,215152, HD\,10700, and HD\,192310, respectively. The improvement given by the {bp} method is less evident in these three cases, as these stars are less active than $\alpha$ Cen B and their time series span a shorter temporal range. Since these time series already appear to be piecewise stationary as a whole, the {bp} and {oc} methods are comparable. However, thanks to its better cleaning performance, only the {bp} method was able to detect both the exoplanets hosted by HD\,192310, following our cv-threshold criterion. In particular, we essentially confirm the orbital parameters estimates already available in the literature, but we sensibly improve the precision in the case of HD\,192310 $c$. Given the key role of the planetary mass when studying the composition of exoplanets, we emphasize that our $M_p\sin{i}$ estimates are affected by an error of $\sim$\,5\% for both planet $b$ and planet $c$. We reached the same precision level of \citet{Pepe-2011} for planet $b$, while we improved the precision of $M_p\sin{i}$ of HD\,192310 $c $ by a factor of approximately four.

To further evaluate the ability of the {bp} and {oc} methods to detect exoplanets, we designed a simulation study starting from $\alpha$ Cen B data. We added several synthetic Keplerian signals to the time series and checked the exoplanet detection effectiveness of both the {bp} and {oc} methods when dealing with RV data points contaminated by solar-like activity signals. We found that the {bp} method lowers the detection threshold (i.e., the minimum $K$ of the Keplerian signal at which a planet is detected from the inspection of the GLS periodogram) with respect to the {oc} method by $74\%$ when considering planets up to an orbital period of $500$ days.
\section{Conclusion} \label{sec:conclu}

Stellar activity, in the form of active regions evolving on a star's photosphere, has so far been the major obstacle for the detection and the characterization of Earth-like exoplanets when using the RV method. Spots and faculae cause variations in the shape and in the width of the CCF, changing the correlations between $\overline{RV}$ and the indicators of stellar activity, such as $A$, $\gamma$, and FWHM$_\text{SN}$. Since an exoplanet would not change the shape of the CCF, but just its barycenter, a common strategy to account for stellar activity is to employ a linear correction of the RV time series involving the activity parameters. In fact, it is well known that variations in the correlations between $\overline{RV}$ and these activity indicators suggest the presence of active regions evolving on the stellar photosphere over time. A simple way to model the changes in $\overline{RV}$ caused by stellar activity is provided by Eq. \eqref{eq:RV:correction}. Since RV surveys often spread over years of measurements, it seems reasonable to assume that the stellar activity level changes multiple times during the observational period. Rather than using an overall correction for stellar activity on the entire time series, we still rely on Eq. \eqref{eq:RV:correction}, but we propose  performing multiple corrections by suitably dividing the overall time series into segments. The number of segments depends on how often the correlations between $\overline{RV}$ and the activity indicators significantly change. 

In order to estimate the time series locations where the dependence of stellar activity upon activity indicators significantly changes, we draw attention to the family of CPD algorithms.

In particular, in this paper we used the CPD-based {bp} method \citep[e.g.,][]{bai2003computation} to properly model the variations of $\overline{RV}$ caused by stellar activity. We compared the effectiveness of the {bp} method with the commonly employed {oc} method, by using real observations taken on four different stars. 
The results show that identifying the locations in the RV data where the correlations between $\overline{RV}$ and the indicators of stellar activity significantly change, produces much cleaner RV time series as we model the stellar activity signals on each of the piecewise stationary segments. The GLS periodograms are then less contaminated by the presence of spurious periodical peaks caused by stellar activity. As demonstrated by our simulation study on $\alpha$ Cen B, the {bp} method was able to detect exoplanets that produce RV amplitudes  $74\%$ smaller than those detected  by the {oc} method.

Finally, we note that the {bp} method is most effective when working with active stars whose RV time series are made of several hundreds of  data points. In fact, the longer the time series, the more likely sensible variations in stellar activity have occurred, suggesting that the {bp} technique is a suitable statistical tool for removing activity-induced variations from the RV data.

\begin{acknowledgements}
The authors are extremely thankful to the CSC--IT Center for Science, Finland, for the computational resources provided to perform the analyses presented in this work.
US was funded by Academy of Finland grant no. 320182.  
JC was funded by the ERC grant no. 742158. 
XD is grateful to The Branco Weiss Fellowship--Society in Science for its financial support.
JCK was partially supported by the National Science Foundation under Grant AST 1616086 and 2009528, and by the National Aeronautics and Space Administration under grant 80NSSC18K0443.
The authors are grateful to all technical and scientific collaborators of the HARPS Consortium, ESO Headquarters and ESO La Silla who have contributed with their extraordinary passion and valuable work to the success of the HARPS project. This project has received funding from the European Research Council (ERC) under the European Union’s Horizon 2020 research and innovation programme (grant agreement SCORE No 851555). This work has been carried out within the framework of the NCCR PlanetS supported by the Swiss National Science Foundation.
\end{acknowledgements}

\bibliographystyle{aa}
\bibliography{mybib.bib}

\begin{appendix}

\section{HD\,215152, HD\,10700, and HD\,192310} \label{appendixStars}
In this Appendix we show the relevant Tables and Figures summarizing the results we obtained for the other stars of our sample: HD\,215152, HD\,10700, and HD\,192310. Similarly to $\alpha$ Cen B, we show the change point locations detected in each RV time series and the covariate correlations within each stationary segment. In addition, we show the effectiveness of both the {bp} and {oc} methods in cleaning the RV time series and the consequent GLS periodograms.

%
\begin{table*}
\caption{Similar to Table \ref{tab:change.points.alphacentb}, but for HD\,215152.}
 \label{tab:change.points.hd215152}
\centering
\begin{tabular}{lcccccccc}
\hline\hline
CPL & JD & Date & Time span &  CCFs &  $\overline{RV}$ & $A$ & $\gamma$ & FWHM$_\text{SN}$ \\
 & & & [d] & [\#] & [\ms] & [$\cdot10^{-1}$] & & [\ms] \\
\hline                        
$l_0$ & $2454699.82311$         &       21 Aug 2008 & \multirow{2}*{$797.74$} & \multirow{2}*{151} &  
\multirow{2}*{$ -0.7_{-1.8}^{+0.6}$} & 
\multirow{2}*{$3.0969_{-0.0027}^{+0.0021}$} & \multirow{2}*{$0.0094_{-0.0021}^{+0.0022}$} & \multirow{2}*{$6.0746_{-0.0093}^{+0.0126}$} \\
$l_1$ & $2455497.56753$         &       28 Oct 2009 & \multirow{2}*{$1398.15$} & \multirow{2}*{122} &  
\multirow{2}*{$0.95_{-0.05}^{+1.92}$} & \multirow{2}*{$3.0946_{-0.0025}^{+0.0025}$} & \multirow{2}*{$0.0071_{-0.0020}^{+0.0029}$} & \multirow{2}*{$6.0670_{-0.0051}^{+0.0067}$}     \\
$l_2$ & $2456895.71456$ &       26 Aug 2014 &  & & & & &        \\
\hline
\end{tabular}
\end{table*}
\begin{table*}
   \caption{Similar to Table \ref{tab:correlations.bp.alphacentb}, but for HD\,215152.}
   \label{tab:correlations.bp.hd215152}
\centering
\begin{tabular}{lcc}
\hline\hline
& $1$ & $2$ \\
\hline                        
$\rho(\overline{RV} ; A)$                  & $-0.09$ & $-0.01$   \\
& $(-0.24,  0.073)$ & $(-0.19,  0.17)$ \\
\hline  
$\rho(\overline{RV}; \gamma)$             & $0.30$ & $0.29$    \\
& $(0.15, 0.44)$ & $(0.12, 0.44)$ \\
\hline  
$\rho(\overline{RV}; \mathrm{FWHM_{SN}})$ & $0.19$ & $0.10$    \\
& $(0.028, 0.34)$ & $(-0.083,  0.27)$ \\
\hline
\end{tabular}
\end{table*}
\begin{table*}
\caption{Similar to Table \ref{tab:bp.vs.oc.alphacentb}, but for HD\,215152.}
  \label{tab:bp.vs.oc.hd215152}
\centering
\begin{tabular}{llcccc}
\hline\hline
\multicolumn{2}{c}{rms} & {bp} segment 1 & {bp} segment 2 &  {bp} overall & {oc} overall  \\
\hline                        
$\overline{RV}$ & [\ms] & $2.03$ & $2.19$ & $2.25$ & $2.25$  \\
$RV^{*}_{\text{activity}}$ & [\ms] & $0.63$ & $0.67$ & $1.56$  & $1.25$ \\
$\Delta RV^*$ & [\ms] & $1.93$  & $2.08$ & $1.80$ & $1.99$\\
\hline
BIC & & & & $1150$  & $1229$\\
\hline
\end{tabular}
\end{table*}
%
%
%
\begin{table*}
\caption{Similar to Table \ref{tab:change.points.alphacentb}, but for HD\,10700.}
\label{tab:change.points.hd10700}
\centering
\begin{tabular}{lcccccccc}
\hline\hline
CPL & JD & Date & Time span & CCFs &  $\overline{RV}$ & $A$ & $\gamma$ & FWHM$_\text{SN}$ \\
 & & & [d] & [\#] & [\ms] & [$\cdot10^{-1}$] & & [\ms] \\
\hline                        
$l_0$ & $2453280.55000$         &       2 Oct 2004 & \multirow{2}*{$1059.20$} & \multirow{2}*{1905} &  
\multirow{2}*{$-1.7_{-1.5}^{+1.6}$} & 
\multirow{2}*{$3.0168_{-0.0052}^{+0.0052}$} & \multirow{2}*{$-0.0578_{-0.0011}^{+0.0011}$} & \multirow{2}*{$6.3026_{-0.0062}^{+0.0044}$} \\
$l_1$ & $2454339.75428$         &       27 Aug 2007 & \multirow{2}*{$438.95$} & \multirow{2}*{1423} & 
\multirow{2}*{$-0.2_{-1.4}^{+0.7}$} & \multirow{2}*{$3.0129_{-0.0011}^{+0.0011}$} & \multirow{2}*{$-0.0587_{-0.0009}^{+0.0012}$} & \multirow{2}*{$6.3071_{-0.0032}^{+0.0028}$}    \\
$l_2$ & $2454778.69248$         &       8 Nov 2008 & \multirow{2}*{$361.84$} & \multirow{2}*{1578} &  
\multirow{2}*{$-0.2_{-1.3}^{+1.1}$} & \multirow{2}*{$3.014_{-0.0008}^{+0.0010}$} & \multirow{2}*{$-0.0585_{-0.0007}^{+0.0010}$} & \multirow{2}*{$6.3107_{-0.0036}^{+0.0037}$}    \\
$l_3$ & $2455140.52938$         &       5 Nov 2009 & \multirow{2}*{$694.11$} & \multirow{2}*{1386} &  
\multirow{2}*{$0.3_{-1.1}^{+1.6}$} & \multirow{2}*{$3.0139_{-0.0012}^{+0.0017}$} & \multirow{2}*{$-0.0584_{-0.0008}^{+0.0011}$} & \multirow{2}*{$6.3135_{-0.0056}^{+0.0064}$}    \\
$l_4$ & $2455834.64038$         &       30 Sep 2011 & \multirow{2}*{$396.16$} & \multirow{2}*{1395} & 
\multirow{2}*{$0.5_{-0.4}^{+1.5}$} & \multirow{2}*{$3.016_{-0.0011}^{+0.0010}$} & \multirow{2}*{$-0.0586_{-0.0007}^{+0.0009}$} & \multirow{2}*{$6.3203_{-0.0033}^{+0.0048}$}    \\
$l_5$ & $2456230.80473$         &       30 Oct 2012 & \multirow{2}*{$778.84$} & \multirow{2}*{1556} & 
\multirow{2}*{$1.5_{-1.5}^{+1.2}$} & \multirow{2}*{$3.0162_{-0.0010}^{+0.0009}$} & \multirow{2}*{$-0.0587_{-0.0007}^{+0.0009}$} & \multirow{2}*{$6.3278_{-0.0043}^{+0.0036}$}    \\
$l_6$ & $2457009.64892$ &       18 Dec 2014 &  & & & & &        \\
\hline
\end{tabular}
\end{table*}
\begin{table*}[htbp]
   \caption{Similar to Table \ref{tab:correlations.bp.alphacentb}, but for HD\,10700.}
   \label{tab:correlations.bp.hd10700}
\centering
\scalebox{1}{%
\begin{tabular}{lcccccc}
\hline\hline
& $1$ & $2$ & $3$ & $4$ & $5$ & $6$  \\
\hline                        
$\rho(\overline{RV} ; A)$ & $-0.29$  & $0.18$ & $-0.21$ & $0.32$  & $0.18$ & $-0.12$  \\
& $(-0.33, -0.25)$ & $(0.13, 0.23)$ & $(-0.26, -0.17)$ & $(0.27, 0.37)$ & $(0.12, 0.23)$ & $(-0.17, -0.068)$ \\
\hline 
$\rho(\overline{RV}; \gamma)$  & $0.11$ & $0.20$ & $0.20$ & $0.07$ & $0.10$ & $0.03$   \\
& $(0.069, 0.16)$ & $(0.15, 0.25)$ & $(0.15, 0.24)$ & $(0.018, 0.12)$ & $(0.045, 0.15)$ & $(-0.020,  0.079)$ \\
\hline 
$\rho(\overline{RV} ; \mathrm{FWHM_{SN}})$ & $-0.08$ & $0.20$ & $0.23$ & $0.48$ & $0.32$ & $0.48$  \\
& $(-0.12, -0.032)$ & $(0.15, 0.25)$ & $(0.18, 0.27)$ & $(0.44, 0.52)$ & $(0.27, 0.37)$ & $(0.44, 0.52)$ \\
\hline
\end{tabular}}
\end{table*}
\begin{table*}
\caption{Similar to Table \ref{tab:bp.vs.oc.alphacentb}, but for HD\,10700.}
  \label{tab:bp.vs.oc.hd10700}
\centering
\scalebox{0.85}{
\begin{tabular}{llcccccccc}
\hline\hline
\multicolumn{2}{c}{rms} & {bp} segment 1 & {bp} segment 2 & {bp} segment 3 & {bp} segment 4 & {bp} segment 5 & {bp} segment 6 & {bp} overall & {oc} overall  \\
\hline                        
$\overline{RV}$ & [\ms] & $1.54$ & $1.36$ & $1.41$ & $1.64$ & $1.47$ & $1.49$ & $1.76$  & $1.76$ \\
$RV^{*}_{\text{activity}}$ & [\ms] & $0.52$ & $0.40$ & $0.59$  & $0.85$ & $0.54$ & 0.83  & $1.14$ & $1.07$ \\
$\Delta RV^*$ & [\ms] & $1.45$  & $1.30$ & $1.28$ & $1.40$  & $1.36$ & $1.24$ & $1.34$  &  $1.40$ \\
\hline
BIC & & &       &  &  &  & & $31958$ & $32461$  \\
\hline
\end{tabular}
}
\end{table*}

%
%
\begin{table*}
\caption{Similar to Table \ref{tab:change.points.alphacentb}, but for HD\,192310.}
\label{tab:change.points.hd192310}
\centering
\begin{tabular}{lcccccccc}
\hline\hline
CPL & JD & Date & Time span & CCFs &  $\overline{RV}$ & $A$ & $\gamma$ & FWHM$_\text{SN}$ \\
 & & & [d] & [\#] & [\ms] & [$\cdot10^{-1}$] & & [\ms] \\
\hline                        
$l_0$ & $2455019.69743$         &       7 Jul 2009 & \multirow{2}*{$477.82$} & \multirow{2}*{748} &  
\multirow{2}*{$ -1.3_{-3.1}^{+0.1}$} & 
\multirow{2}*{$2.9536_{-0.0009}^{+0.0010}$} & \multirow{2}*{$0.0130_{-0.0013}^{+0.0013}$} & \multirow{2}*{$6.1023_{-0.0050}^{+0.0043}$} \\
$l_1$ & $2455497.51422$         &       28 Oct 2010 & \multirow{2}*{$621.32$} & \multirow{2}*{300} &  
\multirow{2}*{$1.5_{-1.1}^{+2.6}$} & \multirow{2}*{$2.9537_{-0.0010}^{+0.0011}$} & \multirow{2}*{$0.0135_{-0.0014}^{+0.0033}$} & \multirow{2}*{$6.110_{-0.007}^{+0.018}$}        \\
$l_2$ & $2456118.83597$         &       10 Jul 2012 & \multirow{2}*{$755.86$} & \multirow{2}*{300} &  
\multirow{2}*{$3.5_{-3.4}^{+2.6}$} & \multirow{2}*{$2.9539_{-0.0014}^{+0.0014}$} & \multirow{2}*{$0.0159_{-0.0017}^{+0.0018}$} & 
\multirow{2}*{$6.127_{-0.007}^{+0.026}$}        \\
$l_3$ & $2456874.69449$ &       5 Aug 2014 &  & & & & & \\
\hline
\end{tabular}
\end{table*}
\begin{table*}[htbp]
   \caption{Similar to Table \ref{tab:correlations.bp.alphacentb}, but for HD\,192310.}
   \label{tab:correlations.bp.hd192310}
\centering
\scalebox{1}{%
\begin{tabular}{lcccc}
\hline\hline
& $1$ & $2$  & $3$ \\
\hline                        
$\rho(\overline{RV}; A)$                  & $-0.07$  & $-0.43$ & $-0.11$  \\
& $(-0.15, -0.0033)$ & $(-0.52, -0.34)$ & $(-0.22,  0.0065)$ \\
\hline 
$\rho(\overline{RV}; \gamma)$             & $0.08$   & $0.65$ & $0.22$  \\
& $(0.0054, 0.15)$ & $(0.58, 0.71)$ & $(0.11, 0.33)$ \\
\hline 
$\rho(\overline{RV} ; \mathrm{FWHM_{SN}})$ & $-0.09$  & $0.54$  & $0.24$  \\
& $(-0.16, -0.017)$ & $(0.46, 0.62)$ & $(0.13, 0.34)$ \\
\hline 
\end{tabular}}
\end{table*}

\begin{table*}
\caption{Similar to Table \ref{tab:bp.vs.oc.alphacentb}, but for HD\,192310.}
  \label{tab:bp.vs.oc.hd192310}
\centering
\begin{tabular}{llccccc}
\hline\hline
\multicolumn{2}{c}{rms} & {bp} segment 1 & {bp} segment 2 & {bp} segment 3 & {bp} overall & {oc} overall  \\
\hline                        
$\overline{RV}$ & [\ms]  & $1.47$ & $2.94$ & $2.13$ & $2.28$  & $2.28$ \\
$RV^{*}_{\text{activity}}$ & [\ms]  & $0.61$ & $2.55$ & $2.13$  & $1.79$ & $1.71$ \\
$\Delta RV^*$ & [\ms] & $1.34$  & $1.46$ & $1.52$ & $1.40$  & $1.50$ \\
\hline
BIC & & & & & $4784$ & $4958$  \\
\hline
\end{tabular}
\end{table*}
\begin{table}[htbp]
   \caption{Normalized power $\mathcal{P}$ and cv values inferred from the {bp}-corrected GLS ($\mathcal{P}_{bp}$, cv$_{bp}$) and {oc}-corrected GLS ($\mathcal{P}_{oc}$, cv$_{oc}$) for HD\,192310. They are evaluated at the stellar rotation period ($P_{\mathrm{rot}}$) and at the at the orbital periods ($P_{\mathrm{orb}}$) of the hosted exoplanets found by \citet{Pepe-2011}. The first two rows show $\mathcal{P}$ inferred from the $\Delta RV^*$ signals, while in the last row $\mathcal{P}$ are computed after removing the Keplerian signal of $P_{\mathrm{orb,1}}=75$ days from the $\Delta RV^*$ signals. Both the two planets would be visible in the GLS after applying the {bp} method and the iterative removal of planetary signals, as $\mathcal{P}_{bp}$ are always above the respective cv$_{bp}$ values.
   }
   \label{tab:peaks.hd192310}
\centering
\begin{tabular}{lccccc}
\hline\hline
& $P$ [d] & $\mathcal{P}_{bp}$ & cv$_{bp}$ & $\mathcal{P}_{oc}$ & cv$_{oc}$ \\
\hline                        
Stellar $P_{\mathrm{rot}}$ & 48 & $0.064$ & 0.48 & $0.064$ & 0.54 \\
Planet $1$ $P_{\mathrm{orb}}$ & 75 & $0.52$   & 0.48 & $0.49$ & 0.54   \\
Planet $2$ $P_{\mathrm{orb}}$ & 526 &$0.28$ & 0.087 & $0.40$ & 0.091   \\
\hline
\end{tabular}
\end{table}

\begin{figure*}
\centering
\includegraphics[width=\columnwidth]{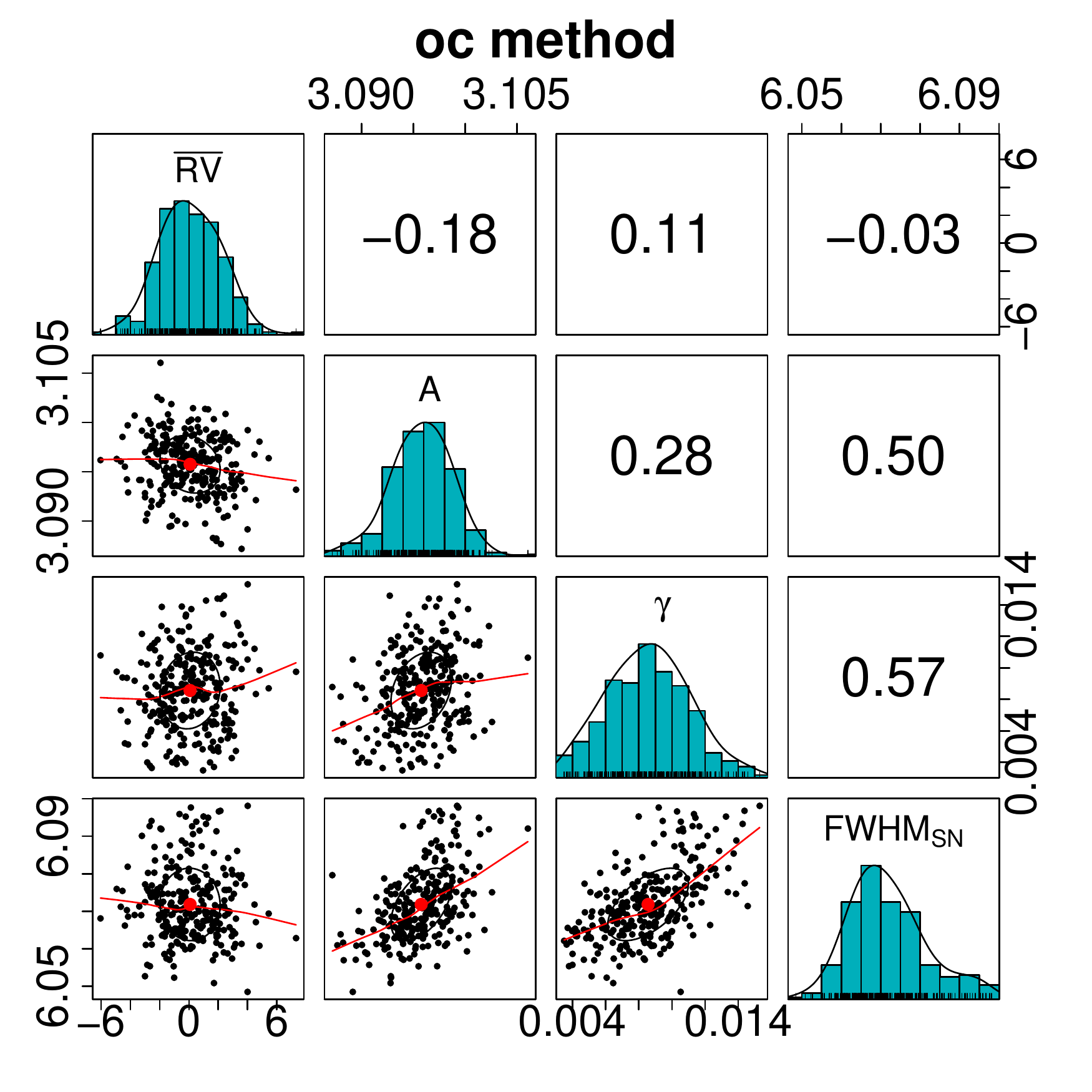}
\includegraphics[width=\columnwidth]{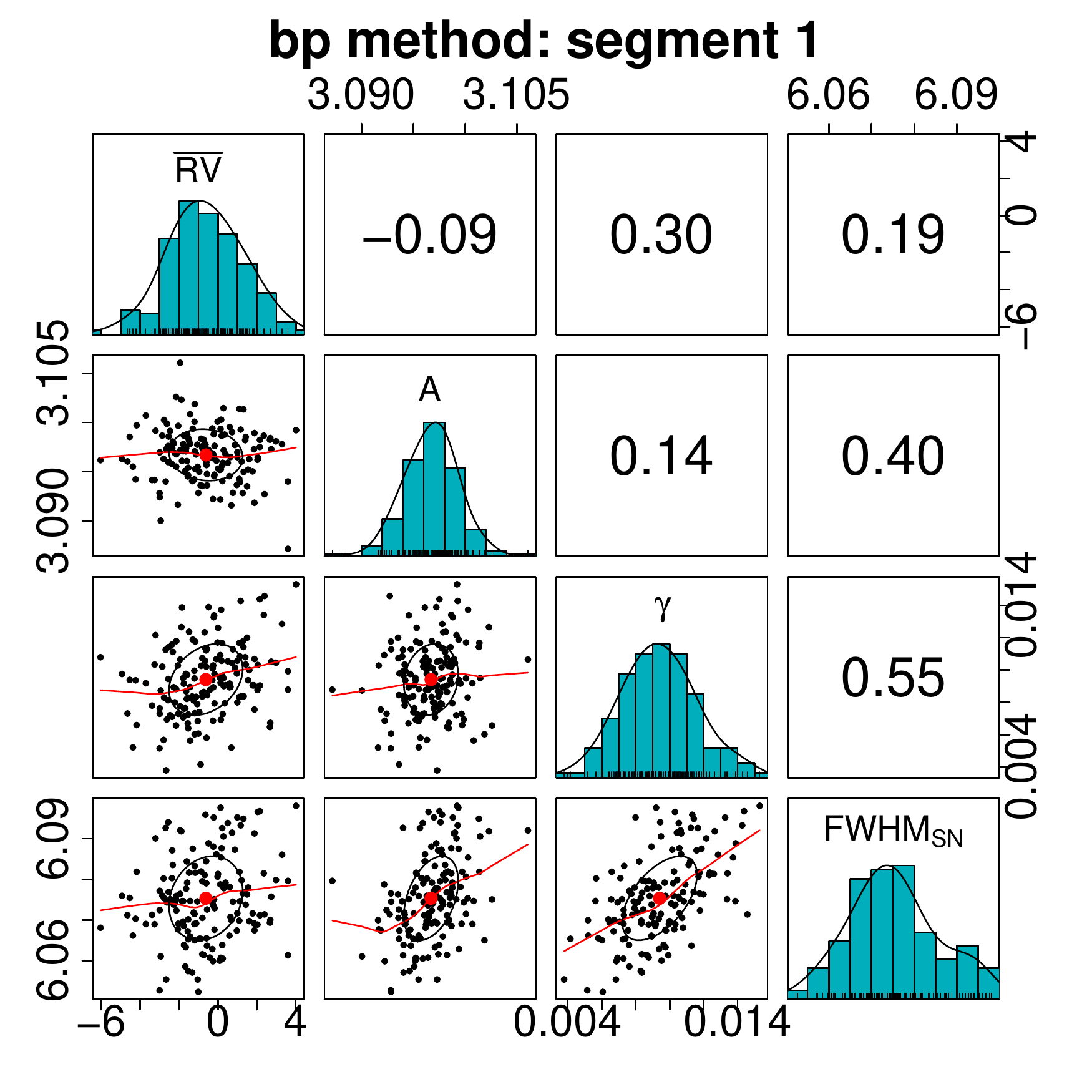}
\includegraphics[width=\columnwidth]{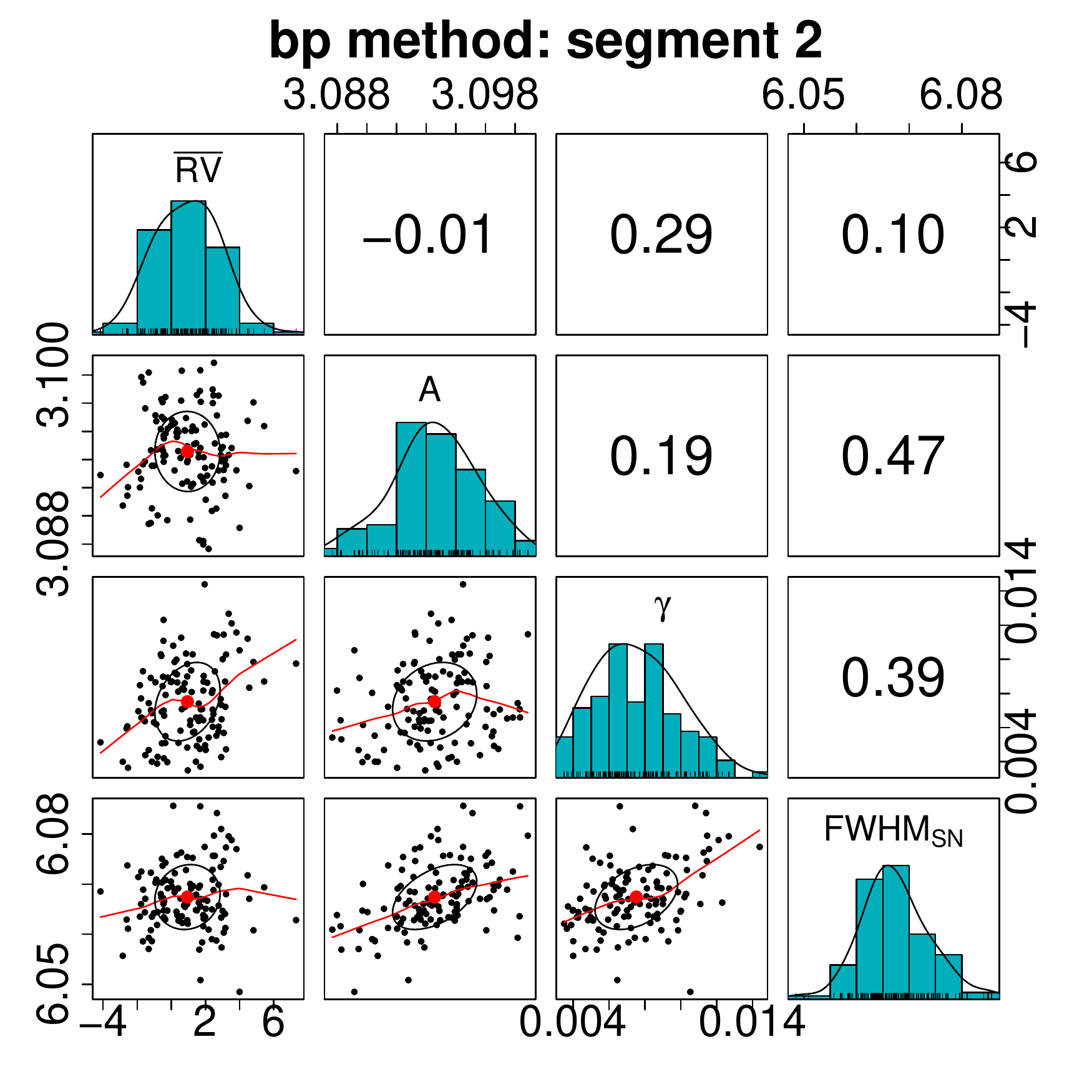}
\caption{Similar to Fig. \ref{fig:alphacentb_pairsx}, but for HD\,215152.}
\label{fig:hd215152_pairsx}
\end{figure*}
\begin{figure*}
\centering
\includegraphics[width=\textwidth]{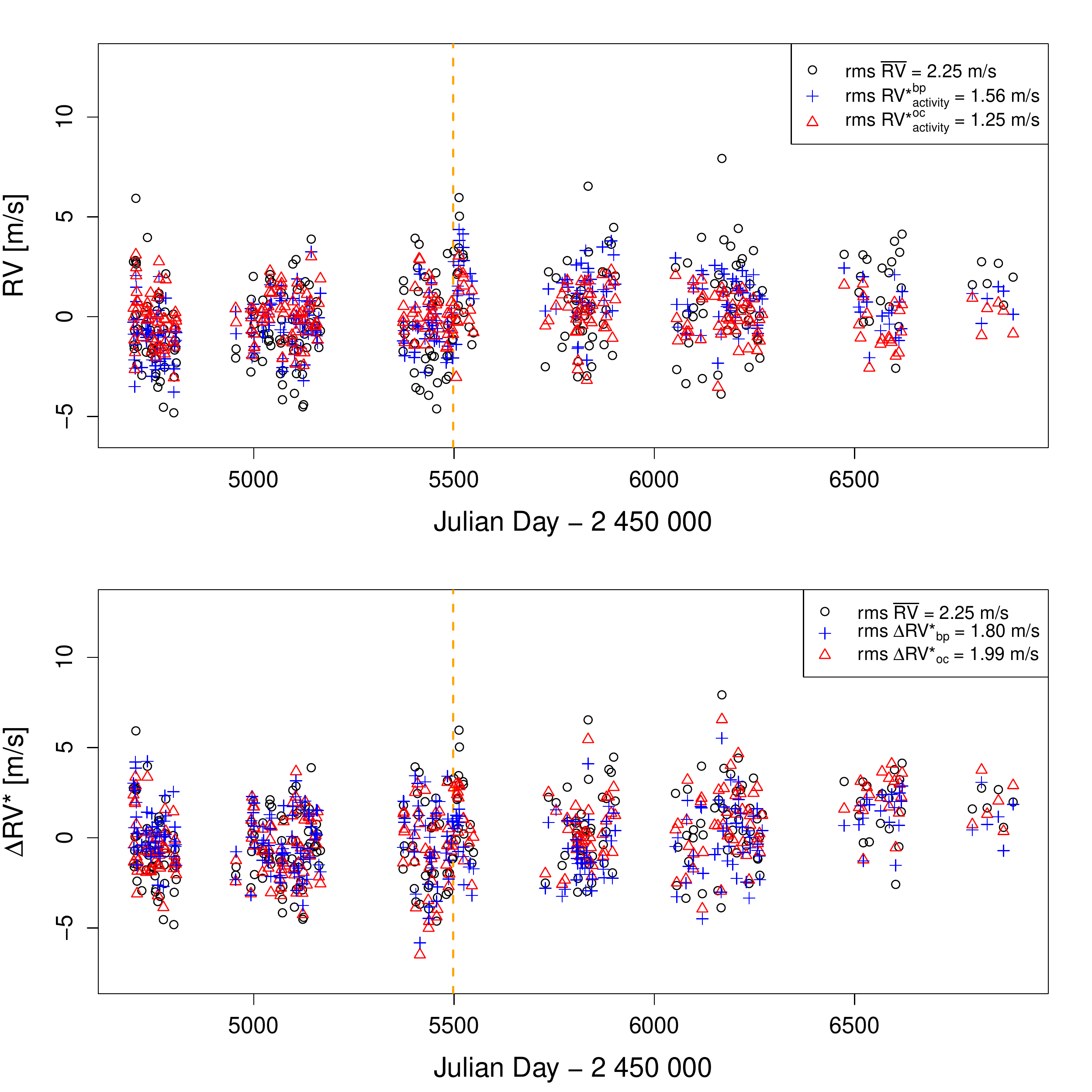}
\caption{Similar to Fig. \ref{fig:alphacentb}, but for HD\,215152.
}
\label{fig:hd215152}
\end{figure*}
\begin{figure*}
\centering
\includegraphics[width=\textwidth]{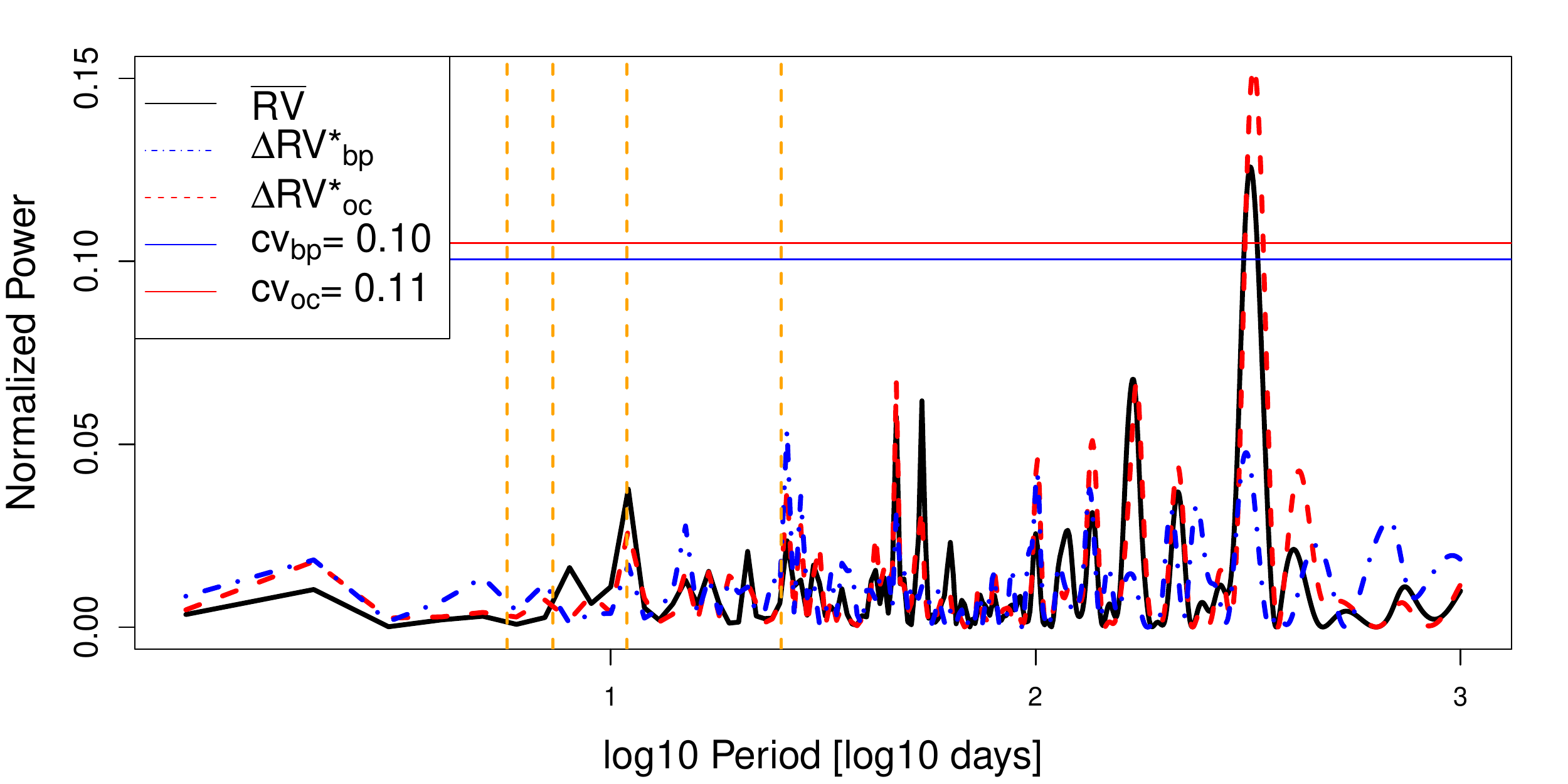}
\caption{Similar to Fig. \ref{fig:alphacentb_GLS}, but for HD\,215152. The orbital periods of the already discovered exoplanets \citep{Delisle:2018aa} are displayed as orange dashed lines. The planetary signals have not been removed from the time series, because they are at the level of the instrumental precision.
}
\label{fig:hd215152_GLS_detection}
\end{figure*}

\begin{figure*}
\centering
\includegraphics[width=0.66882\columnwidth]{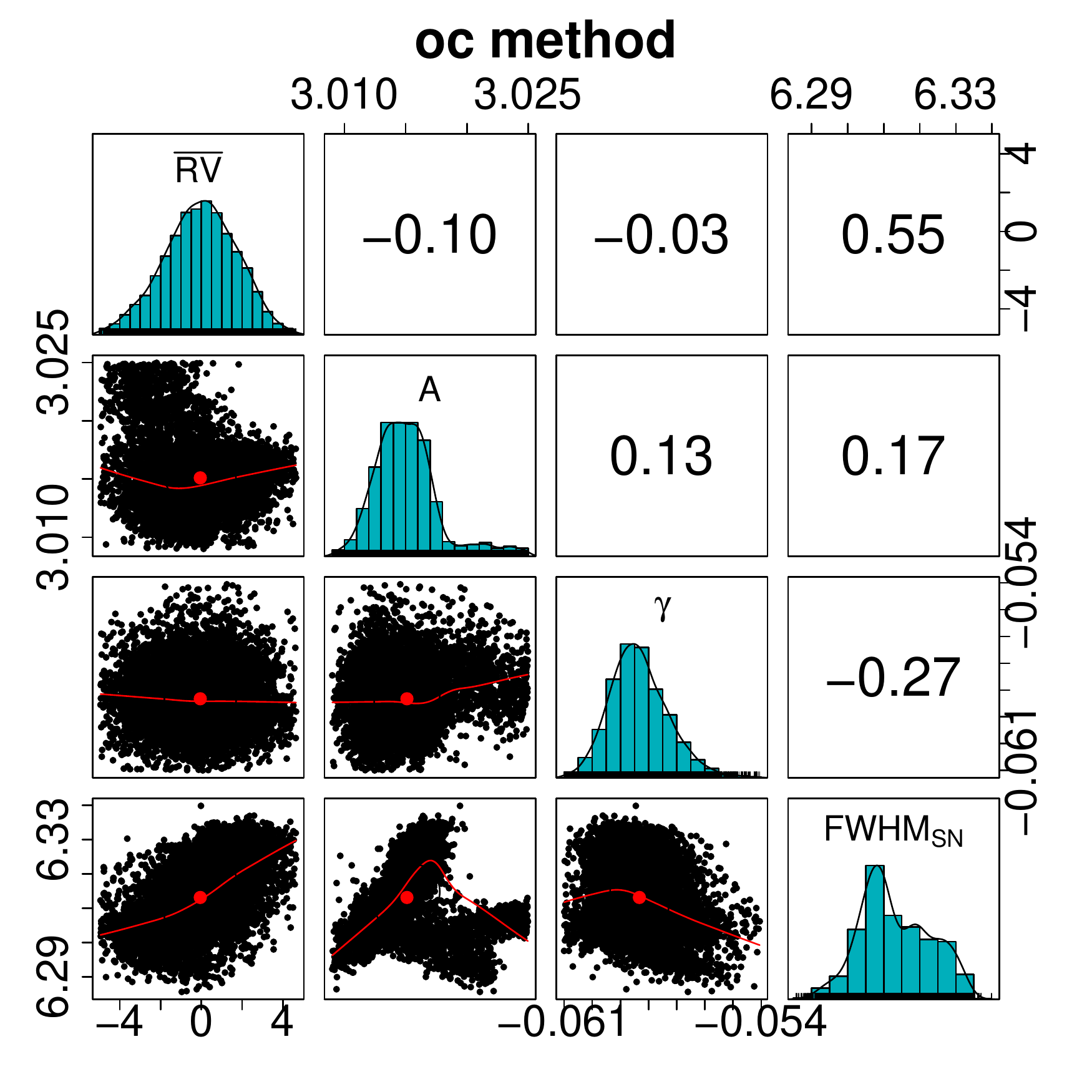}
\includegraphics[width=0.66882\columnwidth]{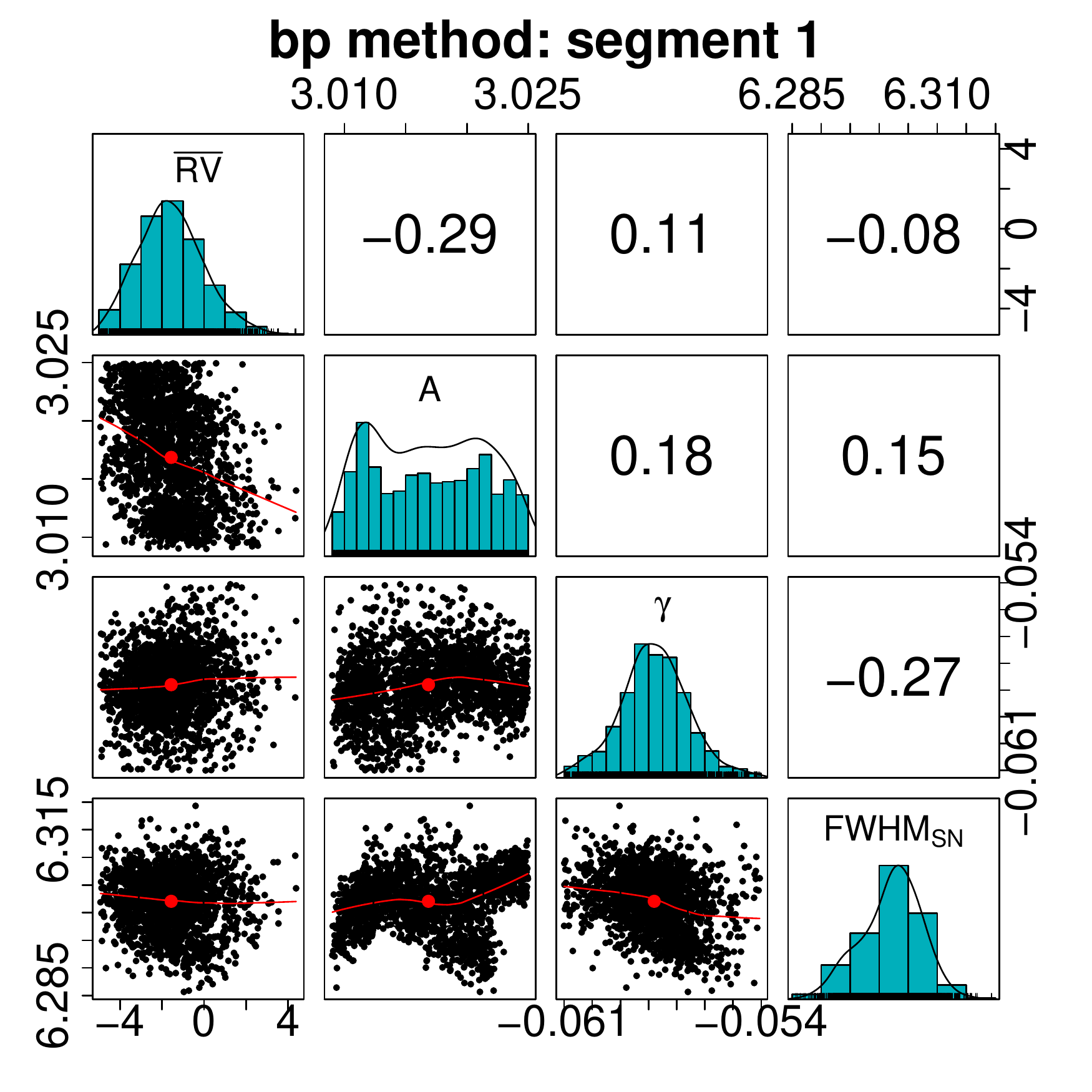}
\includegraphics[width=0.66882\columnwidth]{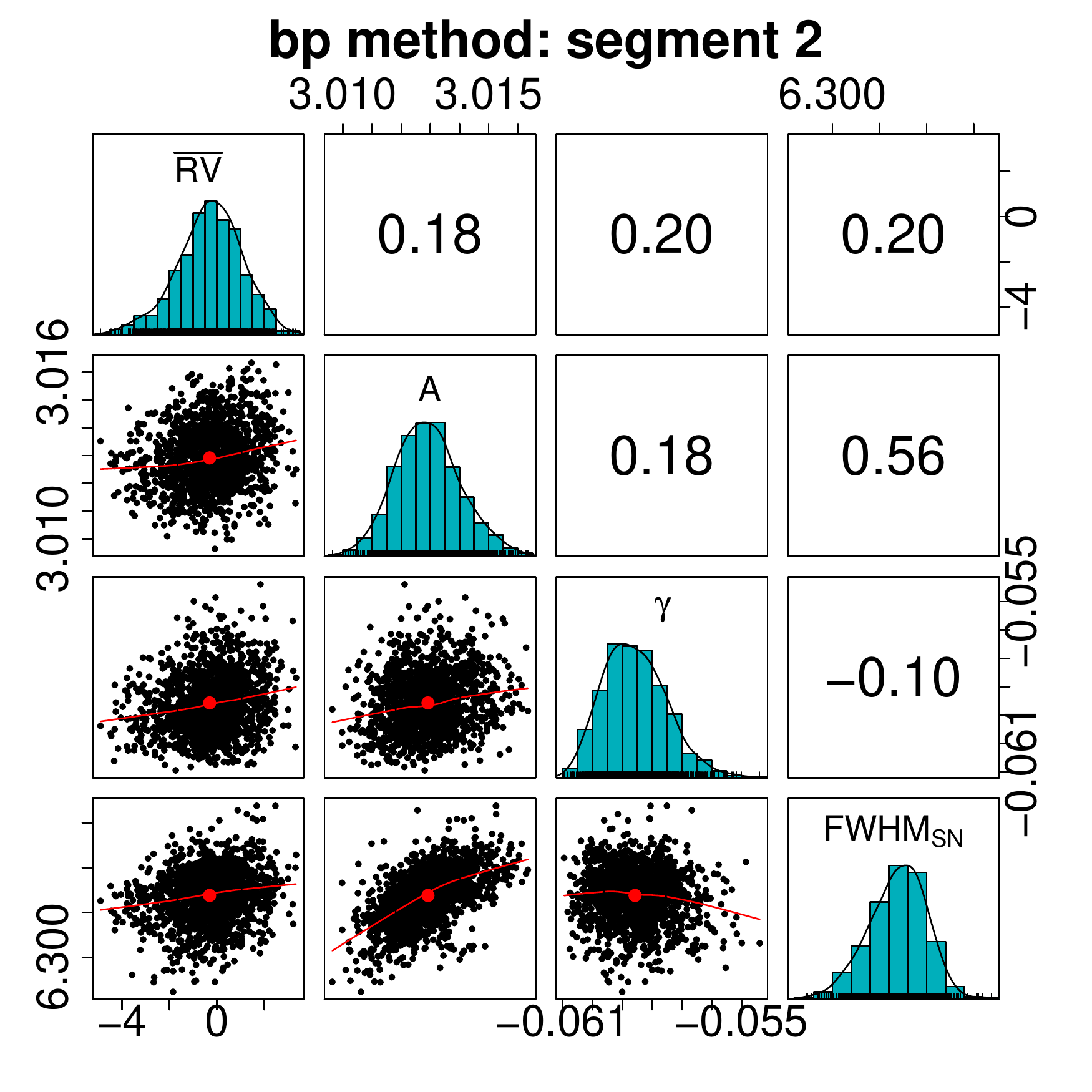}
\includegraphics[width=0.66882\columnwidth]{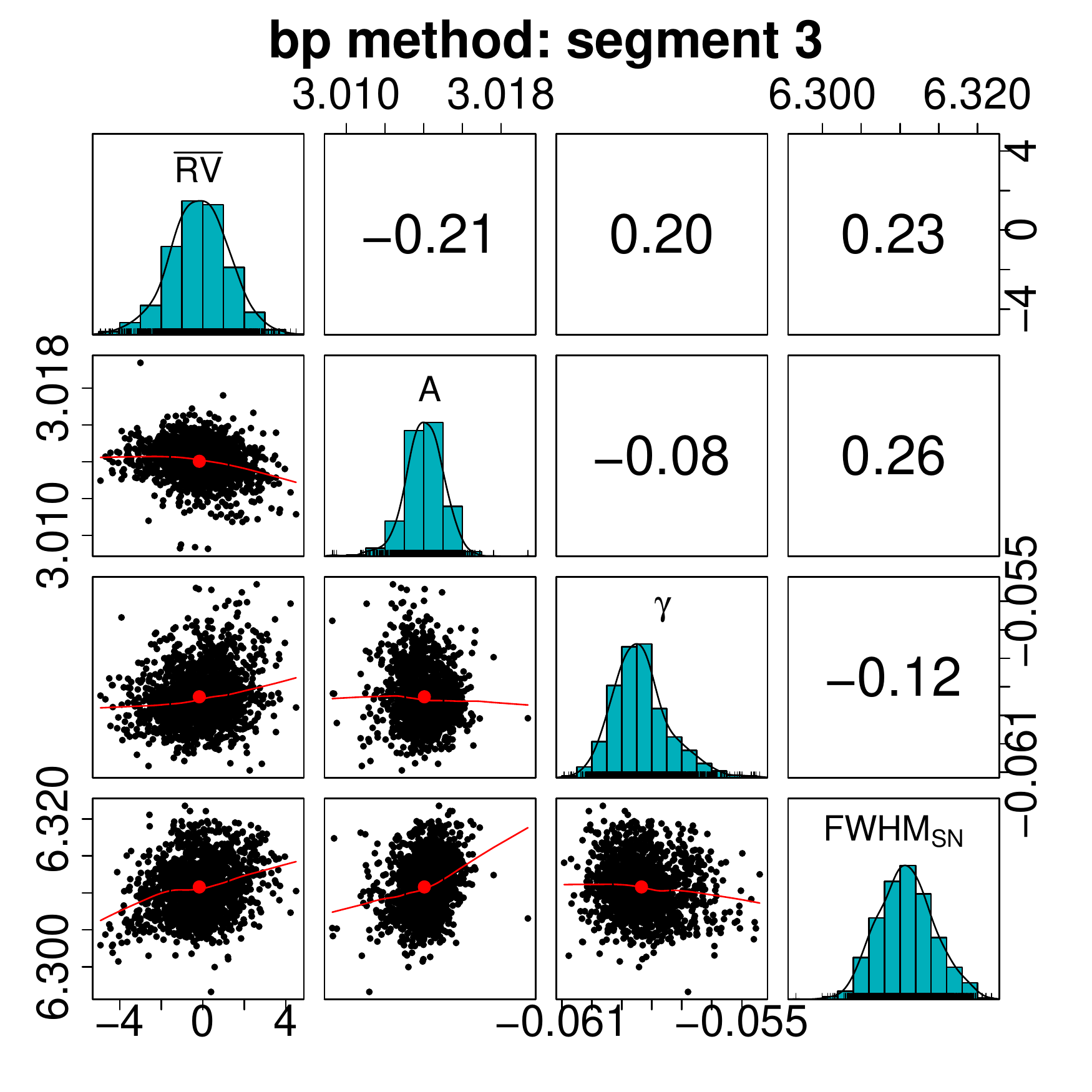}
\includegraphics[width=0.66882\columnwidth]{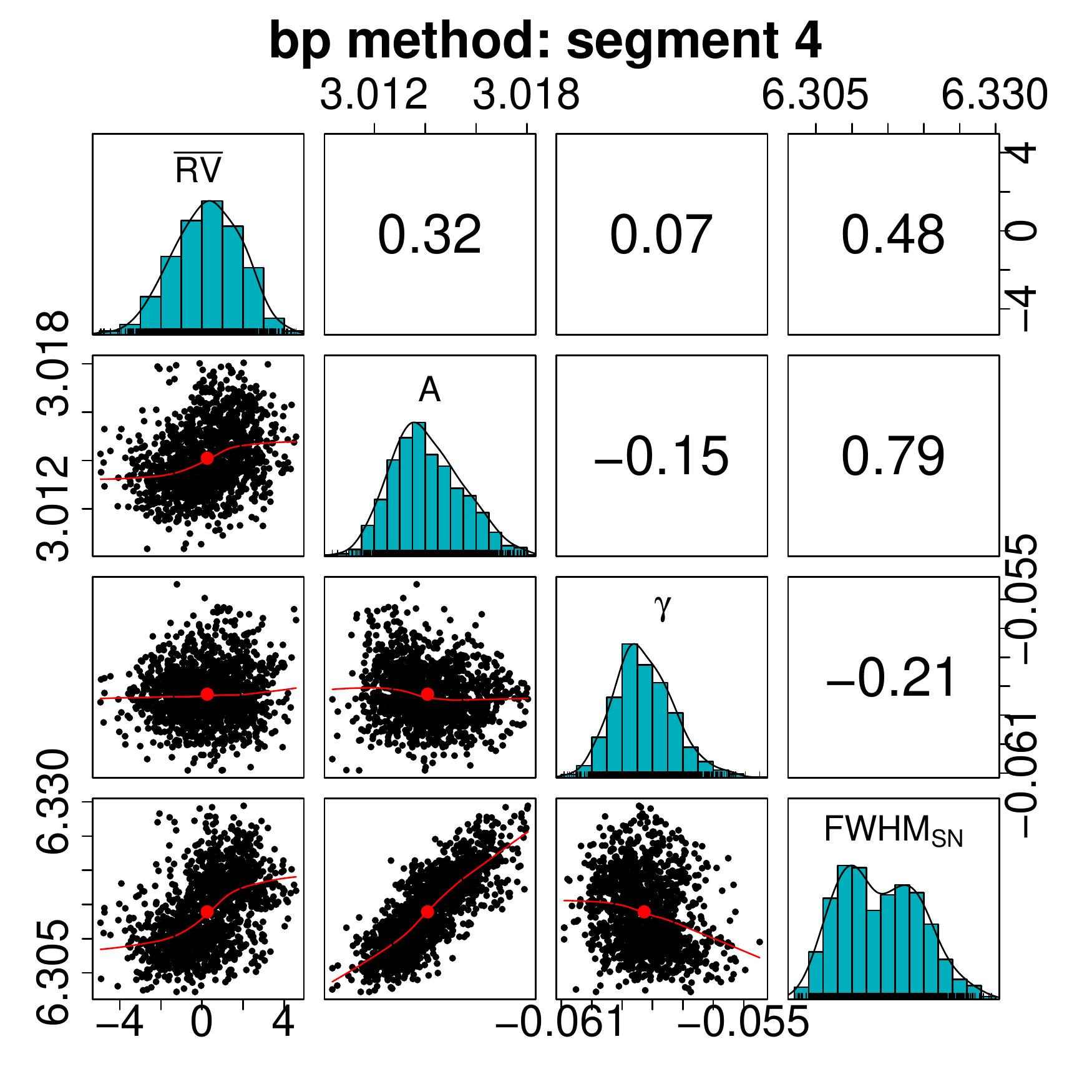}
\includegraphics[width=0.66882\columnwidth]{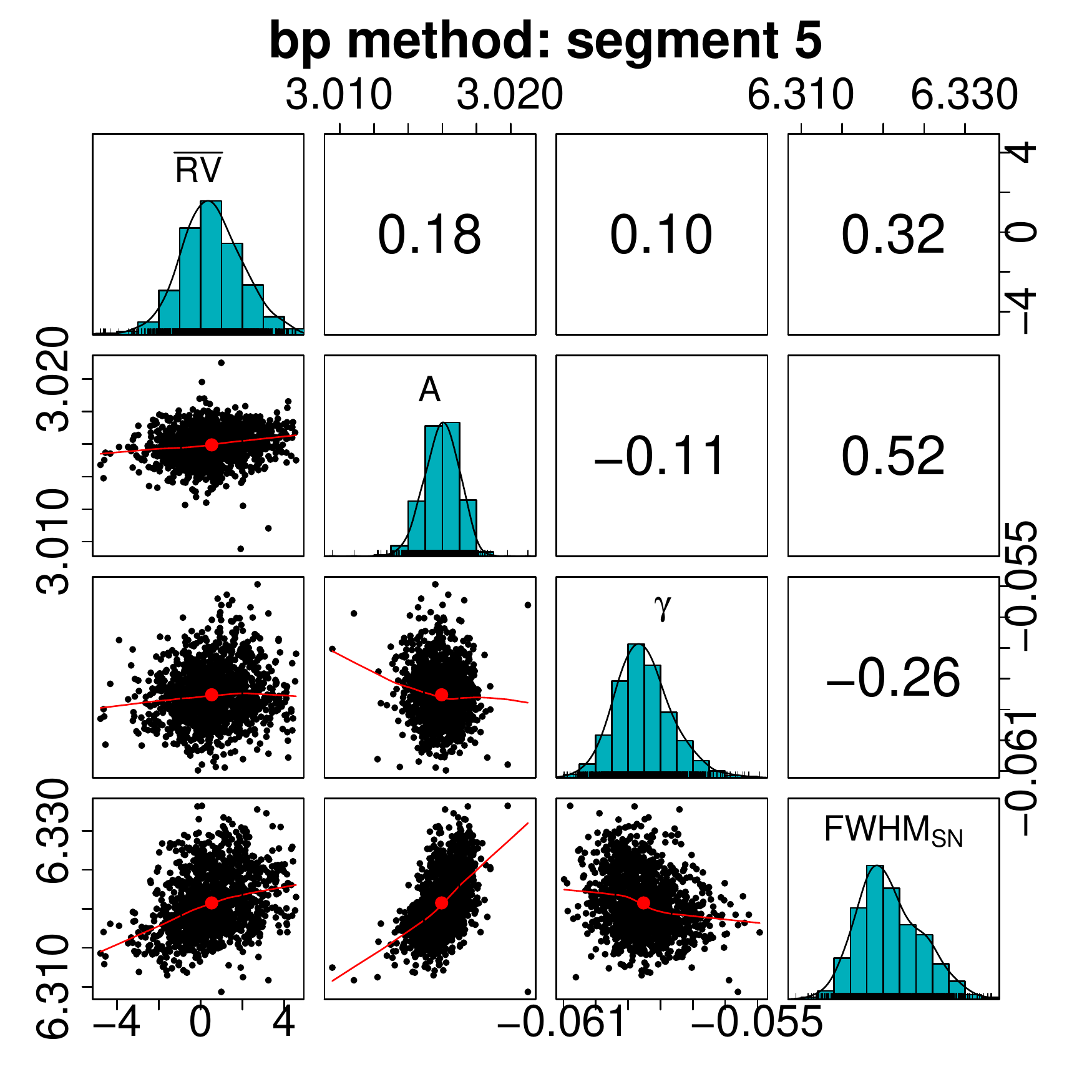}
\includegraphics[width=0.66882\columnwidth]{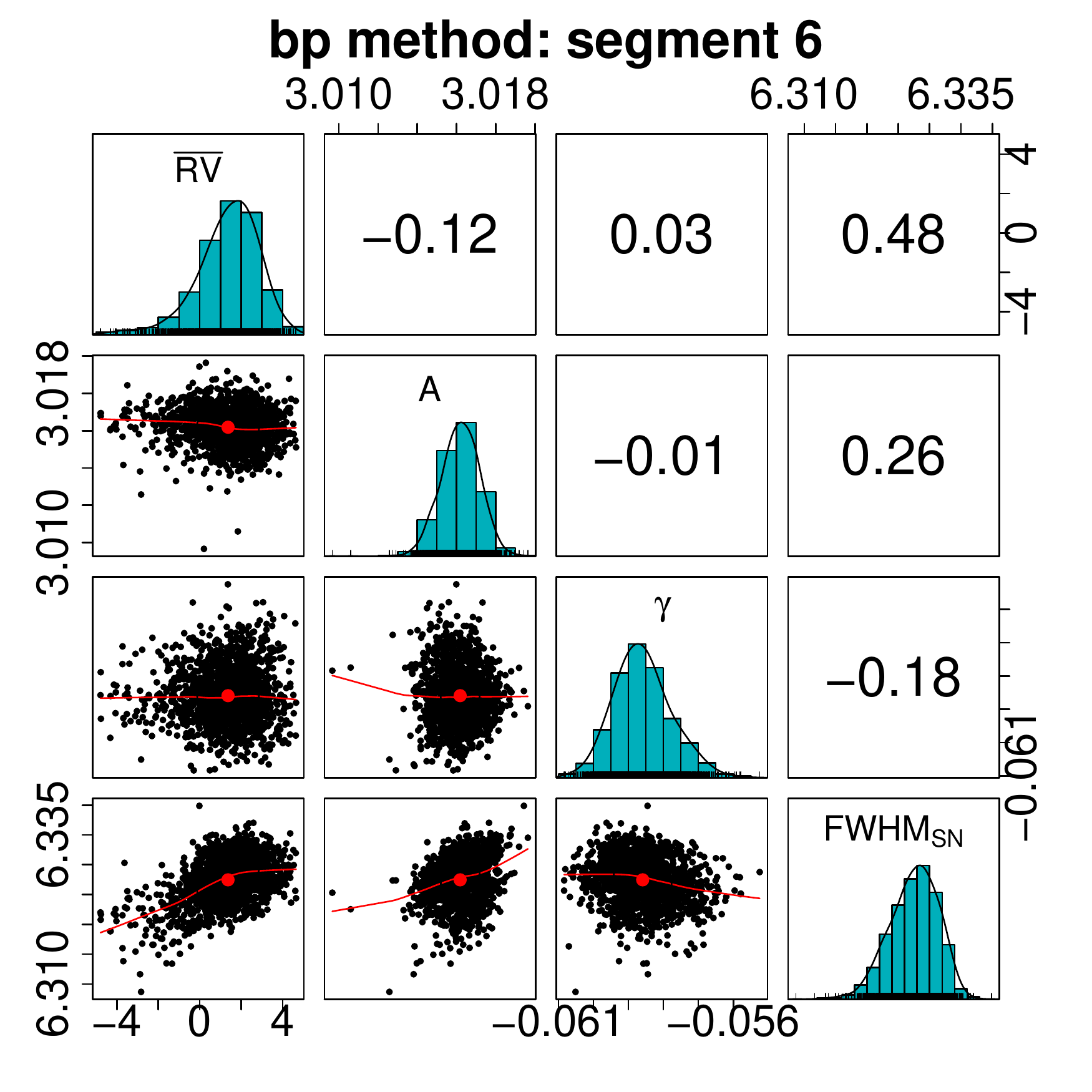}
\caption{Similar to Fig. \ref{fig:alphacentb_pairsx}, but for HD\,10700.}
\label{fig:tauceti_pairsx}
\end{figure*}
\begin{figure*}
\centering
\includegraphics[width=\textwidth]{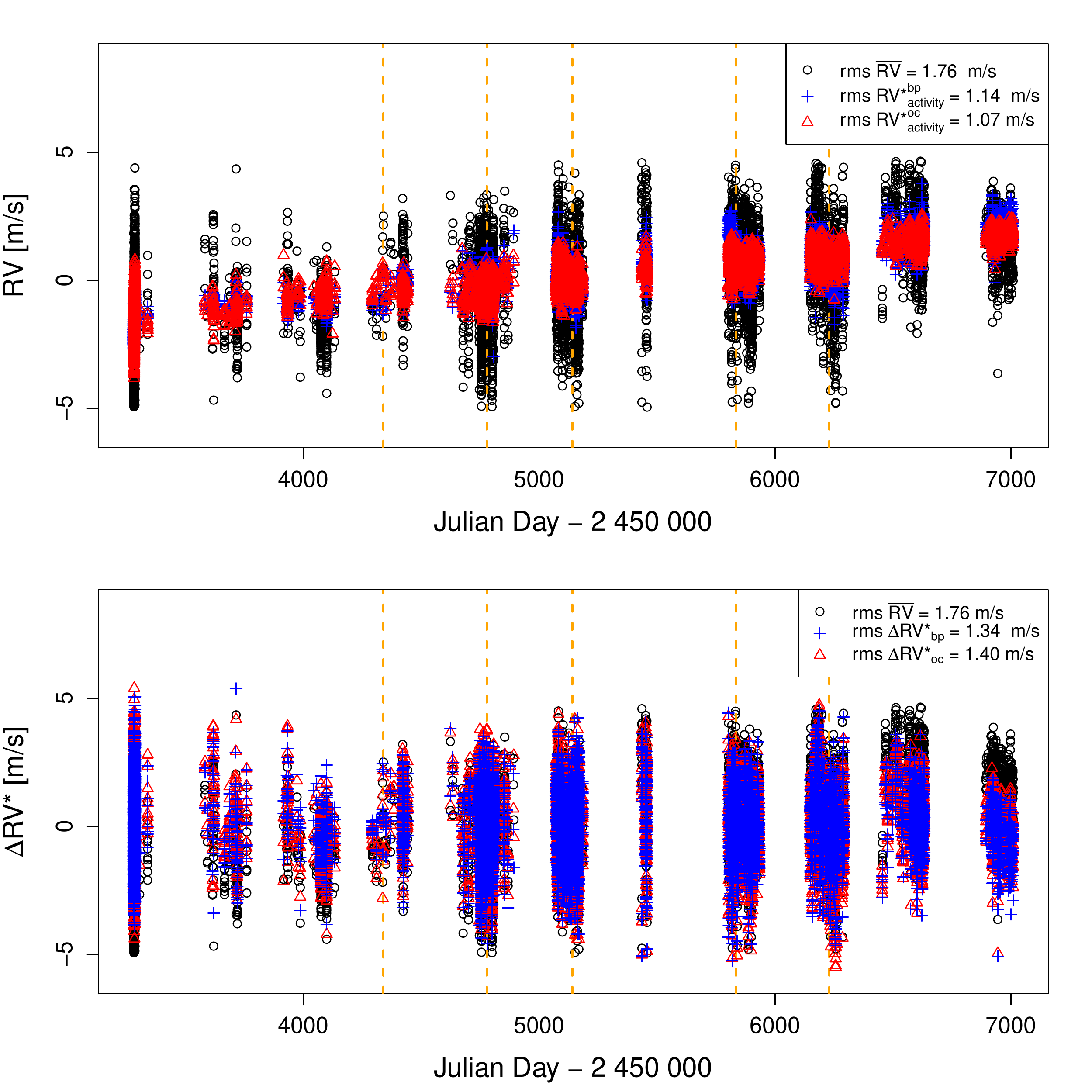}
\caption{Similar to Fig. \ref{fig:alphacentb}, but for HD\,10700.}
\label{fig:hd10700}
\end{figure*}
\begin{figure*}
\centering
\includegraphics[width=\textwidth]{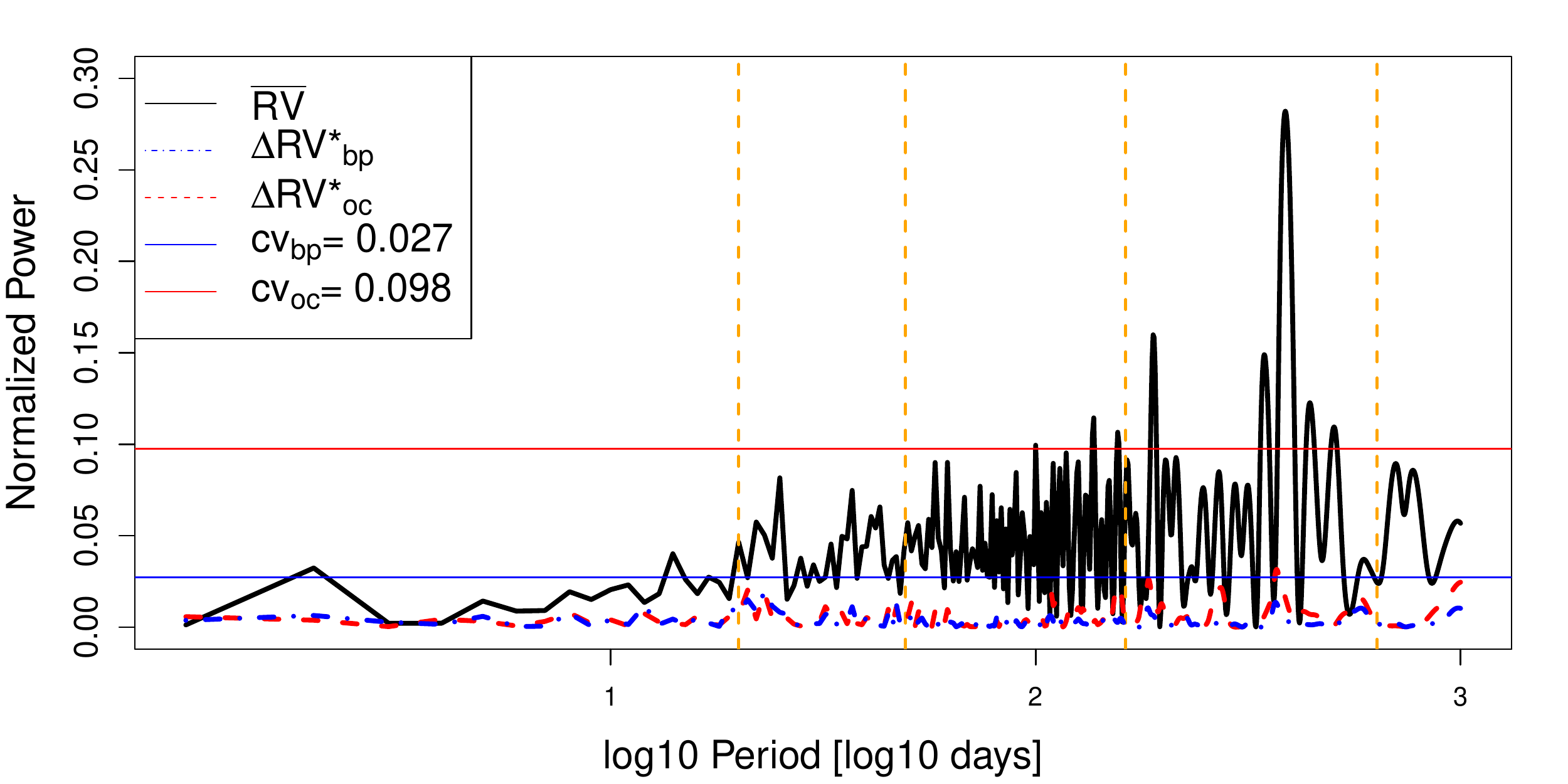}
\caption{Similar to Fig. \ref{fig:alphacentb_GLS}, but for HD\,10700. The orbital periods of the already discovered exoplanets \citep{Feng:2017ac} are displayed as orange dashed lines. The planetary signals have not been removed from the time series, because they are at the level of the instrumental precision.
}
\label{fig:hd10700.GLS_detection}
\end{figure*}

\begin{figure*}
\centering
\includegraphics[width=\columnwidth]{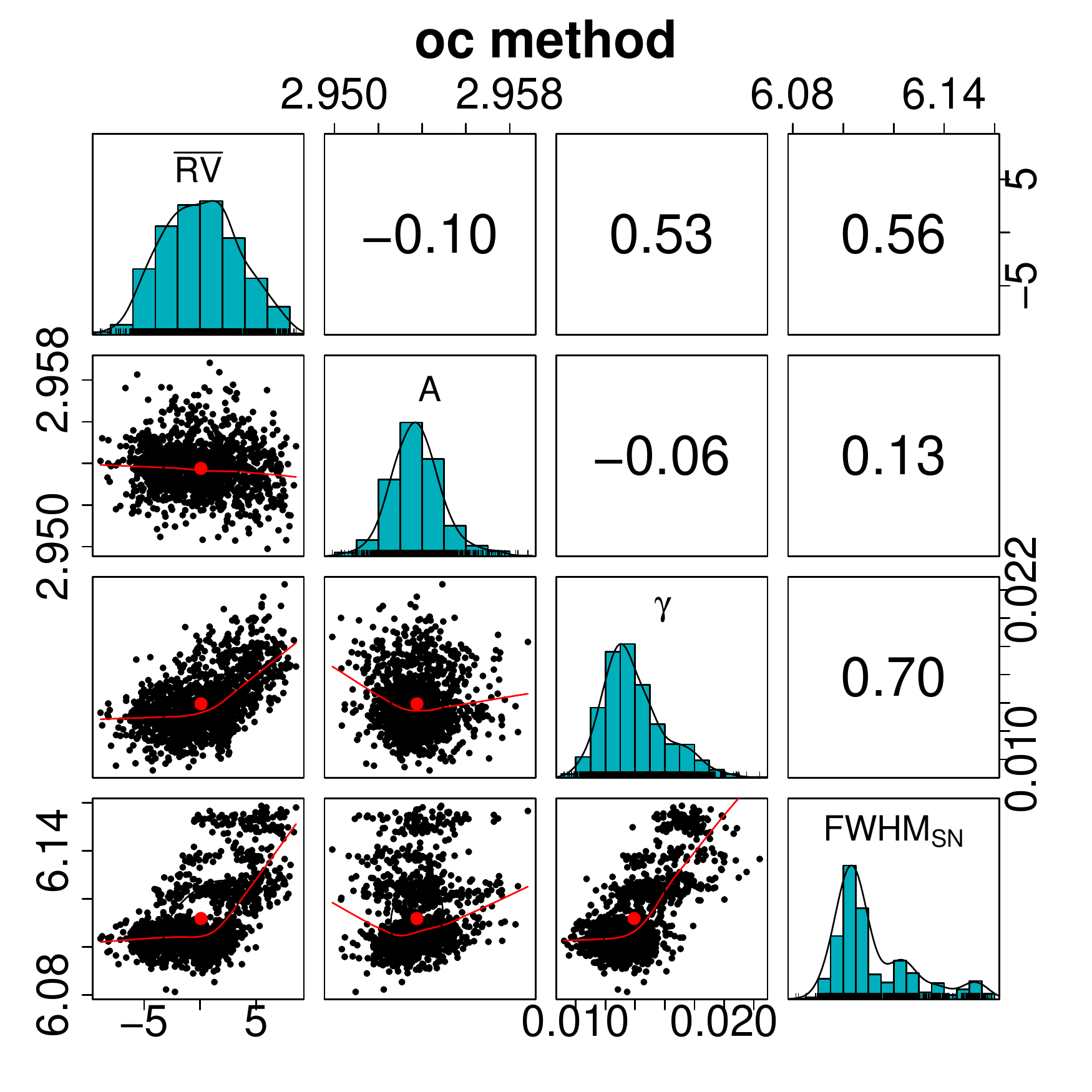}
\includegraphics[width=\columnwidth]{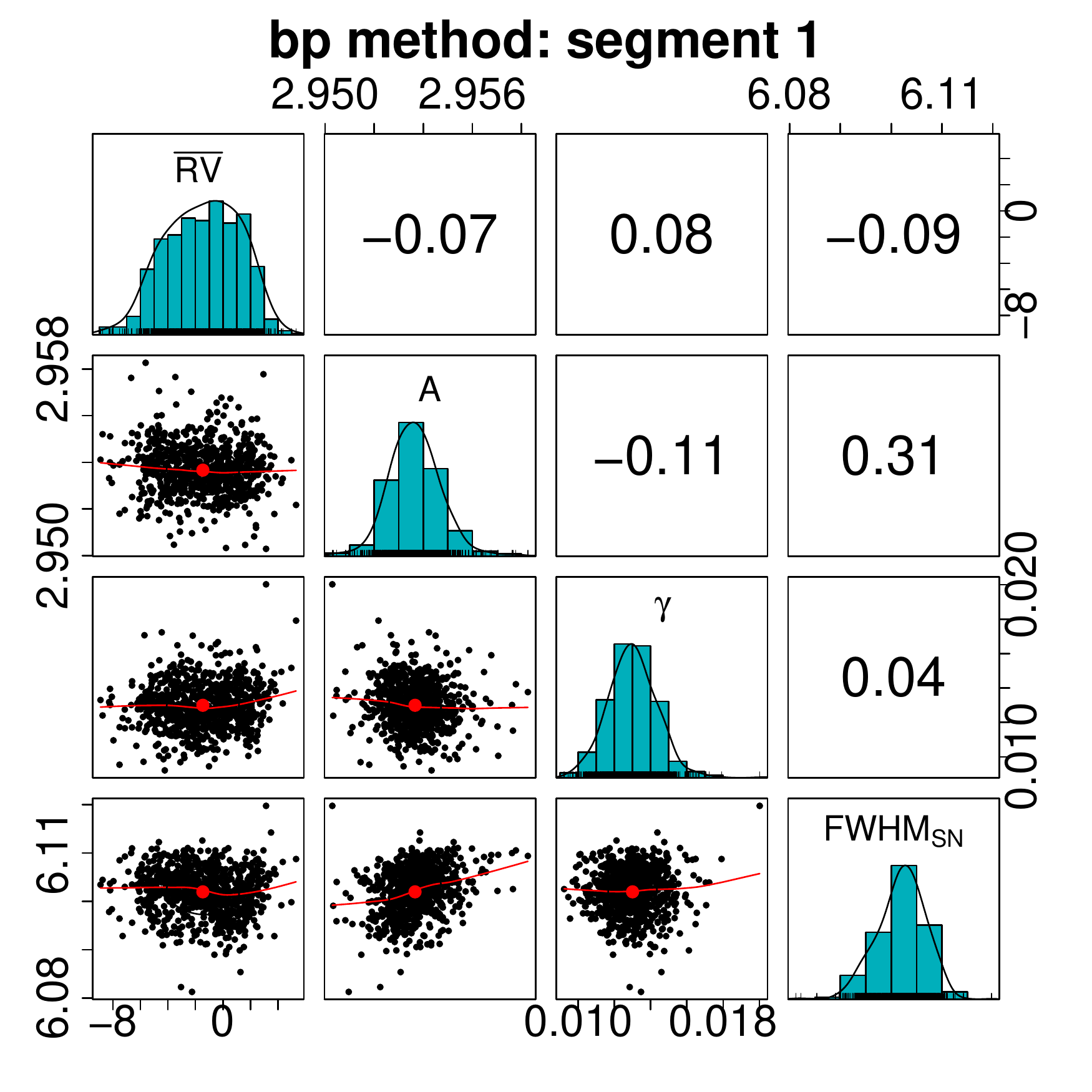}
\includegraphics[width=\columnwidth]{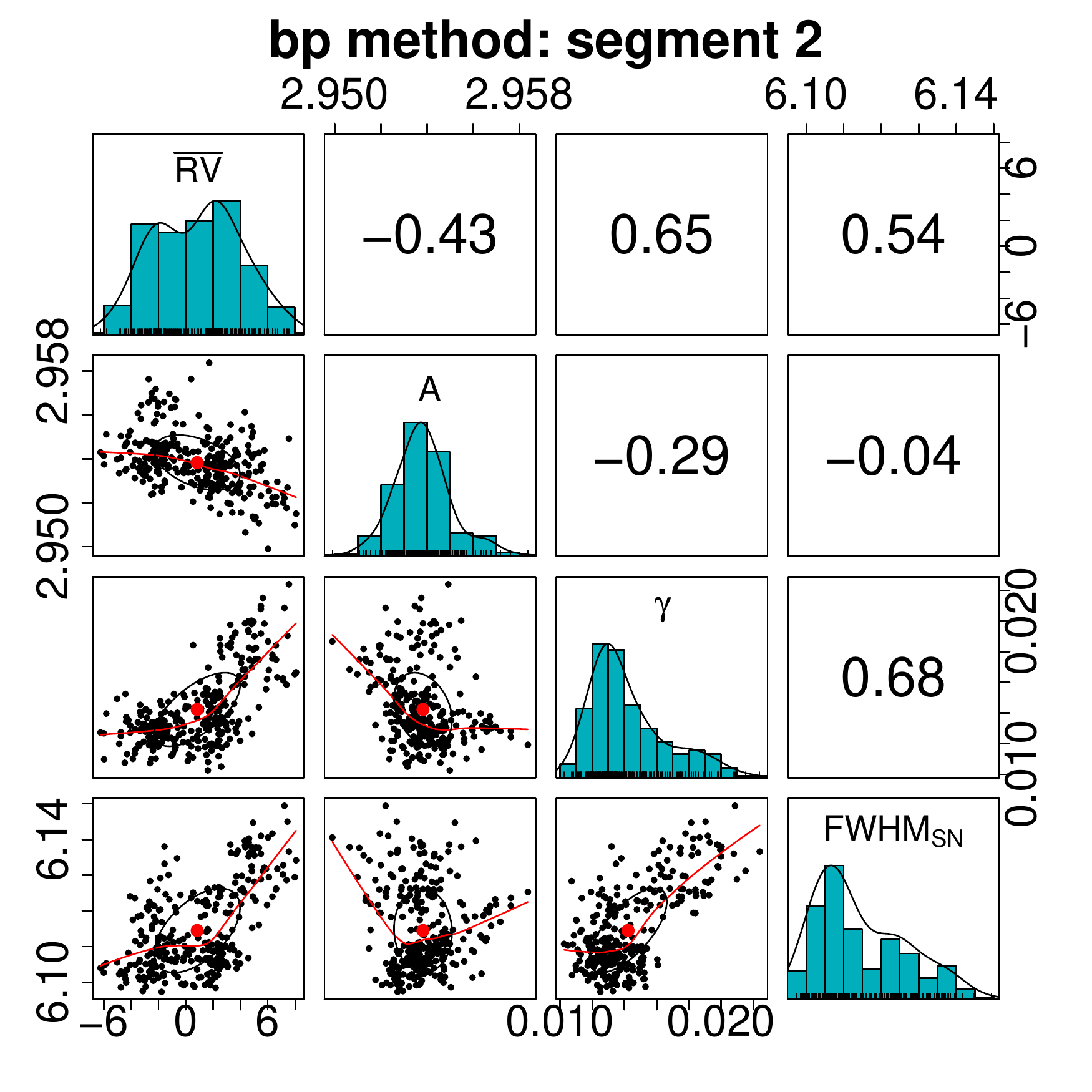}
\includegraphics[width=\columnwidth]{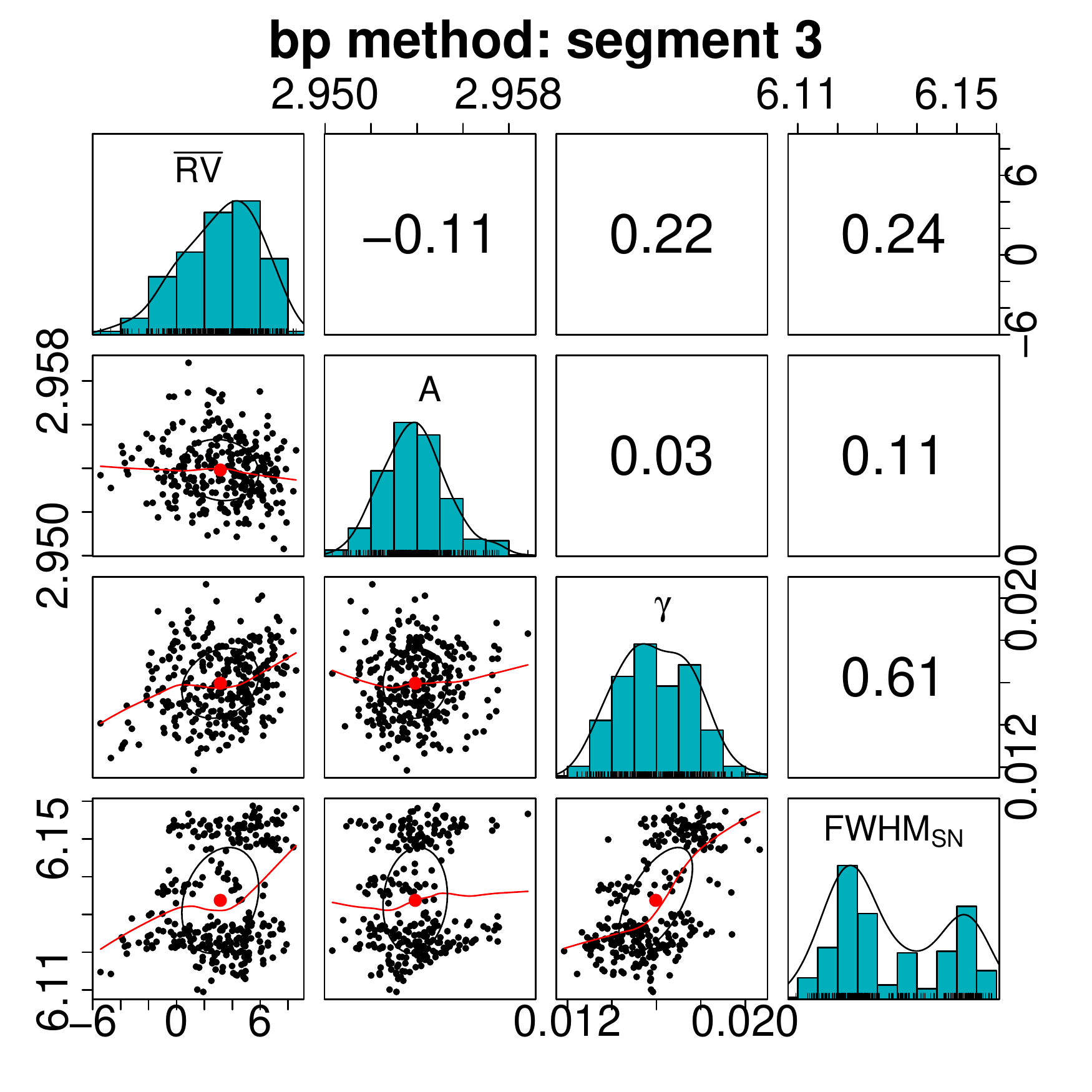}
\caption{Similar to Fig. \ref{fig:alphacentb_pairsx}, but for HD\,192310.}
\label{fig:hd192310_pairsx}
\end{figure*}
\begin{figure*}
\centering
\includegraphics[width=\textwidth]{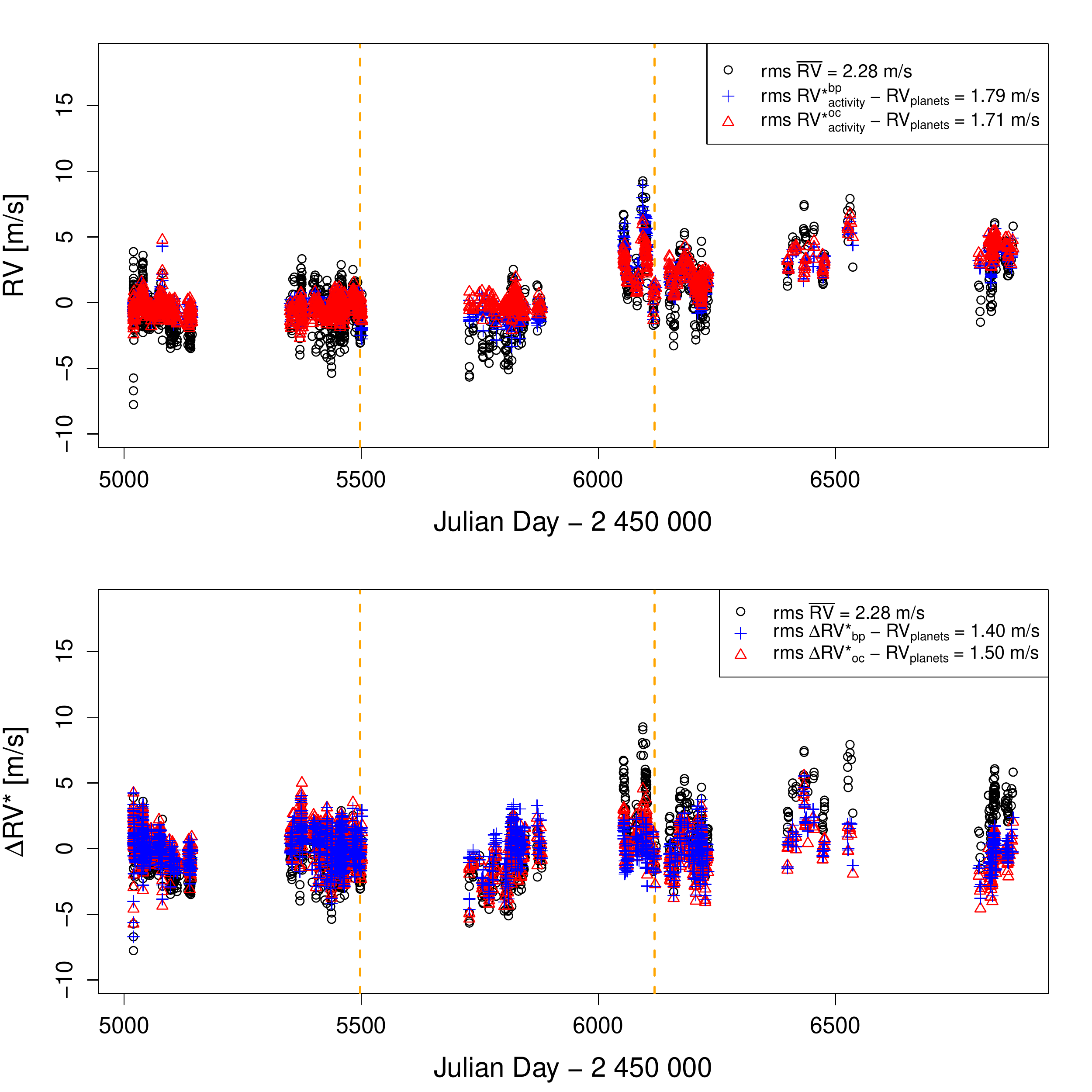}
\caption{Similar to Fig. \ref{fig:alphacentb}, but for HD\,192310. We recall that the two planetary signals were removed from our analyses after the {bp} method and the {oc} method were used, in order to better focus on the residuals of the two methods.}
\label{fig:hd192310}
\end{figure*}

\end{appendix}

\end{document}